\documentclass[12pt,A4paper,twoside]{article}
\usepackage{graphicx}
\usepackage{epsfig}
\usepackage{cite}
\usepackage{slashed}
\usepackage{bbold}
\usepackage{tabls}
\usepackage{textcomp}
\usepackage{amsmath}
\usepackage{amssymb}
\usepackage{physics}
\usepackage{extarrows}
\usepackage{stackengine}
\usepackage[compat=1.1.0]{tikz-feynman}
\usepackage{breqn}
\usepackage{doi}
\usepackage{array}
\usepackage[export]{adjustbox}

\makeatletter
\AtBeginDocument{
    \catcode`_=12
    \begingroup\lccode`~=`_
    \lowercase{\endgroup\let~}\sb
    \mathcode`_="8000
    \immediate\write\@auxout{\catcode`_=12 }
    \immediate\write\@auxout{\catcode`^=12 }
}
\makeatother
\makeatletter
\patchcmd\eq@setnumber{\stepcounter}{\refstepcounter}{}{%
  \errmessage{Patching \noexpand\eq@setnumber failed}%
}
\makeatother

\usetikzlibrary{external}
\immediate\write18{mkdir -p pgf-img}
\usepackage{hyperref}

\newcommand\brabar{\scalebox{.3}{(}\raisebox{-1.7pt}{--}\scalebox{.3}{)}}

\begin{document}
\thispagestyle{empty}
\begin{flushright}
 CERN-TH-2019-060
\end{flushright}
\vskip1truecm
\centerline{\LARGE{\textbf{Effective Theories for Quark Flavour Physics}} 
   \footnote{\noindent
Lectures given at the Les Houches summer school: EFT in Particle Physics and Cosmology,
3-28 July 2017, Les Houches, France. 
   }}
\vskip1.5truecm
\centerline{\large\textbf{Luca Silvestrini}}
\vskip0.5truecm
\centerline{\sl INFN, Sez. di Roma}
\centerline{\sl P.le A. Moro, 2, I-00185 Roma, Italy} 
\centerline{and}
\centerline{\sl Theoretical Physics Department, CERN}
\centerline{\sl 1211 Geneva 23, Switzerland}
\vskip1truecm
\centerline{\bf Abstract}

The purpose of these lectures is to provide the reader with an idea of
how we can probe New Physics with quark flavour observables using effective
theory techniques. After giving a concise review of the quark flavour
structure of the Standard Model, we introduce the effective
Hamiltonian for quark weak decays. We then consider the effective
Hamiltonian for $\Delta F=2$
transitions in the Standard Model and beyond. We discuss how
meson-antimeson mixing and CP violation can be described in terms of
the $\Delta F=1$ and $\Delta F=2$ effective Hamiltonians. Finally we
present the Unitarity Triangle Analysis and discuss how very stringent
constraints on New Physics can be obtained from $\Delta F=2$ processes.


\newpage

\thispagestyle{empty}

\mbox{}

\newpage

\pagenumbering{roman}

\tableofcontents

\newpage

\pagenumbering{arabic}

\setcounter{page}{1}

\section{Introduction}

\begin{figure}[htbp]
\begin{center}      
  \includegraphics[width=0.9\textwidth]{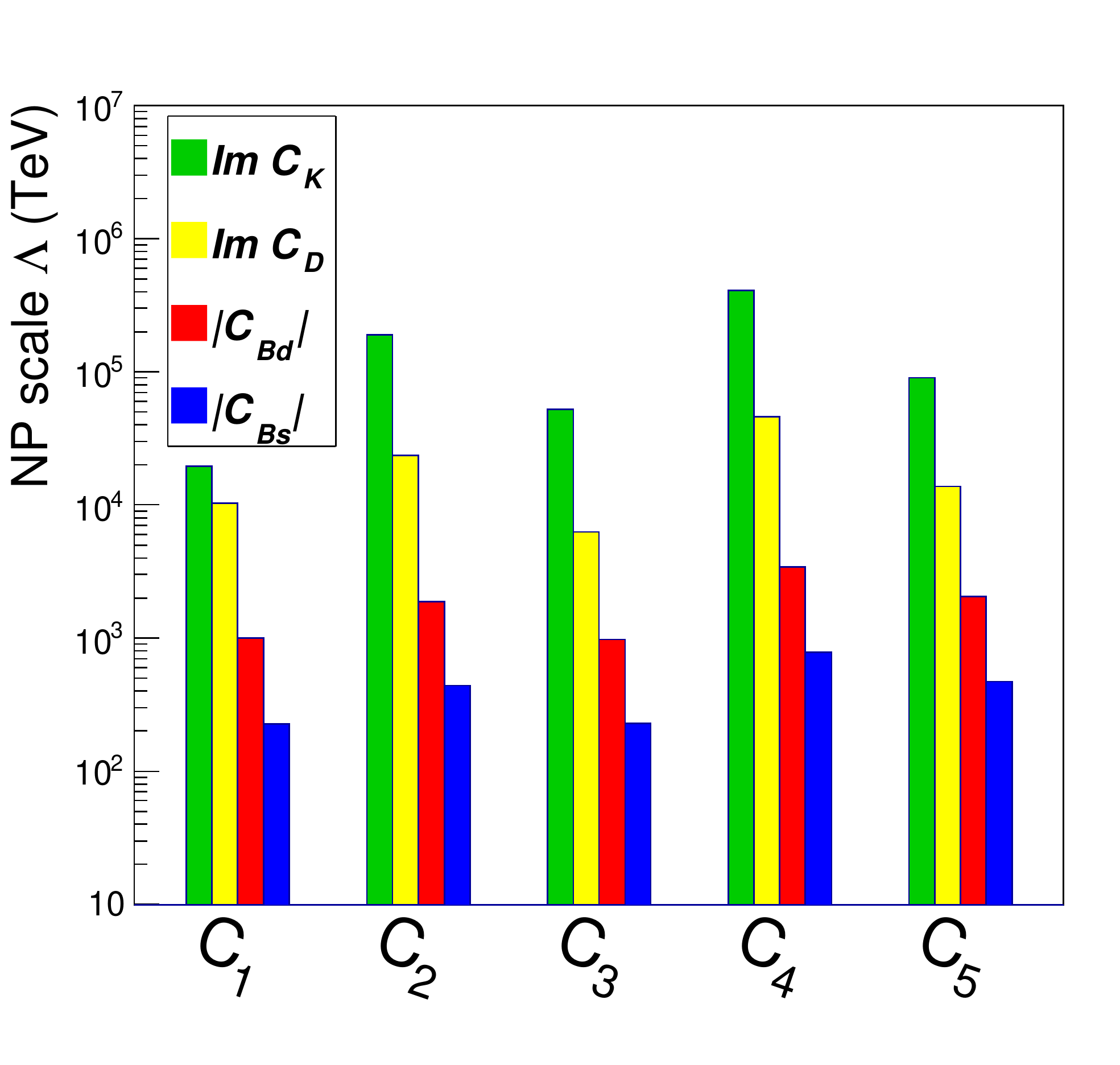}
\end{center}
\caption{Summary of the $95\%$ probability lower bound on the NP
    scale $\Lambda$. See the text for details}
\label{fig:df2bounds} 
\end{figure}

Quark flavour physics is among the most powerful probes of New Physics
(NP) beyond the Standard Model (SM) of electroweak and strong
interactions. The sensitivity to NP in the flavour sector stems from a
few peculiarities of the SM: first of all, the absence of Flavour
Changing Neutral Currents (FCNC) at the tree level, which makes FCNC
processes finite and therefore predictable; second, the
Glashow-Iliopoulos-Maiani (GIM) suppression at the loop
level~\cite{Glashow:1970gm}; third, the hierarchical structure of
quark masses and mixing angles, resulting in the smallness of Jarlskog
commutator~\cite{Jarlskog:1985ht}. Thanks to these suppression
factors, NP contributions to FCNC processes generated by the exchange
of heavy new particles can compete with SM amplitudes, leading to
stringent bounds on the NP mass scale. As an example, in
Fig.~\ref{fig:df2bounds} we report the bounds on the NP scale
$\Lambda$ obtained from $\Delta F=2$ processes (\emph{i.e.} FCNC
$\bar{q}_{i} q_{j} \leftrightarrows q_{i} \bar{q}_{j}$ transitions),
assuming NP contributes at tree level with coupling equal to one in
all possible chiral structures. We will return to this plot at the end
of these lectures, after working out the basic ingredients of the
phenomenological analysis leading to these results; we can however
already see that, under the above assumptions, scales up to
$\mathcal{O}(10^{5})$ TeV can be probed, demonstrating the
extraordinary NP sensitivity of FCNC processes.

We have stated above that FCNC processes are calculable in the SM, in
the sense that a prediction can be obtained (at least in principle)
once all the parameters in the SM are known. However, in practice the
computation of FCNC processes in the quark sector is in general a very
complicated problem, for several reasons. First of all, what can be
measured are transitions between a hadronic initial state (for example
a $K$, $D_{(s)}$, $B_{(s)}$ meson or a baryon) and a leptonic,
semileptonic or nonleptonic final state. Thus, nonperturbative QCD
effects connected to quark confinement are always involved, at least
in the form of meson decay constants, form factors or other hadronic
matrix elements. Furthermore, for nonleptonic final states we must
include final state interactions, another very difficult
task. Finally, the energy scales involved span several orders of
magnitude, from the strong interaction scale $\Lambda_{\mathrm{QCD}}$
to the weak interaction scale $M_W$ to even larger energies if NP is
involved. Effective theories are then the best tool to cope with such
multi-scale processes, allowing for a systematic expansion in small
ratios of widely different scales, and providing the scale separation
needed to disentangle perturbative and nonperturbative strong
interaction effects.

The absence of tree-level FCNC in the SM implies that NP contributions
to FCNC transitions must appear as higher-dimensional operators
suppressed by the NP scale $\Lambda$. If the NP scale is much larger
than the weak scale, then these higher-dimensional operators will be
invariant under the SM gauge group, leading to the so-called Standard
Model Effective Theory (SMEFT) (see the lectures by A. Manohar
\cite{Manohar:2018aog} and A. Pich \cite{Pich:2018ltt} at this school
for a detailed discussion of the SMEFT). Indeed, the bounds presented
in Fig.~\ref{fig:df2bounds} can be interpreted as bounds on the
coefficients of SMEFT operators \cite{Silvestrini:2018dos}.

The impressive bounds on the NP scale reported in
Fig.~\ref{fig:df2bounds} correspond to a generic flavour
structure. This suggests that any NP close to the EW scale must have a
flavour structure either identical or very similar to the SM one. This
observation leads to the formulation of effective theories based on
the hypothesis of Minimal Flavour Violation (MFV)
\cite{DAmbrosio:2002vsn} or on approximate flavour symmetries
\cite{Barbieri:1995uv,Barbieri:1997tu,Roberts:2001zy,Dudas:2013pja,Linster:2018avp},
which can be viable at scales within the LHC reach.

The goal of these lectures is to give the reader a basic idea of how
the stringent bounds in Fig.~\ref{fig:df2bounds} are obtained. After
giving a concise review of the flavour structure of the SM in
Sec.~\ref{sec:SMFlavour}, we introduce in Sec.~\ref{sec:Heff} the
effective Hamiltonian for quark weak decays. We then consider
$\Delta F=2$ processes in Sec.~\ref{sec:DF2}, generalising to the case
of NP. In Sec.~\ref{sec:mixCP} we discuss how meson-antimeson mixing
and CP violation can be described in terms of the effective
Hamiltonians introduced in the previous Sections. Finally, we put
everything together in the context of the Unitarity Triangle Analysis
in Sec.~\ref{sec:UTA}. Sec.~\ref{sec:concl} contains suggestions for
further reading. A few useful formul{\ae} are collected in Appendix
\ref{sec:loops}.

\section{The flavour structure of the Standard Model}
\label{sec:SMFlavour}

To set the stage for our discussion, and to fix the notation, let us
quickly review the flavour structure of the SM. The SM is described by
the most general renormalizable
$\mathit{SU}(3)_c \otimes \mathit{SU}(2)_L \otimes U(1)_Y$
gauge-invariant Lagrangian involving three generations of leptons and
quarks and one Higgs doublet:\footnote{We neglect the QCD $\theta$
  term since it is irrelevant for our discussion.}
\begin{dgroup*}
  \begin{dmath}[label={eq:LSM}]
    \mathcal{L}_{\mathrm{SM}}= \mathcal{L}_{\mathrm{gauge}} +
    \mathcal{L}_{\mathrm{fermionic}} + \mathcal{L}_{\mathrm{Higgs}} +
    \mathcal{L}_{\mathrm{Yukawa}} \,,
  \end{dmath}
  \begin{dmath}[label={eq:Lgauge}]
    \mathcal{L}_{\mathrm{gauge}} = \frac{1}{4} G_{\mu\nu}^{a}
    G^{a\mu\nu} + \frac{1}{4} W_{\mu\nu}^{\alpha} W^{\alpha\mu\nu} +
    \frac{1}{4} B_{\mu\nu} B_{\mu\nu}\,,
  \end{dmath}
  \begin{dmath}[label={eq:Lfermionic}]
    \mathcal{L}_{\mathrm{fermionic}} = \sum_{f} \overline{\psi}_f i
    D^{\mu}\gamma_{\mu} \psi_f\,,
  \end{dmath}
  \begin{dmath}[label={eq:LHiggs}]
    \mathcal{L}_{\mathrm{Higgs}} = \left( D_{\mu} \phi \right)^{\dagger}
    \left( D^{\mu}\phi \right) + \mu^{2} \phi^{\dagger} \phi -
    \frac{\lambda}{4} \left( \phi^{\dagger} \phi \right)^{2} \,,
  \end{dmath}
  \begin{dmath}[label={eq:LY}]
    \mathcal{L}_{\mathrm{Yukawa}} = Y^{u}_{ij}\overline{Q}_L^{i}u_R^{j}\phi +
    Y^{d}_{ij}\overline{Q}_L^{i}d_R^{j} \tilde{\phi} + \mathrm{H.c.} + \ldots
  \end{dmath}
\end{dgroup*}
with $a = 1, \ldots, 8$ and $\alpha = 1,2,3$ indices in the adjoint
representation of $\mathit{SU}(3)_c$ and $\mathit{SU}(2)_L$
respectively, $f = \{Q_L^{i}, u_R^{i}, d_R^{i}, L_L^{i}, \ell_R^{i}\}$, $i$ and
$j$ generation indices and the ellipse in the last equation denotes
lepton Yukawa couplings. $Q_L^{i}$, $u_R^{i}$, $d_R^{i}$, $L_L^{i}$ and
$\ell_R^{i}$ represent left-handed $\mathit{SU}(2)_L$ quark doublets,
right-handed up- and down-quarks, left-handed lepton
$\mathit{SU}(2)_L$ doublets and right-handed charged leptons
respectively. $\phi$ denotes the Higgs boson doublet, with
$\tilde{\phi}^{i} = \epsilon^{ij} \phi^{*}_j$.  $G_{\mu\nu}$, $W_{\mu\nu}$
and $B_{\mu\nu}$ represent the field strength tensors for
$\mathit{SU}(3)_c$, $\mathit{SU}(2)_L$ and $U(1)_Y$ respectively.

Let us now focus on the flavour quantum numbers. The first three terms
in eq.~(\ref{eq:LSM}) are invariant under global
$U(3)_{Q_{L}}\otimes U(3)_{u_{R}} \otimes U(3)_{d_{R}} \otimes U(3)_{L_{L}} \otimes
U(3)_{\ell_{R}}$ transformations acting on generation indices. From now on,
we concentrate on quarks. The Yukawa couplings break the
$U(3)_{Q_{L}}\otimes U(3)_{u_{R}} \otimes U(3)_{d_{R}}$ symmetry to $U(1)_{B}$,
corresponding to baryon number conservation, an accidental symmetry of
the SM. The top Yukawa coupling provides an $\mathcal{O}(1)$ breaking
of the $U(3)^{3}$ flavour symmetry in the quark sector, while an
approximate $U(2)^{3}$ symmetry remains valid up to terms of
$\mathcal{O}(Y_c)\sim 10^{-2}$. Due to the $U(3)^{3}$ invariance of
$\mathcal{L}_{\mathrm{gauge}}+\mathcal{L}_{\mathrm{fermionic}}+
\mathcal{L}_{\mathrm{Higgs}}$, the SM Yukawa couplings are defined up to
an $SU(3)^{3} \otimes U(1)^{2}$ transformation (they are invariant under
$U(1)_B$ transformations), which allows to eliminate nine real
parameters and seventeen phases from $Y^{u,d}$, leaving us with nine
observable real parameters and one phase. Since the Lagrangian in
eq.~(\ref{eq:LSM}) is CP-invariant only for real Yukawa couplings, we
see that the observable phase in the Yukawa couplings is responsible
for CP violation in weak interactions. For two generations of
fermions, the Yukawa couplings would contain $8-(9-1)=0$ observable
phases, leading to CP conservation. Thus, the presence of three
generations of fermions is crucial to allow for CP violation in weak
interactions \cite{Kobayashi:1973fv}. This strongly restricts the
number of processes in which we can observe CP violation in weak
interactions: CP violation can occur only in processes where all the
three generations are involved, either as interfering real states or
as virtual ones.

\subsection{The Cabibbo-Kobayashi-Maskawa mixing matrix}
\label{sec:ckm}

Let us now take into account electroweak symmetry breaking induced by
the vacuum expectation value (vev) of the neutral component of the
Higgs doublet $\phi$:
\begin{equation}
  \label{eq:hvev}
  \langle\phi\rangle = \frac{1}{\sqrt{2}} 
  \left(
    \begin{array}{c}
      v \\
      0
    \end{array}
  \right)\,.
\end{equation}
In the SM, the Higgs vev generates masses for the $W^{\pm}$ and $Z^{0}$ bosons
through electroweak interactions as well as for the fermions through
Yukawa interactions. For the latter we obtain
\begin{dmath}[label={eq:fmassesnd}]
  \mathcal{L}_{m} = m^{u}_{ij}\overline{u}^{i}_{L}u_{R}^{j} + m^{d}_{ij}\overline{d}^{i}_{L}d_{R}^{j}
  + H.c.
\end{dmath}
with $m^{u,d}_{ij}\equiv Y^{u,d}_{ij} v/\sqrt{2}$. The complex mass
matrices $m^{u,d}_{ij}$ can be brought to diagonal form via a
biunitary transformation:
\begin{dgroup*}
  \begin{dmath}[label={eq:UUdrot},compact]
    U_{u_L} m^{u} m^{u\dagger} U_{u_L}^{\dagger} = U_{u_R} m^{u\dagger}
    m^{u} U_{u_R}^{\dagger} = (m^{u}_D)^{2}\,,
  \end{dmath}
  \begin{dmath}[label={eq:Urot}]
    U_{u_L} m^{u} U_{u_R}^{\dagger} = m^{u}_D\,,
  \end{dmath}
  \begin{dmath}[label={eq:DDdrot},compact]
    U_{d_L} m^{d} m^{d\dagger} U_{d_L}^{\dagger} = U_{d_R} m^{d\dagger}
    m^{d} U_{d_R}^{\dagger} = (m^{d}_D)^{2}\,,
  \end{dmath}
  \begin{dmath}[label={eq:Drot}]
    U_{d_L} m^{d} U_{d_R}^{\dagger} = m^{d}_D\,,
  \end{dmath}
\end{dgroup*}
with $m^{d}_D$ ($m^{u}_D$) a diagonal matrix with the masses of down,
strange and bottom (up, charm and top) quarks on the diagonal. We can
go to the mass eigenstate basis for quarks defining
\begin{equation}
  \label{eq:massrot}
  u^{\prime}_{L,R} = U_{u_{L,R}} u_{L,R}\,, \qquad
  d^{\prime}_{L,R} = U_{d_{L,R}} d_{L,R}\,. 
\end{equation}
Given the $U(3)_{u_R} \otimes U(3)_{d_R}$ invariance of
$\mathcal{L}_{\mathrm{gauge}}+\mathcal{L}_{\mathrm{fermionic}}+
\mathcal{L}_{\mathrm{Higgs}}$,
switching from unprimed to primed right-handed quarks has no
effect. For left-handed fermions, it is convenient to rewrite the
transformations in eq.~(\ref{eq:massrot}) in the form of a
transformation on $Q_L$ followed by an additional transformation on
$u_L$:
\begin{equation}
  \label{eq:massrotinv}
  d^{\prime}_{L} = U_{d_{L}} d_{L} \,, \qquad
  u^{\prime}_{L} = V U_{d_{L}} u_{L}\,,
\end{equation}
where $V \equiv U_{u_L}^{\phantom{*}} U_{d_L}^{\dagger}$ is the
Cabibbo-Kobayashi-Maskawa (CKM) matrix
\cite{Cabibbo:1963yz,Kobayashi:1973fv}.

The CKM matrix can be parameterized in terms of three angles
and one phase, as in the so-called ``standard'' parameterization:
\begin{equation}
  \label{eq:CKMstandard}
  V=\left(
    \begin{array}{ccc}
      c_{12}c_{13} & s_{12}c_{13} & s_{13}e^{-i\delta}\\
      -s_{12}c_{23} -c_{12}s_{23}s_{13}e^{i\delta}
                   & c_{12}c_{23} -s_{12}s_{23}s_{13}e^{i\delta}
                                  & s_{23}c_{13} \\
      s_{12}s_{23} -c_{12}c_{23}s_{13}e^{i\delta}
                   & -c_{12}s_{23} -s_{12}c_{23}s_{13}e^{i\delta}
                                  & c_{23}c_{13} 
    \end{array}
\right)\,,
\end{equation}
where we have introduced the shorthand notation $s_{ij} =
\sin(\theta_{ij})$, $c_{ij} = \cos(\theta_{ij})$.

Given that $s_{13}\ll s_{23} \ll s_{12} \ll 1$, a
perturbative expansion in powers of the sine of the Cabibbo angle $s_{12}$ can be performed
\cite{Wolfenstein:1983yz}, defining
\begin{equation}
  \label{eq:Wolfenstein-Buras}
  \lambda \equiv s_{12}\,, \qquad A \equiv s_{23}/\lambda^{2}\,,\qquad
  \left( \rho + i \eta\right) \equiv s_{13} e^{i \delta}/(A \lambda^{3})
\end{equation}
and imposing the unitarity constraint at the desired order
\cite{Buras:1994ec}. In particular, expanding all matrix elements up
to $\mathcal{O}(\lambda^{5})$, one obtains
\begin{displaymath}
  \label{eq:CKMWB}
  V=\left(
    \begin{array}{ccc}
      1-\frac{\lambda^{2}}{2} -\frac{\lambda^{4}}{8} & \lambda
      & A \lambda^{3} \left(\rho - i \eta\right)\\
      -\lambda + A^{2} \lambda^{5} \left(\frac{1}{2} - \rho - i \eta \right)
                   & 1-\frac{\lambda^{2}}{2} -\frac{\lambda^{4} \left(1 +
                     4 A^{2}\right) }{8} 
                                  & A \lambda^{2} \\
      A \lambda^{3} \left(1 - \overline{\rho} - i \overline{\eta}\right)
                   & -A \lambda^{2} \left(1 - \frac{\lambda^{2}}{2}\right) - A
                     \lambda^{4} \left(\rho + i \eta\right)
                                  & 1 - \frac{A^{2} \lambda^{4}}{2}
    \end{array}
\right)\,,
\end{displaymath}
with
\begin{equation}
  \label{eq:rhobaretabar}
  \overline{\rho} = \rho 
  \left(
    1 - \frac{\lambda^{2}}{2}
  \right)\,, \qquad \overline{\eta} = \eta 
  \left(
    1 - \frac{\lambda^{2}}{2}
  \right)\,.
\end{equation}

\begin{figure}[ht]
  \centering
  \includegraphics[width=0.7\textwidth]{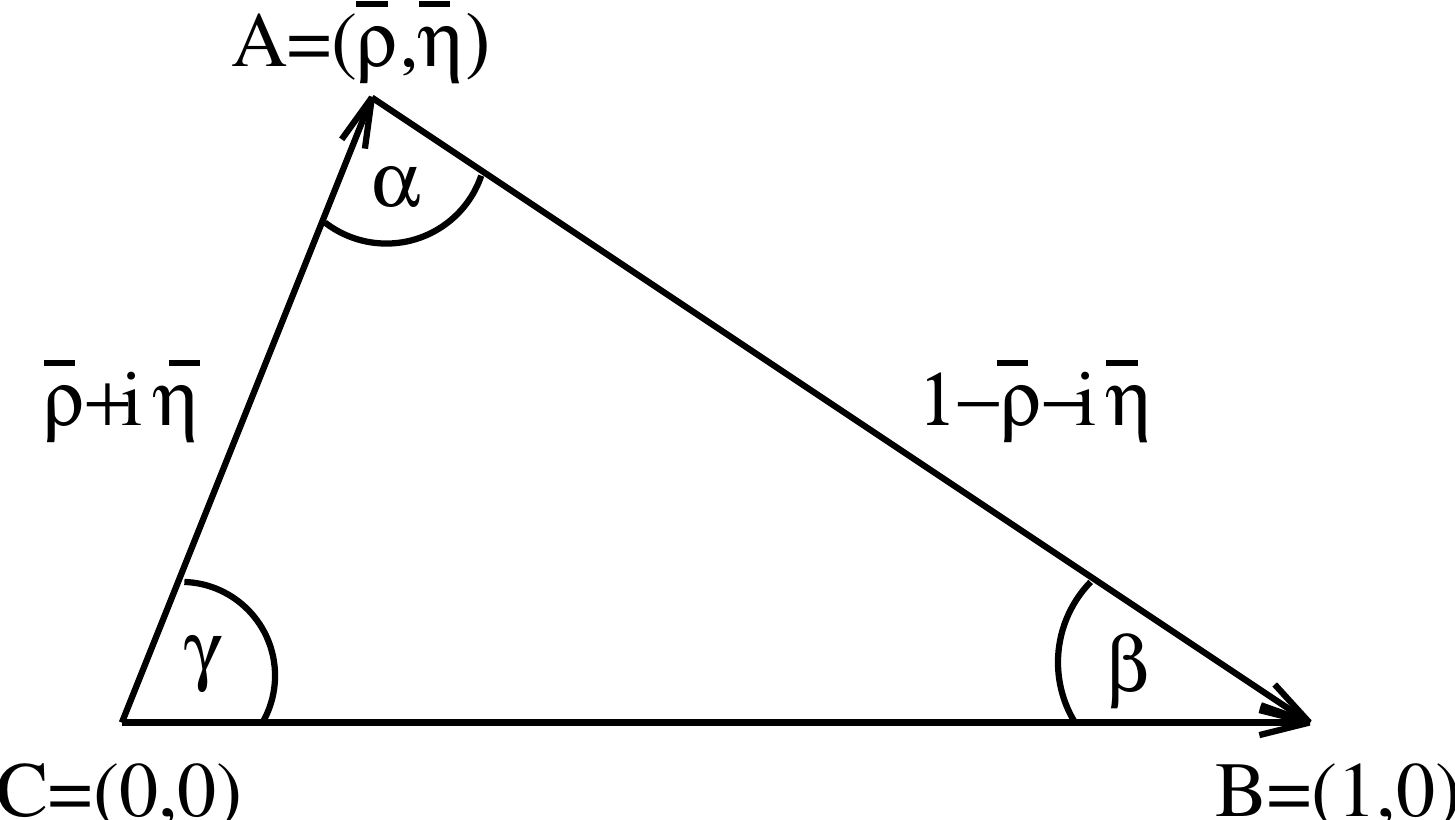}
  \caption{The Unitarity Triangle.}
  \label{fig:UT}
\end{figure}

The unitarity of the CKM matrix implies triangular relations, which
however involve sides of very different lengths, except for the
ones corresponding to transitions between the first and third
families, namely:
\begin{align}
  \label{eq:Unitarity}
 & V^{\phantom{*}}_{ud}V_{ub}^{*} + V^{\phantom{*}}_{cd}V_{cb}^{*} +
   V^{\phantom{*}}_{td}V_{tb}^{*} =0\,, \\
 & V^{\phantom{*}}_{ud}V_{td}^{*} + V^{\phantom{*}}_{us}V_{ts}^{*} +
   V^{\phantom{*}}_{ub}V_{tb}^{*} =0\,,
   \label{eq:uselessut}
\end{align}
where all sides are of $\mathcal{O}(\lambda^{3})$. Let us focus on the
relation in eq~(\ref{eq:Unitarity}) and divide it by the last term,
defining the so-called Unitarity Triangle (UT):
\begin{equation}
  \label{eq:UT}
  -\frac{ V^{\phantom{*}}_{ud}V_{ub}^{*}}{V^{\phantom{*}}_{cd}V_{cb}^{*}} -
  \frac{V^{\phantom{*}}_{td}V_{tb}^{*}}{V^{\phantom{*}}_{cd}V_{cb}^{*}}  = R_b e^{i\gamma} + R_t e^{-i
    \beta} = 1 \simeq 
  \left(
    \overline{\rho} + i \overline{\eta}
  \right) + \left(
    1 - \overline{\rho} - i \overline{\eta}
  \right) \,, 
\end{equation}
where
\begin{equation}
  \label{eq:UTdef}
  R_b \equiv 
  \left\vert
    \frac{V^{\phantom{*}}_{ud}V_{ub}^{*}}{V^{\phantom{*}}_{cd}V_{cb}^{*}}
  \right\vert, \quad
  R_t \equiv 
  \left\vert
    \frac{V^{\phantom{*}}_{td}V_{tb}^{*}}{V^{\phantom{*}}_{cd}V_{cb}^{*}}
  \right\vert, \quad
  \gamma \equiv 
  \mathrm{arg}\left(
    -\frac{V^{\phantom{*}}_{ud}V_{ub}^{*}}{V^{\phantom{*}}_{cd}V_{cb}^{*}}
  \right), \quad
  \beta \equiv 
  \mathrm{arg}\left(
    -\frac{V^{\phantom{*}}_{cd}V_{cb}^{*}}{V^{\phantom{*}}_{td}V_{tb}^{*}}
  \right).
\end{equation}
The UT can then be represented as a triangle in the complex
$ \left( \overline{\rho}, \overline{\eta} \right)$ plane, see
Fig.~\ref{fig:UT}. It is useful to define also
\begin{equation}
  \label{eq:angdef}
  \alpha \equiv 
  \mathrm{arg}\left(
    -\frac{V^{\phantom{*}}_{td}V_{tb}^{*}}{V^{\phantom{*}}_{ud}V_{ub}^{*}}
  \right)\,, \qquad
  \beta_s \equiv 
  \mathrm{arg}\left(
    -\frac{V^{\phantom{*}}_{ts}V_{tb}^{*}}{V^{\phantom{*}}_{cs}V_{cb}^{*}}
  \right)\,.
\end{equation}
The latter angle enters the ``squashed'' UT corresponding to the
$ \left( b,s \right) $ unitarity relation.

\subsection{Weak interactions below the EW scale}
\label{sec:weak}

After electroweak symmetry breaking, going to the quark mass
eigenstate basis through the transformations in
eq.~(\ref{eq:massrot}) and dropping primes for simplicity, we
are left with the following couplings of gauge
bosons with fermionic currents:
\begin{dgroup*}
  \begin{dmath}[label={eq:Lint}]
    \mathcal{L}_{\mathrm{int}} = -\frac{g_2}{\sqrt{2}} \left(W_{\mu}^{+}
      J^{\mu}_{\mathrm{ch}} + W_{\mu}^{-} J^{\mu^{\dagger}}_{\mathrm{ch}}
    \right) - g_1 \cos \theta_W A_{\mu} J^{\mu}_{\mathrm{em}} -
    \frac{g_2}{\cos \theta_W} Z_{\mu} J^{\mu}_{Z}\,,
  \end{dmath}
  \begin{dmath}[label={eq:Jem}]
    J^{\mu}_{\mathrm{em}} = \sum_{f=\ell,u,d} Q_f \overline{f}_i \gamma^{\mu}
    f_i \,,
  \end{dmath}
  \begin{dmath}[label={eq:JZ}]
    J^{\mu}_{Z} = \sum_{f=\ell,u,d} \left(I^{3}_f - Q_f
      \sin^{2} \theta_W\right) \overline{f}_L^{i} \gamma^{\mu} f_L^{i} - Q_f \sin^{2}
    \theta_W \overline{f}_R^{i} \gamma^{\mu} f_R^{i} \,,
  \end{dmath}
  \begin{dmath}[label={eq:JW}]
    J^{\mu}_{\mathrm{ch}} = \overline{u}_L^{i} V^{ij}\gamma^{\mu} d_L^{j} +
    \overline{\nu}_L^{i} \gamma^{\mu} \ell_L^{i}\,,
  \end{dmath}
\end{dgroup*}
with $g_{1}$ and $g_{2}$ the $U(1)_{Y}$ and $SU(2)_{L}$ gauge
couplings respectively, $A_{\mu}$ the photon field, $Z_{\mu}$ the
$Z^{0}$ field, $\nu_{L}$ left-handed neutrinos, $\theta_W$ the weak
mixing angle, $e = g_1 \cos \theta_W = g_2 \sin \theta_W$,
$Q_{\ell} = -1$, $Q_u = 2/3$, $Q_d = -1/3$ and $I^{3}$ the third
component of weak isospin.  In the 't-Hooft-Feynman gauge, we have the
following Feynman rules:
\begin{align}
  \label{eq:Wphiprops}
  &  \includegraphics[width=0.18\linewidth, valign=c]{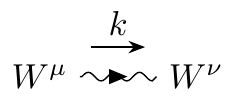} = \frac{-i
      g^{\mu\nu}}{k^{2}-M_W^{2}+i\epsilon}\,, \qquad \includegraphics[width=0.18\linewidth, valign=c]{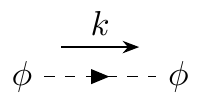} =
    \frac{i}{k^{2}-M_W^{2}+i\epsilon}\,, \\
  \label{eq:Wvertices}
  &   \includegraphics[width=0.18\linewidth, valign=c]{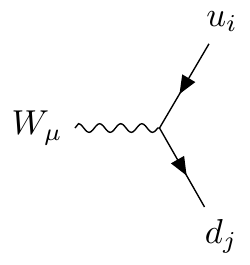}
   = \frac{i g_2}{\sqrt{2}} \gamma_{\mu} P_L V_{u_id_j}^{*}\,,\qquad
   \includegraphics[width=0.18\linewidth, valign=c]{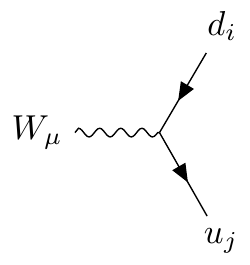}
   = \frac{i g_2}{\sqrt{2}} \gamma_{\mu} P_L V^{\phantom{*}}_{u_id_j}\,,
\end{align}
\begin{align}
 \label{eq:phivertices}
 &     \includegraphics[width=0.18\linewidth, valign=c]{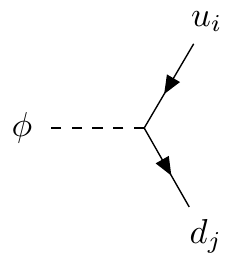}
   = \frac{-i g_2}{\sqrt{2}M_W} 
    \left[
    m_{d_j} P_L - m_{u_i} P_R
    \right] V_{u_id_j}^{*}\,,\\
 \label{eq:phivertices2}
 & \includegraphics[width=0.18\linewidth, valign=c]{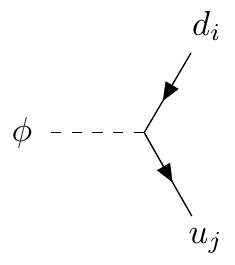}
   = \frac{-i g_2}{\sqrt{2} M_W} \left[
    m_{d_j} P_R - m_{u_i} P_L
  \right] V^{\phantom{*}}_{u_id_j}\,,
\end{align}
with $P_{L,R} = (1\mp\gamma_{5})/2$ the left- and right-handed chiral
projectors. 

Having fixed the notation, in the next Sections we introduce the
effective Hamiltonians relevant for quark flavour physics.

\section{Effective Hamiltonians for quark weak decays}
\label{sec:Heff}

Let us start by considering the amplitude for the
$u \bar{d} \to \nu_{\ell} \bar{\ell}$ transition, which is generated
at lowest order by the following Feynman diagrams:
\begin{equation}
\includegraphics[width=0.25\linewidth, valign=c]{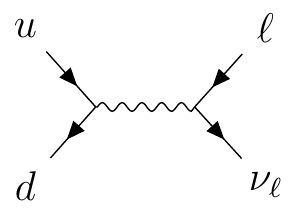} +
\includegraphics[width=0.25\linewidth, valign=c]{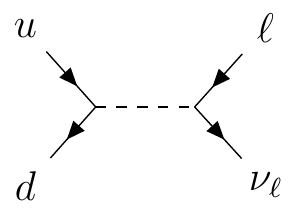}
\label{eq:semildiag}
\end{equation}

The Goldstone boson exchange can be neglected here since its couplings are
proportional to light fermion masses. The amplitude mediated by the
$W$ reads
\begin{equation}
  \label{eq:Wtreeampfull}
  i\mathcal{A}_W  =  \left(\frac{i g_2}{\sqrt{2}}\right)^{2} V_{ud}^{*}
  \left(\overline{u}_{\nu_{\ell}}(p_{\nu_{\ell}}) \gamma_{\nu} P_L v_\ell(p_\ell)\right)
  \left(\overline{v}_{d}(p_{d}) \gamma_\mu P_L u_u(p_u)\right)
                      \frac{-i g^{\mu\nu}}{k^{2}-M_W^{2}+i\epsilon}\,, 
\end{equation}
with $k=(p_u+p_d)=(p_\ell+p_\nu)$. Now, if we are interested in
low-energy processes such as pion leptonic or semileptonic decays, we
should consider external momenta of the order of the pion mass,
therefore much lower than $M_W$. Thus, we can perform an expansion of
the $W$ propagator in powers of the momentum $k$, leading to
\begin{dmath}
  \label{eq:Wtreeampexp}
  i\mathcal{A}_W = -i \frac{V_{ud}^{*} g_2^{2}}{2 M_W^{2}}
  \left(\overline{u}_{\nu_{\ell}}(p_{\nu_{\ell}}) \gamma^{\mu} P_L v_\ell(p_\ell)\right)
  \left(\overline{v}_{d}(p_{d}) \gamma_\mu P_L u_u(p_u)\right)
  \sum_{n=0}^{\infty} 
  \left(
    \frac{k^{2}}{M_W^{2}}
                      \right)^{n}  
  \simeq - i \frac{4 G_F}{\sqrt{2}} V_{ud}^{*}
  \left(\overline{u}_{\nu_\ell}(p_{\nu_\ell}) \gamma^{\mu} P_L v_\ell(p_\ell)\right)
  \left(\overline{v}_{d}(p_{d}) \gamma_\mu P_L u_u(p_u)\right) +
  \mathcal{O}
  \left(
    \frac{k^{2}}{M_W^{2}}
  \right)\,, 
\end{dmath}
where we have introduced the Fermi constant
\begin{equation}
  \label{eq:GF}
  \frac{G_F}{\sqrt{2}} \equiv \frac{g_2^{2}}{8 M_W^{2}}\,,
\end{equation}
with $G_{F} = 1.166 378 7(6) \cdot 10^{-5}$ GeV$^{-2}$
\cite{Tanabashi:2018oca}. The dominant term in
eq.~(\ref{eq:Wtreeampexp}) corresponds to the matrix element of the
following local operator:
\begin{equation}
  \label{eq:Wtreelocal}
  Q^{\overline{d}u\overline{\nu}_\ell \ell} \equiv \overline{d}_L
\gamma^{\mu} u_L \overline{\nu}_{\ell L} \gamma_\mu \ell_L\,,
\end{equation}
while the terms of order $n>0$ in the expansion in powers of $k^{2}/M_W^{2}$
correspond to the matrix elements of higher dimensional local operators
containing $2n$ derivatives. Keeping only the dimension six operator
in this Operator Product Expansion (OPE), we obtain
\begin{equation}
  \label{eq:Wtreematching}
  \mathcal{A}_W = \langle - \mathcal{H}_{\mathrm{eff}} \rangle
  + \mathcal{O}\left(
    \frac{k^{2}}{M_W^{2}}
  \right)\,,\qquad  \mathcal{H}_{\mathrm{eff}} = \frac{G_F}{\sqrt{2}}
  V_{ud}^{*} Q^{\overline{d}u\overline{\nu}_\ell \ell}\,,
\end{equation}
where we have introduced the effective Hamiltonian for
$\overline{d} u \to \overline{\ell} \nu_\ell$ transitions. The effects of the
exchange of the heavy $W$ boson are encoded in the so-called Wilson
coefficient, \emph{i.e.} the coefficient in front of the local
operator $Q ^{\overline{d}u\overline{\nu}_\ell \ell}$ in
$\mathcal{H}_{\mathrm{eff}}$. External momenta are irrelevant in the
matching between the full and effective theory performed in
eq.~(\ref{eq:Wtreematching}), since the dynamics
at scales much lower than $M_W$ is identical in the full and effective
theory, up to the desired order in the OPE.

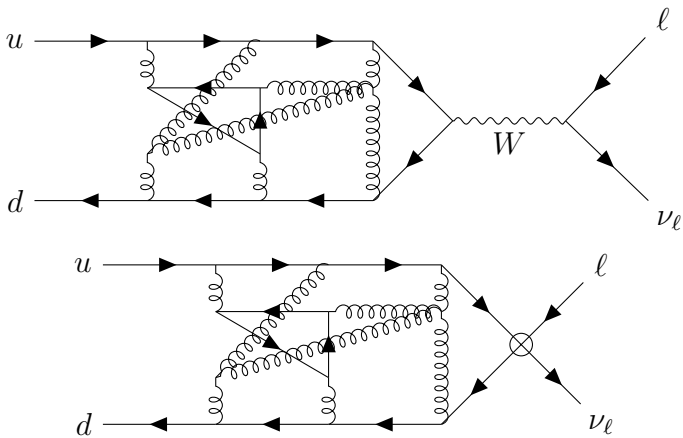
\begin{figure}[t]
  \centering
  \begin{tikzpicture}
  \begin{feynman}
    \vertex (a);
    \vertex [above left=of a] (a1);
    \vertex [left=of a1] (a2);
    \vertex [left=of a2] (a3);
    \vertex [left=of a3] (q1) {\(u\)};
    \vertex [below left=of a] (b1);
    \vertex [left=of b1] (b2);
    \vertex [left=of b2] (b3);    
    \vertex [left=of b3] (q2) {\(d\)};
    \vertex [right=of a] (b);
    \vertex [below=of a1] (c1);
    \vertex [below=of a2] (c2);
    \vertex [below=of a3] (c3);
    \vertex [above=of b1] (d1);
    \vertex [above=of b2] (d2);
    \vertex [above=of b3] (d3);
    \vertex [above right=of b] (l1) {\(\ell\)};
    \vertex [below right=of b] (l2) {\(\nu_{\ell}\)};
    \diagram* {
      (a) -- [anti fermion] (a1),
      (a1) -- [anti fermion] (a2),
      (a2) -- [anti fermion] (a3),
      (a3) -- [anti fermion] (q1),
      (a) -- [fermion] (b1),
      (b1) -- [fermion] (b2),
      (b2) -- [fermion] (b3),
      (b3) -- [fermion] (q2),
      (a) -- [boson, edge label'=\(W\)] (b),
      (b) -- [anti fermion] (l1),
      (b) -- [fermion] (l2),
      (a2) -- [gluon] (c3),
      (b2) -- [gluon] (c2),
      (d1) -- [gluon] (c3),
      (b1) -- [gluon] (d1),
      (d1) -- [gluon] (d2),
      (d1) -- [gluon] (a1),
      (b3) -- [gluon] (c3),
      (d2) -- [fermion] (d3),
      (d3) -- [fermion] (c2),
      (c2) -- [fermion] (d2),
      (d3) -- [gluon] (a3)
    };
  \end{feynman}
\end{tikzpicture}\\
\begin{tikzpicture}
  \begin{feynman}
    \vertex [crossed dot] (a) {};
    \vertex [above left=of a] (a1);
    \vertex [left=of a1] (a2);
    \vertex [left=of a2] (a3);
    \vertex [left=of a3] (q1) {\(u\)};
    \vertex [below left=of a] (b1);
    \vertex [left=of b1] (b2);
    \vertex [left=of b2] (b3);    
    \vertex [left=of b3] (q2) {\(d\)};
    \vertex [below=of a1] (c1);
    \vertex [below=of a2] (c2);
    \vertex [below=of a3] (c3);
    \vertex [above=of b1] (d1);
    \vertex [above=of b2] (d2);
    \vertex [above=of b3] (d3);
    \vertex [above right=of a] (l1) {\(\ell\)};
    \vertex [below right=of a] (l2) {\(\nu_{\ell}\)};
    \diagram* {
      (a) -- [anti fermion] (a1),
      (a1) -- [anti fermion] (a2),
      (a2) -- [anti fermion] (a3),
      (a3) -- [anti fermion] (q1),
      (a) -- [fermion] (b1),
      (b1) -- [fermion] (b2),
      (b2) -- [fermion] (b3),
      (b3) -- [fermion] (q2),
      (a) -- [anti fermion] (l1),
      (a) -- [fermion] (l2),
      (a2) -- [gluon] (c3),
      (b2) -- [gluon] (c2),
      (d1) -- [gluon] (c3),
      (b1) -- [gluon] (d1),
      (d1) -- [gluon] (d2),
      (d1) -- [gluon] (a1),
      (b3) -- [gluon] (c3),
      (d2) -- [fermion] (d3),
      (d3) -- [fermion] (c2),
      (c2) -- [fermion] (d2),
      (d3) -- [gluon] (a3)
    };
  \end{feynman}
\end{tikzpicture}
\caption{An example of strong interaction corrections to the diagram
  in (\ref{eq:semildiag}) (top) and its counterpart in the effective
  theory (bottom). The blob denotes the insertion of the four-fermion
  operator in eq.~(\ref{eq:Wtreelocal}).}
  \label{fig:semildiagstrong}
\end{figure}

Of course, we should now worry about the effects of strong
interactions. Given the low scale at which pion decays occur, we
cannot invoke any argument to suppress strong corrections to the
diagram in (\ref{eq:semildiag}) such as the one in the first row of
Fig.~\ref{fig:semildiagstrong}. However, such corrections are
identical in the full theory and in the effective one, \emph{i.e.} the
diagrams in the first and second row of Fig.~\ref{fig:semildiagstrong}
are identical. Therefore, in this example we do not need to take
strong corrections into account in the matching; all strong
interactions will be captured by the matrix element of
$Q^{\overline{d}u\overline{\nu}_\ell \ell}$ between the relevant
initial and final states.

\begin{figure}[htp]
  \centering
  \begin{tabular}{cc}
   \begin{tikzpicture}
   \begin{feynman}[small]
     \vertex (a);
     \vertex [above left=of a] (a1);
     \vertex [above left=of a1] (q1) {\(c\)};
     \vertex [below left=of a] (b1);
     \vertex [below left=of b1] (q2) {\(s\)};
     \vertex [right=of a] (b);
     \vertex [above right=of b] (c1);
     \vertex [below right=of b] (d1);
     \vertex [above right=of c1] (q3) {\(u\)};
     \vertex [below right=of d1] (q4) {\(d\)};
     \diagram* {
       (a) -- [plain] (a1),
       (a1) -- [anti fermion] (q1),
       (a) -- [plain] (b1),
       (b1) -- [fermion] (q2),
       (a) -- [boson, edge label'=\(W\)] (b),
       (b) -- [plain] (c1),
       (c1) -- [fermion] (q3),
       (b) -- [plain] (d1),
       (d1) -- [anti fermion] (q4)
     };
   \end{feynman}
 \end{tikzpicture} &  \begin{tikzpicture}
   \begin{feynman}[small]
     \vertex (a);
     \vertex [above left=of a] (a1);
     \vertex [above left=of a1] (q1) {\(c\)};
     \vertex [below left=of a] (b1);
     \vertex [below left=of b1] (q2) {\(s\)};
     \vertex [right=of a] (b);
     \vertex [above right=of b] (c1);
     \vertex [below right=of b] (d1);
     \vertex [above right=of c1] (q3) {\(u\)};
     \vertex [below right=of d1] (q4) {\(d\)};
     \diagram* {
       (a) -- [plain] (a1),
       (a1) -- [anti fermion] (q1),
       (a) -- [plain] (b1),
       (b1) -- [fermion] (q2),
       (a) -- [boson, edge label'=\(W\)] (b),
       (b) -- [plain] (c1),
       (c1) -- [fermion] (q3),
       (b) -- [plain] (d1),
       (d1) -- [anti fermion] (q4),
       (a1) -- [gluon] (b1)
     };
   \end{feynman}
 \end{tikzpicture} \\
     (a) & (b) \\
\begin{tikzpicture}
  \begin{feynman}[small]
    \vertex (a);
    \vertex [above left=of a] (a1);
    \vertex [above left=of a1] (q1) {\(c\)};
    \vertex [below left=of a] (b1);
    \vertex [below left=of b1] (q2) {\(s\)};
    \vertex [right=of a] (b);
    \vertex [above right=of b] (c1);
    \vertex [below right=of b] (d1);
    \vertex [above right=of c1] (q3) {\(u\)};
    \vertex [below right=of d1] (q4) {\(d\)};
    \diagram* {
      (a) -- [plain] (a1),
      (a1) -- [anti fermion] (q1),
      (a) -- [plain] (b1),
      (b1) -- [fermion] (q2),
      (a) -- [boson, edge label'=\(W\)] (b),
      (b) -- [plain] (c1),
      (c1) -- [fermion] (q3),
      (b) -- [plain] (d1),
      (d1) -- [anti fermion] (q4),
      (a1) -- [gluon] (c1)
    };
  \end{feynman}
\end{tikzpicture} &
\begin{tikzpicture}
  \begin{feynman}[small]
    \vertex (a);
    \vertex [above left=of a] (a1);
    \vertex [above left=of a1] (q1) {\(c\)};
    \vertex [below left=of a] (b1);
    \vertex [below left=of b1] (q2) {\(s\)};
    \vertex [right=of a] (b);
    \vertex [above right=of b] (c1);
    \vertex [below right=of b] (d1);
    \vertex [above right=of c1] (q3) {\(u\)};
    \vertex [below right=of d1] (q4) {\(d\)};
    \diagram* {
      (a) -- [plain] (a1),
      (a1) -- [anti fermion] (q1),
      (a) -- [plain] (b1),
      (b1) -- [fermion] (q2),
      (a) -- [boson, edge label'=\(W\)] (b),
      (b) -- [plain] (c1),
      (c1) -- [fermion] (q3),
      (b) -- [plain] (d1),
      (d1) -- [anti fermion] (q4),
      (a1) -- [gluon, half left] (d1)
    };
  \end{feynman}
  \end{tikzpicture} \\
    (c) & (d) \\
   \begin{tikzpicture}
   \begin{feynman}[small]
     \vertex [crossed dot] (a) {};
     \vertex [above left=of a] (a1);
     \vertex [above left=of a1] (q1) {\(c\)};
     \vertex [below left=of a] (b1);
     \vertex [below left=of b1] (q2) {\(s\)};
     \vertex [above right=of a] (c1);
     \vertex [below right=of a] (d1);
     \vertex [above right=of c1] (q3) {\(u\)};
     \vertex [below right=of d1] (q4) {\(d\)};
     \diagram* {
       (a) -- [plain] (a1),
       (a1) -- [anti fermion] (q1),
       (a) -- [plain] (b1),
       (b1) -- [fermion] (q2),
       (a) -- [plain] (c1),
       (c1) -- [fermion] (q3),
       (a) -- [plain] (d1),
       (d1) -- [anti fermion] (q4)
     };
   \end{feynman}
 \end{tikzpicture} &
   \begin{tikzpicture}
   \begin{feynman}[small]
     \vertex [crossed dot] (a) {};
     \vertex [above left=of a] (a1);
     \vertex [above left=of a1] (q1) {\(c\)};
     \vertex [below left=of a] (b1);
     \vertex [below left=of b1] (q2) {\(s\)};
     \vertex [above right=of a] (c1);
     \vertex [below right=of a] (d1);
     \vertex [above right=of c1] (q3) {\(u\)};
     \vertex [below right=of d1] (q4) {\(d\)};
     \diagram* {
       (a) -- [plain] (a1),
       (a1) -- [anti fermion] (q1),
       (a) -- [plain] (b1),
       (b1) -- [fermion] (q2),
       (a) -- [plain] (c1),
       (c1) -- [fermion] (q3),
       (a) -- [plain] (d1),
       (d1) -- [anti fermion] (q4),
       (a1) -- [gluon] (b1)
     };
   \end{feynman}
 \end{tikzpicture} \\
     (e) & (f) \\
   \begin{tikzpicture}
   \begin{feynman}[small]
     \vertex [crossed dot] (a) {};
     \vertex [above left=of a] (a1);
     \vertex [above left=of a1] (q1) {\(c\)};
     \vertex [below left=of a] (b1);
     \vertex [below left=of b1] (q2) {\(s\)};
     \vertex [above right=of a] (c1);
     \vertex [below right=of a] (d1);
     \vertex [above right=of c1] (q3) {\(u\)};
     \vertex [below right=of d1] (q4) {\(d\)};
     \diagram* {
       (a) -- [plain] (a1),
       (a1) -- [anti fermion] (q1),
       (a) -- [plain] (b1),
       (b1) -- [fermion] (q2),
       (a) -- [plain] (c1),
       (c1) -- [fermion] (q3),
       (a) -- [plain] (d1),
       (d1) -- [anti fermion] (q4),
      (a1) -- [gluon] (c1)
     };
   \end{feynman}
 \end{tikzpicture} &
   \begin{tikzpicture}
   \begin{feynman}[small]
     \vertex [crossed dot] (a) {};
     \vertex [above left=of a] (a1);
     \vertex [above left=of a1] (q1) {\(c\)};
     \vertex [below left=of a] (b1);
     \vertex [below left=of b1] (q2) {\(s\)};
     \vertex [above right=of a] (c1);
     \vertex [below right=of a] (d1);
     \vertex [above right=of c1] (q3) {\(u\)};
     \vertex [below right=of d1] (q4) {\(d\)};
     \diagram* {
       (a) -- [plain] (a1),
       (a1) -- [anti fermion] (q1),
       (a) -- [plain] (b1),
       (b1) -- [fermion] (q2),
       (a) -- [plain] (c1),
       (c1) -- [fermion] (q3),
       (a) -- [plain] (d1),
       (d1) -- [anti fermion] (q4),
      (a1) -- [gluon, half left] (d1)
     };
   \end{feynman}
 \end{tikzpicture} \\
    (g) & (h)
  \end{tabular}                    
\caption{Diagrams relevant for $c \to s u \overline{d}$ transitions in the
  full and effective theory, including leading order QCD
  corrections. See the text for details.}
\label{fig:dc1}
\end{figure}
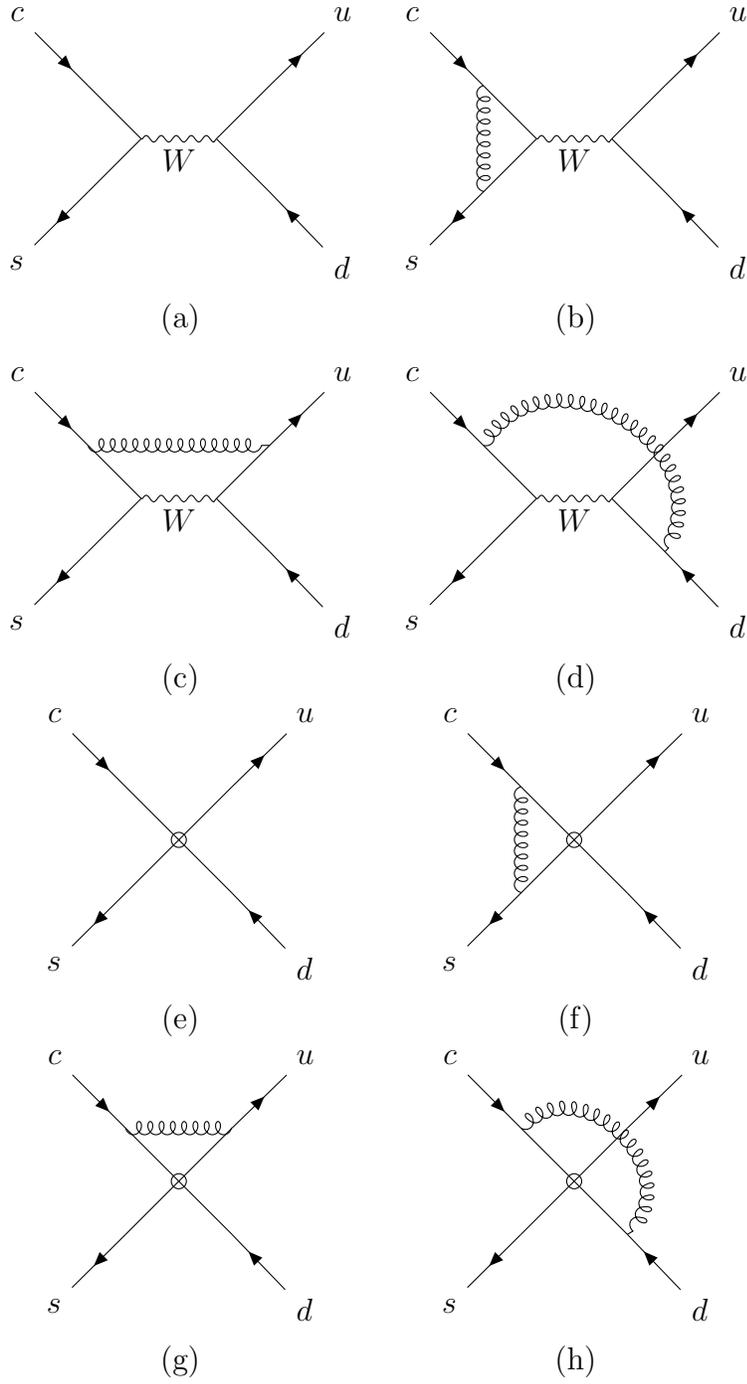

\subsection{Four-quark current-current operators}
\label{sec:current-current}

The situation changes dramatically if we now turn to nonleptonic
decays. Consider for example $c \overline{s}\to u \overline{d}$
transitions. Neglecting Goldstone boson exchange and QCD corrections,
in the SM these are described by diagram (a) in
Fig.~\ref{fig:dc1}. Just as in the case of $u \overline{d} \to
\nu_\ell \overline{\ell}$ transitions discussed above, since the energy scale at
which charm decays take place is much lower than $M_W$, the $W$ boson
will propagate over very short distances, so we can perform
an OPE and consider dimension six operators only. In this case, the
amplitude we obtain expanding diagram (a) at the lowest order in
$k^{2}/M_W^{2}$ is proportional to the one generated by diagram (e) with the
insertion of operator
\begin{equation}
  \label{eq:cdeclocal}
  Q_1^{\overline{s}c\overline{u}d} \equiv \overline{s}_L
\gamma^{\mu} c_L \overline{u}_{L} \gamma_\mu d_L\,.
\end{equation}
Imposing that the two amplitudes be equal,
\begin{equation}
  \label{eq:dc1treematching}
  \mathcal{A}_W = \langle - \mathcal{H}_{\mathrm{eff}} \rangle
  + \mathcal{O}\left(
    \frac{k^{2}}{M_W^{2}}
  \right)\,,
\end{equation}
we obtain the corresponding Wilson coefficient:
\begin{equation}
  \label{eq:Heffdc1tree}
  \mathcal{H}_{\mathrm{eff}} = \frac{4 G_F}{\sqrt{2}}
  V^{\phantom{*}}_{ud}V_{cs}^{*} C_1 Q_1^{\overline{s}c\overline{d} u}\,, \qquad C_1=1\,.
\end{equation}

As in the case of leptonic and semileptonic decays, the
corrections to the SM amplitude generated by the exchange of gluons
between the $c$ and $s$ quarks, such as the one in diagram (b) of
Fig.~\ref{fig:dc1}, are identical in the full and in the effective
theory, represented in this case by diagram (f) of Fig.~\ref{fig:dc1},
so they do not enter the matching. They will be taken into account
in the evaluation of the relevant hadronic matrix element of the
effective Hamiltonian. The same argument applies to the exchange of
gluons between $u$ and $d$ quarks, which has not been explicitly
reported in Fig.~\ref{fig:dc1}.

The situation is however totally different for the exchange of gluons
between the two currents coupled to the $W$ boson, such as in diagrams
(c) and (d). In these diagrams, the $W$ propagator has momentum
$k - \ell$, where $\ell$ is the loop momentum; in the region
$(k - \ell)^{2} \sim M_{W}^{2}$ the $W$ propagator opens up and it
falls as $\ell^{-2}$ for $\ell^{2} \gg M_{W}^{2}$, making the loop
integral convergent. Thus, $M_W$ acts as an ultraviolet regulator in
diagrams (c) and (d). Indeed, evaluating the amplitude explicitly,
putting the quarks off-shell with $p^{2}<0$ to avoid infrared
divergences, one finds a term proportional to
$\alpha_s \log \left( \frac{M_W^{2}}{-p^{2}} \right)$. Taken at face
value, such term implies the breakdown of perturbation theory, since
the effective expansion parameter
$\alpha_s \log \left( \frac{M_W^{2}}{-p^{2}} \right)$ becomes of
$\mathcal{O}(1)$ for quark momenta of $\order{\Lambda_{\mathrm{QCD}}}$,
due to the large logarithm. Fortunately, the effective theory can save
us from this disaster, as we shall see below.

\subsubsection{General considerations}
\label{sec:generalc}

The effective theory counterpart of diagrams (c) and (d) in
Fig.~\ref{fig:dc1} is given by diagrams (g) and (h). Having removed
the $W$ propagator in the effective theory, the latter diagrams are
divergent, so their ultraviolet behaviour is very different from the
corresponding SM diagrams. This is no surprise, since we worked out
the effective theory as an OPE expanding in powers of $k^{2}/M_W^{2}$, so
we expect it to be valid up to a cutoff $\Lambda$ of
$\mathcal{O}(M_W)$; above this cutoff the contribution of all higher
dimensional operators becomes unsuppressed and the expansion breaks
down. Regulating diagrams (g) and (h) with the introduction of a
cutoff $\Lambda$, one would obtain terms proportional to
$\alpha_s \log \left( \frac{\Lambda^{2}}{-p^{2}} \right) $. Since we have
seen that $M_W$ acts as a regulator in the SM amplitude, and since the
infrared logs of external momenta must be identical in the full and
effective theory, it is clear that the coefficients of the log terms
in the SM and in the effective theory are equal.

From the technical point of view, rather than introducing an explicit
cutoff, it is much more convenient to subtract the divergences and
work in the renormalized theory. In this case, the cutoff is removed
but a renormalization scale $\mu$ is introduced, so that after
renormalization $\log\left(
    \frac{\Lambda^{2}}{-p^{2}}
  \right)$ terms are replaced by $\log\left(
    \frac{\mu^{2}}{-p^{2}}
  \right)$.

We can match the amplitudes obtained
from diagrams (c) and (d) with the ones obtained from diagrams (g) and
(h) after subtracting the divergence in the effective theory via a
renormalization constant $Z$. The
infrared logs cancel and we are left with terms proportional to
\begin{equation}
  \alpha_s \log \left( \frac{M_W^{2}}{-p^{2}} \right) -
  \alpha_s \log \left( \frac{\mu^{2}}{-p^{2}} \right) =
  \alpha_s \log \left( \frac{M_W^{2}}{\mu^{2}} \right)\,.
  \label{eq:lmatching}
\end{equation}
Choosing a renormalization scale $\mu_W \sim M_W$, we can therefore
get rid of large logs in the matching procedure. In this way, we can
go from the full theory to the effective one using ordinary
perturbation theory. The Wilson coefficient obtained from the matching
now carries an explicit dependence on the renormalization scale
$\mu$, which cancels against the renormalization scale dependence of
the matrix element of the renormalized operator, since the amplitude
in the full theory does not depend on $\mu$:
\begin{equation}
  \label{eq:muindep}
  \mu \frac{\mathrm{d}}{\mathrm{d}\mu} \mathcal{A}_{\mathrm{full}} = 0 =
  \mu \frac{\mathrm{d}}{\mathrm{d}\mu} 
  \left(
    C(\mu) \langle
  Q^{\mathrm{ren}}(\mu) \rangle
\right) = \mu \frac{\mathrm{d}}{\mathrm{d}\mu} C(\mu) \langle
  Q^{\mathrm{ren}}(\mu) \rangle + C(\mu) \mu \frac{\mathrm{d}}{\mathrm{d}\mu} \langle
  Q^{\mathrm{ren}}(\mu) \rangle\,.
\end{equation}

Performing the matching at $\mu_W \sim M_W$ we got rid of large logs
in the matching procedure, but we actually just shifted them into the
effective theory. Computing the matrix element in the effective
theory, large logs of $ \left( \mu_W^{2}/-p^{2} \right)$ would arise
again, bringing us back into trouble. However, as discussed in detail
in M. Neubert's lectures at this School, the renormalization scale
dependence of a renormalized operator is governed by the
Renormalization Group Equations (RGE) in terms of its anomalous
dimension $\gamma_Q$ (which is nothing else but the coefficient of the
divergent terms, \emph{i.e.}  the coefficient of the
$\log \left( \frac{M_W^{2}}{-p^{2}} \right)$ terms in the full theory):
\begin{equation}
  \label{eq:rge1}
  \mu \frac{\mathrm{d}}{\mathrm{d}\mu} Q^{\mathrm{ren}} = - \gamma_Q Q^{\mathrm{ren}}\,,
  \qquad \gamma_Q = \frac{\mathrm{d} \log Z}{\mathrm{d} \log \mu}\,.
\end{equation}
Combining eqs.~(\ref{eq:muindep}) and (\ref{eq:rge1}) we obtain the
RGE for the renormalization scale dependence of the Wilson coefficient:
\begin{equation}
  \label{eq:rge2}
  \mu \frac{\mathrm{d}}{\mathrm{d}\mu} C(\mu) =  \gamma_Q C(\mu)\,,
\end{equation}
which allows us to obtain the Wilson coefficient for any
$\mu$, starting from its value at $\mu_W$:
\begin{equation}
  \label{eq:cmu}
  C(\mu) =  U(\mu,\mu_W) C(\mu_W)\,, \qquad U(\mu,\mu_W) =
  e^{\int_{g_s(\mu_W)}^{g_s(\mu)} \mathrm{d}g_s^{\prime
}    \frac{\gamma_Q(g_s^{\prime})}{\beta(g_s^{\prime})}}\,,
\end{equation}
where the $\beta$ function governs the running of the strong
coupling constant with the renormalization scale:
\begin{equation}
  \label{eq:beta}
  \beta(g_s) = - g_s \frac{\mathrm{d} \log Z_{g_{s}}}{\mathrm{d} \log
    \mu}\,,
\end{equation}
with $Z_{g_{s}}$ the renormalization constant of the $SU(3)_{c}$
coupling $g_{s}$.  We can now run down from $\mu_W \sim M_W$ to a low
renormalization scale $\mu_h$ close to the physical scale at which the
process we are interested in computing occurs, and then compute the
relevant matrix element (between an initial state $i$ with momenta
$p_i$ and a final state $f$ with momenta $p_f$) without encountering
large logs, since $\mu_h \sim p_i \sim p_f$:
\begin{equation}
  \label{eq:heffamp}
  \langle f(p_f) \lvert \mathcal{H}_{\mathrm{eff}} \rvert i(p_i) \rangle =
  C(\mu_h) \langle f(p_f) \lvert Q(\mu_h) \rvert i(p_i) \rangle\,.
\end{equation}
Where have the large logs gone? They have been resummed via the
renormalization group evolution! Thus, the effective theory allows us
to perform the matching using perturbation theory in the strong
interactions and to resum large logs using the RGE. The evaluation of
the relevant matrix elements of the local operators in
$\mathcal{H}_{\mathrm{eff}}$ can then be performed, if necessary (and
possible), with a nonperturbative method such as Lattice QCD.

The calculation of the $\beta$ function and of the anomalous
dimensions is particularly simple in mass-independent renormalization
schemes such as modified Minimal Subtraction
($\overline{\mathrm{MS}}$) \cite{tHooft:1973mfk,Bardeen:1978yd}. In
dimensional regularization, logarithmic divergences appear as
singularities as the number of space-time dimensions tends to four:
\begin{equation}
  \label{eq:cutofftoeps}
  \log 
  \left(
    \frac{\Lambda^{2}}{-p^{2}}
  \right) \Leftrightarrow \frac{1}{\overline{\epsilon}} +  \log 
  \left(
    \frac{\mu^{2}}{-p^{2}}
  \right)\,,
\end{equation}
where
\begin{dmath}[compact,label={eq:epsbar}]
\frac{1}{\overline{\epsilon}} = \frac{2}{4-D} - \gamma_E + \log(4\pi)
\end{dmath}
and $\mu$ is the renormalization scale. In the
$\overline{\mathrm{MS}}$ scheme we renormalize the operator by
subtracting the $\frac{1}{\overline{\epsilon}}$ divergence. Dropping
the bar for simplicity, and writing the renormalization constant $Z$
as a series in inverse powers of $\epsilon$,
\begin{equation}
  \label{eq:zeps}
  Z = 1 + \sum_{k} \frac{1}{\epsilon^{k}}Z_{k}(g_s)\,,
\end{equation}
and
\begin{equation}
  \label{eq:beps}
  \beta(g_s,\epsilon) = \frac{\mathrm{d} g_s(\mu)}{\mathrm{d} \log\mu}
  = -\epsilon g_s + \beta(g_s)\,,
\end{equation}
one obtains from eq.~(\ref{eq:rge1})
\begin{dmath}
  \gamma_Q (1 + \frac{1}{\epsilon}Z_{1}(g_s) + \ldots) =
  \frac{1}{\epsilon} \frac{\mathrm{d}Z_{1}}{\mathrm{d}\log
    \mu} + \ldots = \frac{1}{\epsilon} \frac{\mathrm{d}Z_{1}}{\mathrm{d}
    g_s} \frac{\mathrm{d} g_s}{\mathrm{d}\log
  \mu} + \ldots = 
  \frac{1}{\epsilon} \frac{\mathrm{d}Z_{1}}{\mathrm{d}
    g_s} ( -\epsilon g_s + \beta(g_s)) + \ldots\,,
  \label{eq:gammaeps}
\end{dmath}
where the ellipses denote higher terms in the $1/\epsilon$
expansion. The finiteness of $\gamma_Q$ implies
\begin{equation}
  \label{eq:gammaeps1}
  \gamma_Q = - 2 \alpha_s \frac{\mathrm{d}Z_{1}}{\mathrm{d} \alpha_s}\,,
\end{equation}
so the anomalous dimension is directly obtained from the
$\frac{1}{\epsilon}$ terms in the renormalization constant of the operator
$Q$.

If we are interested in the dominant, log-enhanced QCD corrections, we
can drop gluonic corrections to the matching (since no large logs arise at
$\mu_W \sim M_W$) and compute the anomalous dimension of the operators
in $\mathcal{H}_{\mathrm{eff}}$ at the first order in
$\alpha_s$. In general, if we expand the anomalous dimension matrix
and the Wilson coefficients in a series in $\alpha_s$,
\begin{dgroup*}
  \begin{dmath}[label={eq:alsseries}]
    C(\mu) = \sum_{n=0}^{n} \left( \frac{\alpha_s}{4\pi} \right)^{n}
    C^{(n)}(\mu)\,,
  \end{dmath}
  \begin{dmath}[label={eq:gammaseries}]
    \gamma = \sum_{n=0}^{n} \left(\frac{\alpha_s}{4\pi}\right)^{(n+1)}
    \gamma_{n}\,,
  \end{dmath}
\end{dgroup*}
we can classify the accuracy of the expansion in $\alpha_s$ as
follows.

\emph{A leading order (LO) calculation resums all terms of
  $\mathcal{O} \left( \alpha_s \log \left( \frac{M_W^{2}}{-p^{2}} \right)
  \right)^{n}$, by computing the anomalous dimensions at
  $\mathcal{O}(\alpha_s)$ and the matching and matrix elements
  neglecting $\alpha_s$ corrections.}\\
Expanding eqs.~(\ref{eq:cmu}), (\ref{eq:beta}) and (\ref{eq:heffamp})
we obtain
\begin{equation}
  \label{eq:LO}
  \mathcal{A}_{\mathrm{LO}} = C^{(0)}(\mu_h) \langle
  Q(\mu_h)\rangle^{(0)}\,,
\end{equation}
where $\langle
  Q(\mu_h)\rangle^{(n)}$ denotes a matrix element computed at
$n$-th order in strong interactions and
\begin{equation}
  \label{eq:LOevol}
  C^{(0)}(\mu_h) =
  U_0(\mu_h,\mu_W)  C^{(0)}(\mu_W)\,,\qquad U_0 = 
  \left(
    \frac{\alpha_s(\mu_W)}{\alpha_s(\mu_h)}
  \right)^{\frac{\gamma_0}{2 \beta_0}}\,.
\end{equation}
The LO evolutor $U_0$ resums all large logs. In general, as
discussed in M.~Neubert's lectures, QCD corrections induce mixing
among different operators, so that in general
$\mathcal{H}_{\mathrm{eff}}$ comprises several operators. The
equations above still apply, provided we consider $C$ and $Q$ as
vectors and $\gamma$ as a matrix; in this case, $U_0 = 
\left(
  \frac{\alpha_s(\mu_W)}{\alpha_s(\mu_h)}
\right)^{\frac{\gamma_0^{T}}{2 \beta_0}}$.

\emph{A (next-to-)$^{m}$leading order (N$^{m}$LO) calculation resums all
  terms of\\
  $\mathcal{O} \left( \alpha_s^{n+m} \log \left( \frac{M_W^{2}}{-p^{2}}
    \right)^{n} \right)$, by computing the anomalous dimensions at
  $\mathcal{O}(\alpha_s^{m+1})$ and the matching and matrix elements
  at $\mathcal{O}(\alpha_s^{m})$.}\\
For example, at NLO we resum all terms of
$\mathcal{O} \left( \alpha_s^{n+1} \log^{n} \left( \frac{M_W^{2}}{-p^{2}}
  \right) \right)$, by computing the anomalous dimension at
$\mathcal{O}(\alpha_s^{2})$ and the matching and matrix elements at
$\mathcal{O}(\alpha_s)$. Explicitly, we have
\begin{equation}
  \label{eq:NLO}
  \mathcal{A}_{\mathrm{NLO}} = C^{(0)}(\mu_h) \langle
  Q(\mu_h)\rangle^{(1)} + \frac{\alpha_s(\mu_h)}{4 \pi} C^{(1)}(\mu_h)  \langle
  Q(\mu_h)\rangle^{(0)}\,,
\end{equation}
where 
\begin{equation}
  \label{eq:NLOevol}
  C^{(1)}(\mu_h) = 
  U_0(\mu_h,\mu_W)  C^{(1)}(\mu_W) + 
  \left(
    J U_0(\mu_h,\mu_W) +  \frac{\alpha_s(M_W)}{\alpha_s(\mu)}
    U_0(\mu_h,\mu_W) J
  \right) C^{(0)}(\mu_W)\,,
\end{equation}
where the matrix $J$ is obtained from $\gamma_1$ and $\beta_1$ as
explained for example in the renowned Les Houches lectures by
A.J. Buras \cite{Buras:1998raa}, where a complete pedagogical
introduction to the subtleties of NLO calculations is presented.
Although in the current lectures we will confine ourselves to LO
calculations, the importance of computing weak Hamiltonians to NLO (or
above) cannot be overemphasized.

Finally, we warn the reader that the $\overline{\mathrm{MS}}$ scheme,
although very convenient for perturbative calculations, is not the
only option. For example, matrix elements computed in Lattice QCD
(LQCD) potentially take into account strong interactions to all orders
in the non-perturbative regime; it is therefore possible (and
desirable) to perform non-perturbative renormalization, subtracting
divergences to all orders in perturbation theory
\cite{Martinelli:1994ty}. To achieve this result, it is convenient to
use the so-called regularization-independent renormalization schemes,
that are defined by fixing the value of a given number of renormalized
Green functions. For example, instead of defining the renormalized
four-quark operator by subtracting the $1/\epsilon$ poles in
dimensional regularization at a given perturbative order, we could
define it by imposing that its matrix element on given initial and
final states be equal to a given number, for example to the tree-level
matrix element of the same operator. This renormalization condition
can be implemented in any regularization at any perturbative order,
making it possible to match the perturbative calculation of the Wilson
coefficient with the nonperturbative calculation of the hadronic
matrix element.

\subsubsection{Current-current operators at LO}
\label{sec:ccLO}

\begin{figure}[htp]
  \centering
  \begin{tabular}{ccc}
   \begin{tikzpicture}
   \begin{feynman}[small]
     \vertex [crossed dot] (a) {};
     \vertex [above left=of a] (a1);
     \vertex [above left=of a1] (q1) {\(c\)};
     \vertex [below left=of a] (b1);
     \vertex [below left=of b1] (q2) {\(s\)};
     \vertex [above right=of a] (c1);
     \vertex [below right=of a] (d1);
     \vertex [above right=of c1] (q3) {\(u\)};
     \vertex [below right=of d1] (q4) {\(d\)};
     \diagram* {
       (a) -- [plain] (a1),
       (a1) -- [anti fermion] (q1),
       (a) -- [plain] (b1),
       (b1) -- [fermion] (q2),
       (a) -- [plain] (c1),
       (c1) -- [fermion] (q3),
       (a) -- [plain] (d1),
       (d1) -- [anti fermion] (q4),
       (c1) -- [gluon] (d1)
     };
   \end{feynman}
 \end{tikzpicture} & 
   \begin{tikzpicture}
   \begin{feynman}[small]
     \vertex [crossed dot] (a) {};
     \vertex [above left=of a] (a1);
     \vertex [above left=of a1] (q1) {\(c\)};
     \vertex [below left=of a] (b1);
     \vertex [below left=of b1] (q2) {\(s\)};
     \vertex [above right=of a] (c1);
     \vertex [below right=of a] (d1);
     \vertex [above right=of c1] (q3) {\(u\)};
     \vertex [below right=of d1] (q4) {\(d\)};
     \diagram* {
       (a) -- [plain] (a1),
       (a1) -- [anti fermion] (q1),
       (a) -- [plain] (b1),
       (b1) -- [fermion] (q2),
       (a) -- [plain] (c1),
       (c1) -- [fermion] (q3),
       (a) -- [plain] (d1),
       (d1) -- [anti fermion] (q4),
      (b1) -- [gluon] (d1)
     };
   \end{feynman}
 \end{tikzpicture} &
   \begin{tikzpicture}
   \begin{feynman}[small]
     \vertex [crossed dot] (a) {};
     \vertex [above left=of a] (a1);
     \vertex [above left=of a1] (q1) {\(c\)};
     \vertex [below left=of a] (b1);
     \vertex [below left=of b1] (q2) {\(s\)};
     \vertex [above right=of a] (c1);
     \vertex [below right=of a] (d1);
     \vertex [above right=of c1] (q3) {\(u\)};
     \vertex [below right=of d1] (q4) {\(d\)};
     \diagram* {
       (a) -- [plain] (a1),
       (a1) -- [anti fermion] (q1),
       (a) -- [plain] (b1),
       (b1) -- [fermion] (q2),
       (a) -- [plain] (c1),
       (c1) -- [fermion] (q3),
       (a) -- [plain] (d1),
       (d1) -- [anti fermion] (q4),
      (b1) -- [gluon, half left] (c1)
     };
   \end{feynman}
 \end{tikzpicture} \\
    (i) & (j) & (k)
  \end{tabular}                    
\caption{``Mirror'' diagrams relevant for $c \to s u \overline{d}$
  transitions in the effective theory. See the text for details.}
\label{fig:dc1m}
\end{figure}
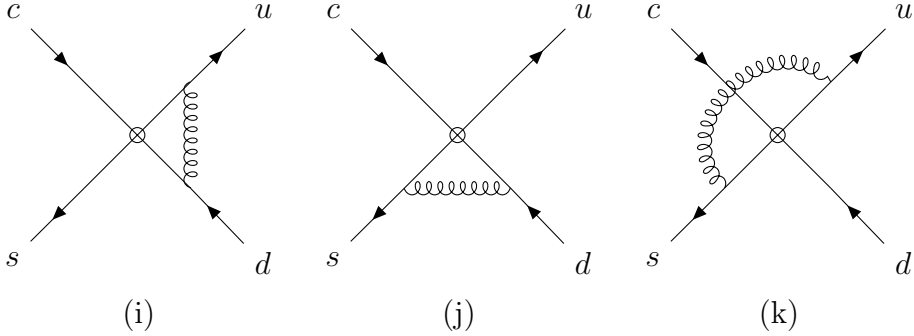

As we have seen above, if we are interested in capturing the dominant,
log-enhanced QCD corrections only, we just need to start at $\mu_W$
with the effective Hamiltonian obtained from tree-level matching,
eq.~(\ref{eq:Heffdc1tree}), and run it down to $\mu_h$ using
eq.~(\ref{eq:rge2}) with the anomalous dimension computed at
$\mathcal{O}(\alpha_s)$. To compute the latter, we need to identify
the $1/\epsilon$ terms generated by diagrams (f)-(h) in
Fig.~\ref{fig:dc1}, plus the ``mirror'' ones reported in
fig.~\ref{fig:dc1m}. We can actually skip diagrams (f) and (i) since
they cancel against the renormalization constants for the quark fields
due to the Ward identity that protects the conserved weak current. Let
us therefore start from diagram (g). Assigning momentum $p$ to the
incoming $c$ and outgoing $u$ quarks, and loop momentum $k$ to the
fermions in the loop, we obtain the following amplitude:
\begin{displaymath}
  i\mathcal{A}_{(g)}= \frac{4 G_F V^{\phantom{*}}_{ud}V_{cs}^{*}}{\sqrt{2}}\int
  \frac{\mathrm{d}^{D}k}{(2\pi)^{D}}
  \overline{u}^{u}_i
  \left(
    i g_s\gamma_\mu T^{A}_{ij}
  \right) \frac{i}{\slashed{k}} \gamma^{\rho} P_L v^{d}_j\,\overline{v}^{s}_k \gamma_\rho
  P_L   \frac{i}{\slashed{k}} \left(
    i g_s\gamma_\nu T^{B}_{kl} 
  \right)u^{c}_l \frac{-i g^{\mu\nu}\delta^{AB}}{(k-p)^{2}} 
\end{displaymath}
where we have used the following Feynman rules for QCD in the Feynman
gauge:
\begin{equation}
  \label{eq:QCDFR}
     \includegraphics[width=0.18\linewidth, valign=c]{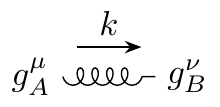}
= \frac{-i g^{\mu\nu}\delta^{AB}}{k^{2}+i\epsilon}\,,\qquad
\includegraphics[width=0.18\linewidth, valign=c]{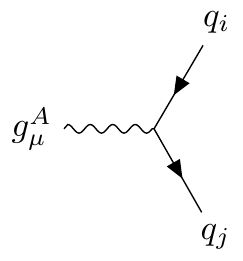}
   = i g_s\gamma_\mu T^{A}_{ij} \,,
\end{equation}
with $A,~B$ colour indices in the adjoint representation,
$i,~j,~k,~l$ colour indices in the fundamental representation and $T$
the $SU(3)$ generators for the fundamental representation. 

Pulling out the Dirac structure we can rewrite the amplitude as
\begin{equation}
  \label{eq:Adiagg}
  i\mathcal{A}_{(g)}= -i \frac{4 G_F }{\sqrt{2}} V^{\phantom{*}}_{ud}V_{cs}^{*}
  g_s^{2} T^{A}_{ij} T^{A}_{kl} 
  \overline{u}^{u}_i
  \gamma_\mu 
  \gamma_\alpha \gamma^{\rho} P_L v^{d}_j\,\overline{v}^{s}_k \gamma_\rho
  P_L   \gamma_\beta \gamma^{\mu} u^{c}_l \mathcal{I}^{\alpha\beta}\,,
\end{equation}
with
\begin{dmath}
  \mathcal{I}^{\alpha\beta} = \int \frac{\mathrm{d}^{D}k}{(2\pi)^{D}}
                                \frac{k^{\alpha} k^{\beta}}{(k^{2})^{2} (k-p)^{2}} = 
                                \int_0^{1} \mathrm{d}x\, 2(1-x) \int
                                \frac{\mathrm{d}^{D}k}{(2\pi)^{D}}
                                \frac{k^{\alpha} k^{\beta}}{[k^{2}(1-x) +
                                (k-p)^{2} x]^{3}} = \int_0^{1} \mathrm{d}x\, 2(1-x) \int
                               \frac{\mathrm{d}^{D}\ell}{(2\pi)^{D}}
                               \frac{(\ell + p x)^{\alpha} (\ell + p
                               x)^{\beta}}{[\ell^{2} + p^{2} x (1-x)]^{3}} = \int_0^{1} \mathrm{d}x\, 2(1-x) \int
                               \frac{\mathrm{d}^{D}\ell}{(2\pi)^{D}}
                               \frac{\ell^{\alpha} \ell^{\beta}}
                               {[\ell^{2} + p^{2} x
                               (1-x)]^{3}}+\mathrm{f.t.,} 
  \label{eq:Ialbe1}
\end{dmath}
where we have introduced the Feynman parameter $x$ using
\begin{equation}
  \label{eq:Fpars1}
  \frac{1}{a^{n}b} = n \int_0^{1} \mathrm{d}x \frac{x^{n-1}}{[(1-x)b + x a]^{n+1}}
\end{equation}
(a particular case of the general parameterization in Appendix
\ref{sec:appFP}), and the last equality holds up to non-divergent
terms.

Computing the integral on $\ell$ using the formulae in Appendix
\ref{sec:appI}, we obtain
\begin{dmath}
  \mathcal{I}^{\alpha\beta}= \int_0^{1} \mathrm{d}x\, 2(1-x)
  \frac{g^{\alpha\beta}}{2} \frac{i}{(4 \pi)^{D/2}}
  \frac{\Gamma(3-D/2-1)}{\Gamma(3)}
  \left(
  \frac{\mu^{2}}{-p^{2}x(1-x)}
  \right)^{3-D/2-1} = \frac{i}{16 \pi^{2}}
  \frac{g^{\alpha\beta}}{4}\frac{1}{\epsilon} + \mathrm{f.t.},
  \label{eq:Ialbe2}
\end{dmath}
again up to irrelevant finite terms.

Since we are only interested in the divergent terms, we can also
perform the Dirac algebra in four dimensions. This greatly simplifies
things since we are then authorized to use the so-called Fierz
identities~\cite{Fierz1937}. Indeed, in four dimensions we can
identify a complete basis for objects carrying two spinor indices. For
example, we might choose
\begin{equation}
  \label{eq:basisnc}
  \mathbb{1},~\gamma_5,~\gamma^{\mu},~\gamma^{\mu} \gamma_5,~\sigma^{\mu\nu}\,,
\end{equation}
but for our purposes it is more convenient to work in a chiral basis
such as
\begin{equation}
  \label{eq:basisc}
  P_L,~P_R,~\gamma^{\mu} P_L,~\gamma^{\mu} P_R,~\sigma^{\mu\nu}\,.
\end{equation}
Thus, any object carrying two spinor indices can be projected on this
basis. Now, looking at the Dirac string in eq.~(\ref{eq:Adiagg}), we
see several Dirac matrices with Lorentz indices contracted across the
two different fermionic lines. Simplifying the Dirac structure would
be simple if those Dirac matrices were in the same fermionic line. We
can bring all those matrices together using a Fierz
transformation,\footnote{I am indebted to R.K. Ellis for pointing this
  trick out to me in an exercise session at the Parma school of
  theoretical physics in September 2001.} by projecting
\begin{equation}
\left(
  P_L v_j^{d} \overline{v}^{s}_k P_R
\right)_{\alpha\beta}
\label{eq:toproj}
\end{equation}
on the basis in eq.~(\ref{eq:basisc}). Since in eq.~(\ref{eq:toproj})
we have left-handed chirality on the left and right-handed chirality
on the right, it is clear that the only structure we can project on is
$\gamma^{\mu} P_R$. To find out the coefficient, we can act on
eq.~(\ref{eq:toproj}) with the operator
$\frac{1}{2}\mathrm{Tr}~\gamma^{\mu} P_L$,
which is a projector on $\gamma^{\mu} P_R$, since
\begin{dgroup*}[noalign]
  \begin{dmath}[compact]
    \label{eq:gammamuPRproj}
    \frac{1}{2}\mathrm{Tr}~\gamma^{\mu} P_L\,P_L = 0\,,\qquad
    \frac{1}{2}\mathrm{Tr}~\gamma^{\mu} P_L\,P_R = 0\,,\qquad
    \frac{1}{2}\mathrm{Tr}~\gamma^{\mu} P_L\,\gamma^{\nu} P_L = 0\,,
  \end{dmath}
  \begin{dmath}[compact]
    \label{eq:gammamuPRproj2}
    \frac{1}{2}\mathrm{Tr}~\gamma^{\mu} P_L\,\gamma^{\nu} P_R =
    g^{\mu\nu}\,,\qquad \frac{1}{2}\mathrm{Tr}~\gamma^{\mu}
    P_L\,\sigma^{\nu\rho} = 0\,.
  \end{dmath}
\end{dgroup*}
We obtain
\begin{equation}
  \label{eq:proj}
  \frac{1}{2}\mathrm{Tr}~\gamma^{\mu} P_L P_L v_j^{d} \overline{v}^{s}_k P_R =
  - \frac{1}{2} \overline{v}^{s}_k \gamma^{\mu} P_L v_j^{d}\,,
\end{equation}
where we have added a minus sign since the $v$ are anticommuting
spinors. We have thus obtained that
\begin{equation}
  \label{eq:Fierz1}
  \left(
    P_L v_j^{d} \overline{v}^{s}_k P_R
  \right)_{\alpha\beta} = - \frac{1}{2} \overline{v}^{s}_k \gamma^{\mu} P_L
  v_j^{d} 
  \left(
    \gamma_\mu P_R
  \right)_{\alpha\beta}\,.
\end{equation}
Substituting eq.~(\ref{eq:Fierz1}) in eq.~(\ref{eq:Adiagg}), and
keeping into account that the divergent part of
$\mathcal{I}^{\alpha\beta}$ is proportional to $g^{\alpha\beta}$, the
Dirac string becomes
\begin{equation}
  \label{eq:dirac2}
  - \frac{1}{2} \overline{v}^{s}_k \gamma^{\nu} P_L
  v_j^{d}\,\overline{u}^{u}_i
  \gamma_\mu 
  \gamma_\alpha \gamma_\rho \gamma_\nu P_R \gamma^{\rho} 
  \gamma^{\alpha} \gamma^{\mu} u^{c}_l\,.
\end{equation}
Using the four-dimensional Dirac algebra rules, we move the chiral
projector to the right and apply thrice the identity $\gamma^{\alpha}
\gamma^{\mu} \gamma_\alpha = -2 \gamma^{\mu}$ to obtain
\begin{equation}
  \label{eq:diracfinal}
  4 \overline{v}^{s}_k \gamma^{\nu} P_L
  v_j^{d}\,\overline{u}^{u}_i\gamma_\nu P_L u^{c}_l\,. 
\end{equation}
Notice that the ordering of the spinors here is different with respect
to the tree-level amplitude generated by $Q_1^{\overline{s} c \overline{d} u}$;
thus, to identify the relevant counterterms, let us use the Fierz
trick again to exchange the $v_j^{d}$ and $u^{c}_l$ spinors. This is
easily achieved:
\begin{equation}
  \label{eq:Fierzagain}
  \overline{v}^{s}_k \gamma^{\nu} P_L
  v_j^{d}\,\overline{u}^{u}_i\gamma_\nu P_L u^{c}_l = -\frac{1}{2}
  \overline{u}^{u}_i\gamma_\mu P_L v_j^{d}\,  \overline{v}^{s}_k \gamma^{\nu} \gamma^{\mu}
  P_R \gamma_\nu u^{c}_l = \overline{u}^{u}_i\gamma_\mu P_L v_j^{d}\,
  \overline{v}^{s}_k \gamma^{\mu} P_L u^{c}_l\,
\end{equation}
where in the first step we used again eq.~(\ref{eq:Fierz1}).  Before
putting all the pieces together, let us look at the colour factor in
eq.~(\ref{eq:Adiagg}). We can simplify it using the $SU(N_c)$ identity
\begin{equation}
  \label{eq:tata}
  T^{A}_{ij}T^{A}_{kl} = \frac{1}{2}
  \delta_{il}\delta_{kj} - \frac{1}{2 N_c}
  \delta_{ij}\delta_{kl}\,.
\end{equation}
Putting everything together, we obtain the amplitude generated by
diagram (g):
\begin{equation}
  \label{eq:Adiaggfinal}
  i\mathcal{A}_{(g)}= \frac{4 G_F}{\sqrt{2}} V^{\phantom{*}}_{ud}V_{cs}^{*}
  \frac{\alpha_s}{4\pi} \frac{1}{2 \epsilon} 
  \left(
    \overline{u}^{u}_i\gamma_\mu P_L v_j^{d}\,
    \overline{v}^{s}_j \gamma^{\mu} P_L u^{c}_i -
    \frac{1}{N_c} \overline{u}^{u} \gamma_\mu P_L v^{d}\,
    \overline{v}^{s} \gamma^{\mu} P_L u^{c}
  \right)\,.
\end{equation}
We see that this diagram generates a divergence proportional to the
matrix element of a new operator,
\begin{equation}
  \label{eq:Q2}
  Q_2^{\overline{s} c \overline{u} d} \equiv \overline{s}_L^{\alpha}
\gamma^{\mu} c_L^{\beta} \overline{u}_{L}^{\beta} \gamma_\mu d_L^{\alpha}\,.
\end{equation}
Thus, the divergence we obtain in the effective theory (or,
equivalently, the large log present in the full theory) is
proportional not only to the operator generated omitting QCD
corrections, but also to a new operator. This forces us to promote the
anomalous dimension to a matrix and the operator basis to a vector. 

Let us now consider diagram (h) in Fig.~\ref{fig:dc1}. For
convenience, let us assign momentum $-p$ to the $\overline{u}d$ quark line,
so that we end up with exactly the same integral as in diagram
(g). Since we are interested in the ultraviolet divergence, the choice
of external momenta is irrelevant. We obtain:
\begin{dmath}
  \label{eq:Adiagh}
  i\mathcal{A}_{(h)}= -i \frac{4 G_F }{\sqrt{2}} V^{\phantom{*}}_{ud}V_{cs}^{*}
  g_s^{2} T^{A}_{ij} T^{A}_{kl} 
  \overline{u}^{u}_i
  \gamma^{\rho} P_L
  \gamma_\alpha \gamma_\mu  v^{d}_j\,\overline{v}^{s}_k \gamma_\rho
  P_L   \gamma_\beta \gamma^{\mu} u^{c}_l (-\mathcal{I}^{\alpha\beta})=
    \frac{4 G_F}{\sqrt{2}} V^{\phantom{*}}_{ud}V_{cs}^{*} T^{A}_{ij} T^{A}_{kl}
     \frac{\alpha_s}{4\pi} \frac{1}{8} \frac{1}{\epsilon}
     \overline{v}^{s}_k \gamma^{\nu} P_Lv_j^{d}\,
     \overline{u}^{u}_i \gamma_\rho \gamma_\alpha \gamma_\mu \gamma_\nu P_R \gamma^{\rho} 
     \gamma^{\alpha} \gamma^{\mu} u^{c}_l = 
  \frac{4 G_F}{\sqrt{2}} V^{\phantom{*}}_{ud}V_{cs}^{*} T^{A}_{ij}
     T^{A}_{kl}\frac{\alpha_s}{4\pi} (-4) \frac{1}{\epsilon}
     \overline{v}^{s}_k \gamma^{\mu} P_Lv_j^{d}\, \overline{u}^{u}_i\gamma_\mu P_L u^{c}_l
     = \frac{4 G_F}{\sqrt{2}} V^{\phantom{*}}_{ud}V_{cs}^{*} \frac{\alpha_s}{4\pi} (-4)
     \frac{1}{\epsilon} 
     \left(
     \frac{1}{2} \overline{v}^{s}_i \gamma^{\mu} P_Lu_j^{c}\,
     \overline{u}^{u}_j\gamma_\mu P_L v^{d}_i - \frac{1}{2N_c} \overline{v}^{s}
     \gamma^{\mu} P_Lu^{c} \,
     \overline{u}^{u}\gamma_\mu P_L v^{d}
     \right)\,,\nonumber
\end{dmath}
where, in addition to eqs.~(\ref{eq:Ialbe2}), (\ref{eq:Fierz1}),
(\ref{eq:Fierzagain}) and (\ref{eq:tata}), we have used the identity
\begin{dmath}[compact]
  \gamma^{\mu} \gamma_\alpha \gamma_\beta \gamma_\gamma \gamma_\mu =
  -2 \gamma_\gamma \gamma_\beta \gamma_\alpha\,,
\end{dmath}
always performing Dirac
algebra in four dimensions.

Let us now turn to the diagrams in Fig.~\ref{fig:dc1m}. Diagram (i)
cancels against quark wave function renormalization, while diagrams
(j) and (k) give the same result as diagrams (g) and (h) respectively.

Putting everything together we obtain
\begin{equation}
  \label{eq:Adiagghjk}
  i\mathcal{A}= \frac{4 G_F}{\sqrt{2}} V^{\phantom{*}}_{ud}V_{cs}^{*}
  \frac{\alpha_s}{4\pi} \frac{-3}{\epsilon} 
  \left( 
    \overline{u}^{u}_i\gamma_\mu P_L v_j^{d}\,
    \overline{v}^{s}_j \gamma^{\mu} P_L u^{c}_i -
    \frac{1}{N_c} \overline{u}^{u} \gamma_\mu P_L v^{d}\,
    \overline{v}^{s} \gamma^{\mu} P_L u^{c}
  \right)\,.
\end{equation}

In order to obtain the two-by-two anomalous dimension matrix we need
to compute the one-loop renormalization of $Q_2^{\overline{s} c \overline{d}
  u}$, by inserting it in diagrams (f) to (k). The only difference in
the calculation is given by the colour factors, which now read
\begin{equation}
  T^{A}_{il}T^{A}_{kj}=
    \frac{1}{2}\delta_{ij}\delta_{kl}- \frac{1}{2N_c}\delta_{il}\delta_{kj}\,.
  \label{eq:tata2}
\end{equation}
Defining
\begin{equation}
  \label{eq:zetaij}
  Q^{\mathrm{bare}}_i = Z_{ij}Q_j^{\mathrm{ren}}
\end{equation}
we thus obtain in $\overline{\mathrm{MS}}$ 
\begin{equation}
  \label{eq:zetacc}
  Z = \mathbb{1} + \frac{\alpha_s}{4\pi}\frac{1}{\epsilon} Z_1
  = \mathbb{1} + \frac{\alpha_s}{4\pi}\frac{1}{\epsilon} 
  \left(
    \begin{array}{cc}
      \frac{3}{N_c} & -3 \\
       -3 & \frac{3}{N_c}
    \end{array}
  \right)\,.
\end{equation}

From this we obtain, using eq.~(\ref{eq:gammaeps1}),
\begin{equation}
  \label{eq:gammacc}
  \gamma_0 =  \left(
    \begin{array}{cc}
      -\frac{6}{N_c} & 6 \\
      6 & -\frac{6}{N_c}
    \end{array}
  \right)\,. 
\end{equation}

The corresponding evolutor can be obtained from
eq.~(\ref{eq:rge2}). In practice, it can be evaluated by going to the
basis in which $\gamma_0$ is diagonal, defining
\begin{equation}
  \label{eq:plusminusbasis}
  Q_{\pm} = \frac{Q_1 \pm Q_2}{2}\,,\qquad C_{\pm}=C_1 \pm
  C_2\,,\qquad \gamma_0^{\pm} = \pm 6 \frac{N_c\mp 1}{N_c}\,, \qquad U_0^{\pm} = 
  \left(
    \frac{\alpha_s(\mu_W)}{\alpha_s(\mu_h)}
  \right)^{\frac{\gamma_0^{\pm}}{2 \beta_0}}\,.
\end{equation}
Notice that $ \beta_0$ depends on the number of active flavours, so if
we want to compute the Wilson coefficients at $\mu_h = 2$ GeV we need
to take into account the bottom quark threshold at a scale $\mu_b \sim
m_b$:
\begin{equation}
  \label{eq:cplusminus2}
  C_\pm(2 \mathrm{GeV}) = \left(
    \frac{\alpha_s(\mu_b)}{\alpha_s(2 \mathrm{GeV})}
  \right)^{\frac{\gamma_0^{\pm}}{2 \beta_0(4)}} \left(
    \frac{\alpha_s(\mu_W)}{\alpha_s(\mu_b)}
  \right)^{\frac{\gamma_0^{\pm}}{2 \beta_0(5)}} C_\pm(\mu_W)\,. 
\end{equation}
At LO, we have $C_\pm(\mu_W)=1$ and the evolution decreases $C_+$
and increases $C_-$, since $\gamma_0^{+}$ is positive and $\gamma_0^{-}$
is negative.

\subsection{Penguin operators}
\label{sec:penguins}

\begin{figure}[htp]
  \centering
  \begin{tabular}{ccc}
  \begin{tikzpicture}
   \begin{feynman}
     \vertex [crossed dot] (a) {};
     \vertex [left=of a] (q1) {\(s\)};
     \vertex [right=of a] (q2) {\(d\)};
     \vertex [below=of a] (a1);
     \vertex [below=of a1] (a2);
     \diagram* {
       (a) -- [fermion, rmomentum'=\(p\)] (q1),
       (a) -- [fermion, half left] (a1),
       (a1) -- [fermion, half left] (a),
       (a) -- [anti fermion, momentum=\(p-q\)] (q2),
       (a1) -- [gluon, momentum=\(q\)] (a2)
     };
   \end{feynman}
 \end{tikzpicture}&
    \begin{tikzpicture}
   \begin{feynman}
     \vertex [crossed dot] (a) {};
     \vertex [left=of a] (q1) {\(s\)};
     \vertex [right=of a] (q2) {\(d\)};
     \vertex [below=of a] (a1);
     \vertex [below=of a1] (a2);
     \vertex [left=of a2] (q3) {\(q_f\)};
     \vertex [right=of a2] (q4) {\(q_f\)};
     \diagram* {
       (a) -- [fermion, rmomentum'=\(p\)] (q1),
       (a) -- [fermion, half left] (a1),
       (a1) -- [fermion, half left] (a),
       (a) -- [anti fermion, momentum=\(p-q\)] (q2),
       (a1) -- [gluon, momentum=\(q\)] (a2),
       (a2) -- [anti fermion] (q3),
       (a2) -- [fermion] (q4)
     };
   \end{feynman}
 \end{tikzpicture} &
    \begin{tikzpicture}
   \begin{feynman}
     \vertex (a);
     \vertex [left=1 cm of a] (q1) {\(s\)};
     \vertex [right=1 cm of a] (b);
     \vertex [right=1 cm of b] (q2) {\(d\)};
     \vertex at ($(a)!0.5!(b)-(0,1.5cm)$) (a1);
     \vertex [below=of a1] (a2);
     \vertex [left=of a2] (q3) {\(q_f\)};
     \vertex [right=of a2] (q4) {\(q_f\)};
     \diagram* {
       (a) -- [fermion, rmomentum'=\(p\)] (q1),
       (a) -- [photon] (b),
       (a) -- [fermion, bend right] (a1),
       (a1) -- [fermion, bend right] (b),
       (b) -- [anti fermion, momentum=\(p-q\)] (q2),
       (a1) -- [gluon, momentum=\(q\)] (a2),
       (a2) -- [anti fermion] (q3),
       (a2) -- [fermion] (q4)
     };
   \end{feynman}
 \end{tikzpicture}  \\
    (a) & (b) & (c)
  \end{tabular}
  \caption{``Penguin'' diagrams relevant for
    $\overline{s} \to \overline{d}$ transitions. See the text for
    details.}
\label{fig:penguins}
\end{figure}
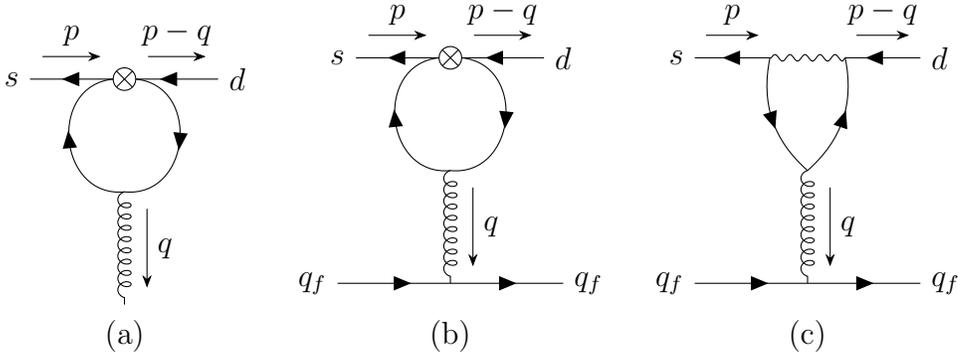

Let us now change the flavour content of our current-current operators
and consider $W$ exchange between a $\overline{s}u$ and a $\overline{u}d$
current. Following exactly the same considerations as in
Sec.~\ref{sec:current-current}, we obtain the following effective
Hamiltonian:
\begin{equation}
  \label{eq:heffds1cc_u}
    \mathcal{H}_{\mathrm{eff}}^{\overline{s}\to \overline{d}} =
    \frac{4 G_F}{\sqrt{2}}
  V^{\phantom{*}}_{ud}V_{us}^{*} 
  \left(
    C_1 Q_1^{\overline{s}u\overline{u} d} + C_2 Q_2^{\overline{s}u\overline{u} d} 
  \right)\,.
\end{equation}
However, when computing the renormalization of the operators in
$\mathcal{H}_{\mathrm{eff}}^{\overline{s}\to \overline{d}}$, additional diagrams
\cite{Vainshtein:1975sv} (called penguin diagrams\footnote{For an
  instructive recollection of how the term ``penguin diagram'' made
  its way in particle physics see ref.~\cite{Shifman:1995hc}.})  arise
by contracting the $u$ and $\bar u$ fields in $Q_{1,2}$ and attaching
a gluon, as in Fig.~\ref{fig:penguins} (a). 

Let us think for a second on the possible structure of the amplitude
induced by these diagrams. It is clearly a FCNC
$\overline{s}_i\Gamma^{\mu} T^{A}_{ij} d_j$ coupling, which as we have seen
cannot arise in the SM Lagrangian. Thus, the amplitude must be
proportional to the matrix element of some higher dimensional
operator. Indeed, gauge invariance forces us to impose
$q^{\mu}\overline{s}\Gamma^{\mu} T^{A}_{ij}  d=0$, so we can identify two possible structures:
\begin{equation}
  \label{eq:ginvfcnc}
  \overline{s}
  \left(
    q^{2} \gamma_\mu - q_\mu \slashed{q}
  \right) T^{A} d\qquad \mathrm{and}\qquad \overline{s}
  \sigma^{\mu\nu} q_\nu T^{A} d\,.
\end{equation}
The second one connects quarks of different helicity, so for massless
quarks it cannot be generated. The first structure corresponds to the
matrix element of operator $\overline{s}\gamma^{\mu} T^{A} d D^{\nu} G^{A}_{\mu\nu}$. Since the
equations of motion imply
\begin{equation}
  D^{\nu} G^{A}_{\mu\nu} = g_s \sum_f \overline{q}_f \gamma_\mu T^{A} q_f \,, 
  \label{eq:eqmot}
\end{equation}
with $f$ any active quark flavour, this is equivalent to operator
\begin{equation}
  \label{eq:pengnc}
  \overline{s}\gamma_\mu T^{A} d \sum_f \overline{q}_f \gamma^{\mu} T^{A} q_f\,.
\end{equation}
A diagrammatic representation of the equation of motion can be
obtained by attaching the gluon line to a quark line of flavour $f$, as in
Fig.~\ref{fig:penguins} (b). When contracted with the gluon
propagator, the $q^{\mu}$ term in the FCNC vertex acts on the fermionic
current giving zero for gauge invariance. The $q^{2}$ term, instead,
cancels the pole of the gluon propagator, yielding precisely the local
amplitude $\overline{v}_s\gamma_\mu T^{a} v_d \sum_f \overline{u}_f \gamma^{\mu} T^{a}
u_f$, \emph{i.e.} the matrix element of the local operator in
eq.~(\ref{eq:pengnc}). By power counting, diagrams (a) or (b) are
logarithmically divergent, so we will find a $1/\epsilon$ term which
must be subtracted, forcing us to enlarge the operator basis
again. Since the new operator in eq.~(\ref{eq:pengnc}) must be
renormalized itself, we must insert it in all the crosses in the
diagrams in Figs.~\ref{fig:dc1}, \ref{fig:dc1m} and
\ref{fig:penguins}. As in the case of the insertion of operator $Q_1$,
gluonic corrections will change the original colour structure;
moreover, the Dirac algebra will be different for the left- and
right-handed components of the $q_f$ current, so we will get different
renormalization constants for the two components. All in all, we need
to add four ``penguin'' operators:
\begin{dgroup*}
  \begin{dmath}[label={eq:Q3}]
    Q_3^{\overline{s}d} = \overline{s}_L \gamma_\mu d_L \sum_f \overline{q}_{fL}
    \gamma^{\mu} q_{fL}\,,
  \end{dmath}
  \begin{dmath}[label={eq:Q4}]
    Q_4^{\overline{s}d} = \overline{s}_L^{\alpha} \gamma_\mu d_L^{\beta} \sum_f
    \overline{q}_{fL}^{\beta} \gamma^{\mu} q_{fL}^{\alpha} \,,
  \end{dmath}
  \begin{dmath}[label={eq:Q5}]
    Q_5^{\overline{s}d} = \overline{s}_L \gamma_\mu d_L \sum_f \overline{q}_{fR}
    \gamma^{\mu} q_{fR}\,,
  \end{dmath}
  \begin{dmath}[label={eq:Q6}]
    Q_6^{\overline{s}d} = \overline{s}_L^{\alpha} \gamma_\mu d_L^{\beta} \sum_f
    \overline{q}_{fR}^{\beta} \gamma^{\mu} q_{fR}^{\alpha} \,.
  \end{dmath}
\end{dgroup*}
When we insert operators $Q_{3-6}$ in the diagram in
Fig.~\ref{fig:penguins}, we should remember that we are using
$\overline{MS}$ which is a mass-independent renormalization scheme, so
we should manually decouple quark flavours at a threshold
$\mu_f\sim m_f$. The anomalous dimensions (as well as the $\beta$
function) depend on the number of active flavours $n_{f}(\mu)$, which
also determines the summation range in operators $Q_{3-6}$.

As an explicit example of how penguin operators are generated, let us
evaluate diagram \emph{(b)} in Fig.~\ref{fig:penguins} with the
insertion of operator $Q_1^{\overline{s}u\overline{u}d}$. We get,
omitting the prefactor
$\frac{4 G_F}{\sqrt{2}} V^{\phantom{*}}_{ud}V^{*}_{us}$ and setting
for simplicity $p=0$,
\begin{dmath}[label={eq:penguin1}]
 \mathcal{A}_{(b)} = \int \frac{\mathrm{d}^{D}k}{(2\pi)^{D}}
  \overline{v}_d^{i} \gamma^{\mu} P_L \frac{i}{\slashed{k}-\slashed{q}}
  (i g_s \gamma^{\nu} T^{a}_{ij}) \frac{i}{\slashed{k}} \gamma_\mu P_L v_s^{j}
     \frac{-i}{q^{2}} \overline{u}_f^{k}(i g_s \gamma_\nu T^{a}_{kl})
     u_f^{l}= -i \frac{g_s^{2}}{q^{2}} \int \frac{\mathrm{d}^{D}k}{(2\pi)^{D}}
  \overline{v}_d^{i} \gamma^{\mu} P_L \frac{\slashed{k}-\slashed{q}}{(k-q)^{2}}
  \gamma^{\nu} T^{a}_{ij} \frac{\slashed{k}}{k^{2}} \gamma_\mu P_L v_s^{j}
     \overline{u}_f^{k} \gamma_\nu T^{a}_{kl} u_f^{l} = -i \frac{g_s^{2}}{q^{2}} I_{\alpha\beta}
  \overline{v}_d^{i} \gamma^{\mu} \gamma^{\alpha}
  \gamma^{\nu} T^{a}_{ij} \gamma^{\beta} \gamma_\mu P_L v_s^{j}
     \overline{u}_f^{k}\gamma_\nu T^{a}_{kl} u_f^{l}\,,
\end{dmath}
with
\begin{dmath}
  \label{eq:intab}
  I_{\alpha\beta}= \int \frac{\mathrm{d}^{D}k}{(2\pi)^{D}}\frac{(k-q)^{\alpha}
    k^{\beta}}{(k-q)^{2}k^{2}} = \int_0^{1} \mathrm{d}x \int \frac{\mathrm{d}^{D}k}{(2\pi)^{D}}\frac{(k-q)^{\alpha}
    k^{\beta}}{
    \left[
      (k-q)^{2} x + k^{2} (1-x)
    \right]^{2}}= \int_0^{1} \mathrm{d}x \int \frac{\mathrm{d}^{D}k}{(2\pi)^{D}}\frac{(k-q)^{\alpha}
    k^{\beta}}{
    \left[
      k^{2} - 2 q\cdot k x + q^{2} x 
     \right]^{2}}= \int_0^{1} \mathrm{d}x \int
    \frac{\mathrm{d}^{D}l}{(2\pi)^{D}}\frac{(l - q (1-x))^{\alpha}
    (l + q x)^{\beta}}{
    \left[
     l^{2} + q^{2} x (1-x)  \right]^{2}} =\int_0^{1} \mathrm{d}x \int
    \frac{\mathrm{d}^{D}l}{(2\pi)^{D}}\frac{l^{\alpha} l^{\beta} - q^{\alpha}
  q^{\beta} x (1-x)}{
    \left[
     l^{2} + q^{2} x (1-x)  \right]^{2} }=\frac{-i}{16 \pi^{2}}\int_0^{1} \mathrm{d}x 
  \left(
  \frac{g^{\alpha\beta}}{2} q^{2} x (1-x) +  q^{\alpha}
  q^{\beta} x (1-x)
  \right) 
  \left(
  \frac{1}{\epsilon} + \mathrm{f.t.}
  \right)= \frac{-i}{16 \pi^{2}} 
  \left(
  \frac{g^{\alpha\beta}}{2}  q^{2}  +  q^{\alpha}
  q^{\beta}
  \right) 
  \left(
  \frac{1}{6 \epsilon}
     \right)\,,
     \label{eq:ipeng}
\end{dmath}
where we have used the Feynman parameterization and loop integrals
reported in Appendices \ref{sec:appFP} and \ref{sec:appI}. Substituting
eq.~(\ref{eq:ipeng}) in eq.~(\ref{eq:penguin1}) we obtain
\begin{dmath}
  \frac{-g_s^{2}}{16 \pi^{2}} \frac{1}{12}\frac{1}{\epsilon}
  \overline{v}_d^{i} 
   T^{a}_{ij} \gamma^{\mu} 
   \left(
     \gamma^{\alpha}
  \gamma^{\nu} \gamma_\alpha + 2 \frac{\slashed{q} \gamma^{\nu} \slashed{q}}{q^{2}}
\right) \gamma_\mu P_L v_s^{j}
  \overline{u}_f^{k}\gamma_\nu T^{a}_{kl} u_f^{l} =\frac{\alpha_s}{4
       \pi}
     \frac{1}{6}\frac{1}{\epsilon} \overline{v}_d^{i} 
   T^{a}_{ij}  
   \left( \gamma_{\alpha}
  \gamma^{\nu} \gamma^{\alpha} + 2 \frac{\slashed{q} \gamma^{\nu} \slashed{q}}{q^{2}}
\right) P_L v_s^{j}
     \overline{u}_f^{k}(q)\gamma_\nu T^{a}_{kl} u_f^{l}  = -\frac{\alpha_s}{4
       \pi}
     \frac{1}{3}\frac{1}{\epsilon} \overline{v}_d^{i} 
   T^{a}_{ij}  
     \left(
  \gamma^{\nu} - \frac{\slashed{q} (2 q^{\nu}- \slashed{q}\gamma^{\nu})}{q^{2}}
\right) P_L v_s^{j}
     \overline{u}_f^{k}\gamma_\nu T^{a}_{kl} u_f^{l}  =-\frac{\alpha_s}{4
       \pi}
     \frac{2}{3}\frac{1}{\epsilon} \overline{v}_d^{i} 
   T^{a}_{ij}  
     \left(
  \gamma^{\nu} - \frac{ q^{\nu} \slashed{q}}{q^{2}}
\right) P_L v_s^{j}
     \overline{u}_f^{k}\gamma_\nu T^{a}_{kl} u_f^{l}  =  -\frac{\alpha_s}{4
       \pi}
     \frac{2}{3}\frac{1}{\epsilon} \overline{v}_d^{i} 
   T^{a}_{ij}  
  \gamma^{\nu} P_L v_s^{j}
     \overline{u}_f^{k}\gamma_\nu T^{a}_{kl} u_f^{l} \,.
  \label{eq:peng2}
\end{dmath}
From eq.~(\ref{eq:peng2}) we see explicitly that the FCNC gluon vertex
is proportional to $q^{2} \gamma^{\nu} - q^{\nu}\slashed{q}$, and that we
obtain in the end a local four-quark operator as implied by the
equations of motion.

Let us now notice that also current-current operators with charm
quarks can mix into penguin operators via the diagrams in
Fig.~\ref{fig:penguins}. Thus, we should add to the effective
Hamiltonian in eq.~(\ref{eq:heffds1cc_u}) the corresponding operators
with charm, leading to
\begin{equation}
  \label{eq:heffds1cc_c}
  \mathcal{H}_{\mathrm{eff}}^{\overline{s}\to \overline{d}} =
  \frac{4 G_F}{\sqrt{2}}
  \left[
    V^{\phantom{*}}_{ud}V_{us}^{*} 
    \left(
      C_1 Q_1^{\overline{s}u\overline{u} d} + C_2 Q_2^{\overline{s}u\overline{u} d} 
    \right) +  V^{\phantom{*}}_{cd}V_{cs}^{*} 
    \left(
      C_1 Q_1^{\overline{s}c\overline{c} d} + C_2 Q_2^{\overline{s}c\overline{c} d} 
    \right)     
  \right]
  \,.
\end{equation}
Now, when inserted in penguin diagrams, operators
$Q_{1,2}^{\overline{s}u\overline{u} d}$ and $Q_{1,2}^{\overline{s}c\overline{c} d}$ give
exactly the same divergent part, since the divergence is independent
on the mass of the quarks running in the loop. Thus, the penguin
operators will be generated with a coefficient proportional to
$V^{\phantom{*}}_{ud}V_{us}^{*} + V^{\phantom{*}}_{cd}V_{cs}^{*} = -  V^{\phantom{*}}_{td}V_{ts}^{*}$. Taking
everything into account, we end up with the following effective
Hamiltonian:
\begin{dmath}[label={eq:heffds1}]
  \mathcal{H}_{\mathrm{eff}}^{\overline{s}\to \overline{d}} =
  \frac{4 G_F}{\sqrt{2}}
  \left\{
    V^{\phantom{*}}_{ud}V_{us}^{*} 
    \left(
      C_1 Q_1^{\overline{s}u\overline{u} d} + C_2 Q_2^{\overline{s}u\overline{u} d} 
                                                    \right)+
   V^{\phantom{*}}_{cd}V_{cs}^{*} 
    \left(
      C_1 Q_1^{\overline{s}c\overline{c} d} + C_2 Q_2^{\overline{s}c\overline{c} d} 
                                                    \right) -  V^{\phantom{*}}_{td}V_{ts}^{*} \sum_{i=3}^{6} C_i Q_i^{\overline{s}d}    \right\}                                                   
  = \frac{G_F}{\sqrt{2}}
  \left\{
    V^{\phantom{*}}_{ud}V_{us}^{*} 
    \left[
      C_1 
      \left(
      Q_1^{\overline{s}u\overline{u} d} - Q_1^{\overline{s}c\overline{c} d}
      \right) + C_2
      \left(
      Q_2^{\overline{s}u\overline{u} d} - Q_2^{\overline{s}c\overline{c} d}
      \right)
     \right] -
     V^{\phantom{*}}_{td}V_{ts}^{*} \left[
      C_1 Q_1^{\overline{s}c\overline{c} d} + C_2
     Q_2^{\overline{s}c\overline{c} d} + \sum_{i=3}^{6} C_i Q_i^{\overline{s}d} \right] 
  \right\}\,,
\end{dmath}
where we have used CKM unitarity to eliminate the
$V^{\phantom{*}}_{cd}V_{cs}^{*}$ term. Now, the
$V^{\phantom{*}}_{ud}V_{us}^{*}$ part contains current-current
operators appearing in the GIM-suppressed difference of up and charm,
which does not mix into penguin operators since the divergent part of
the diagrams in Fig.~\ref{fig:penguins} does not depend on the mass of
the quark running in the loop. Penguin operators $Q_{3-6}$ are instead
generated by the RG running with the top CKM factor, due to the
insertion in the diagrams in Fig.~\ref{fig:penguins} of the operators
$Q_{1,2}^{\overline{s}c\overline{c} d}$ in the last line of
eq.~(\ref{eq:heffds1}). Their anomalous dimension also gets
contributions from the insertion of $Q_{3-6}$ in the diagrams of
Figs.~\ref{fig:dc1}, \ref{fig:dc1m} and \ref{fig:penguins}.

Had we performed the matching at $\mathcal{O}(\alpha_s)$, we would
have encountered the diagram (c) in Fig.~\ref{fig:penguins} in the
full theory, with $u$, $c$ and $t$ quarks running in the
loop. Diagrams (b) and (c) are of course not identical, since diagram
(b) is logarithmically divergent while diagram (c) is finite (keep in
mind the two powers of external momenta required by gauge invariance,
see eq.~(\ref{eq:ginvfcnc})). However, if we differentiate with
respect to external momenta or quark masses, then also diagram (b)
becomes finite, and therefore the loop integration is dominated by
momenta of the order of the external momenta and quark masses. After
differentiating, we are thus allowed to replace the $W$ propagator in
diagram (c) with $1/M_{W}^{2}$. In this way, we obtain the following
relation between the amplitudes generated by diagrams (b) and (c) with
quark $i$ running in the loop:
\begin{dmath}[compact,label={eq:pengmatch}]
  \mathcal{A}_{\mathrm{(c)}}^{(i)} = \mathcal{A}_{\mathrm{(b)}}^{(i)} +
  \order{\frac{p^{2},m_{i}^{2}}{M_{W}^{2}}} + K\,,
\end{dmath}
where $p$ denotes external momenta and $K$ is a constant term,
independent on quark masses or momenta, proportional to the matrix
element of operators $Q_{3-6}$. Eq.~(\ref{eq:pengmatch})
implies that the contribution of $u$ and $c$ quarks cancels in the
matching up to the constant $K$ and to negligible corrections of
$\order{\frac{p^{2},m^{2}_{i}}{M_{W}^{2}}}$, while the top quark
contribution generates a nontrivial contribution to
$C_{3-6}^{1}(\mu_W)$ since in the effective theory we do not have
diagram (b) with top quarks running in the loop
\cite{Inami:1980fz,Buras:1991jm}. 

\subsection{Electroweak penguins}
\label{sec:ewp}

We may wonder what happens if we replace in the diagrams of
Figs. \ref{fig:dc1}-\ref{fig:penguins} gluon exchange with photon
exchange. Electromagnetic corrections will also get a logarithmic
enhancement, making them comparable to NLO QCD corrections, since
$\alpha \log (\mu_W^{2}/\mu_h^{2}) \sim \alpha_s$. While we do not need to
resum these logarithmic terms, we need to include them when working at
NLO in QCD \cite{Buras:1992zv,Ciuchini:1993vr}. QED corrections bring
a novelty: the operator basis must be enlarged, due to the electroweak
penguin operators generated by the diagrams in Fig.~\ref{fig:penguins}
replacing the gluon with a photon \cite{Lusignoli:1988fz}. While the
FCNC photon coupling emerging from diagram (a) in
Fig.~\ref{fig:penguins} is equivalent to the gluonic one, the equation
of motion introduces a quark charge dependence, giving rise to
operators with flavour structure $\overline{s}d \sum e_q \overline{q}q$, with
$e_q$ the electric charge of flavour $q$. As in the case of QCD
penguins, strong interaction corrections will generate a new colour
structure, and the left- and right-handed components of the quark
current will renormalize differently, so we need to add four more
operators to the basis:
\begin{dgroup*}
  \begin{dmath}[label={eq:Q7}]
    Q_7^{\overline{s}d} = \frac{3}{2} \overline{s}_L \gamma_\mu d_L \sum_f e_q
    \overline{q}_{fR} \gamma^{\mu} q_{fR}\,,
  \end{dmath}
  \begin{dmath}[label={eq:Q8}]
    Q_8^{\overline{s}d} = \frac{3}{2}\overline{s}_L^{\alpha} \gamma_\mu d_L^{\beta}
    \sum_f e_q \overline{q}_{fR}^{\beta} \gamma^{\mu} q_{fR}^{\alpha} \,,
  \end{dmath}
  \begin{dmath}[label={eq:Q9}]
    Q_9^{\overline{s}d} = \frac{3}{2}\overline{s}_L \gamma_\mu d_L \sum_f e_q
    \overline{q}_{fL} \gamma^{\mu} q_{fL}\,,
  \end{dmath}
  \begin{dmath}[label={eq:Q10}]
    Q_{10}^{\overline{s}d} = \frac{3}{2}\overline{s}_L^{\alpha} \gamma_\mu
    d_L^{\beta} \sum_f e_q \overline{q}_{fL}^{\beta} \gamma^{\mu} q_{fL}^{\alpha}
    \,.
  \end{dmath}
\end{dgroup*}

Let us now briefly discuss the matching for operators $Q_{7-10}$. Also
in this case, these operators get a contribution from the top loop in
the matching from diagram (c) in Fig.~\ref{fig:penguins} with the
exchange of a photon. However, this is not the only
contribution. Indeed, one should also consider diagram (c) with the
exchange of a $Z^{0}$ instead of a photon, and box diagrams with the
exchange of two $W$ bosons. We do not dwell into the details of the
matching, but there is an important point we would like to stress. It
is instructive to consider diagram (c) with the exchange of a $Z^{0}$
in two steps: first, the evaluation of the FC $Z$ coupling from the
loop integration, and then the evaluation of the $Z$ exchange. While
$SU(3)_{c}\otimes U(1)_{\mathrm{em}}$ gauge invariance forbids
dimension four FCNC gluon or photon couplings, this is not the case
for FCNC $Z$ couplings, which can indeed arise at dimension four: a
$Z_\mu \overline{s}_L \gamma^{\mu} d_L$ coupling can be generated once
$SU(2)_L \otimes U(1)_Y$ is broken. However, for this to happen the
diagram must ``feel'' the EW symmetry breaking, so the FC $Z$ coupling
must vanish linearly with $m_q^{2}/M_W^{2}$ for small $m_q$. Thus, the
loop is dominated by the top quark \cite{Inami:1980fz}.~\footnote{This
  $m_q^{2}/M_W^{2}$ suppression is sometimes called ``hard GIM'', as
  opposed to the logarithmic dependence on quark masses which arises
  for example when matching on the dimension six FC gluon and photon
  couplings, the so-called ``soft GIM''. We will discuss more in
  detail GIM suppression in Sec.~\ref{sec:DF2}.} Having obtained a
top-induced FC $Z$ vertex from the loop integration, we consider the
$Z$ exchange in the lower part of diagram (c). Expanding the $Z$
propagator for small momenta in the same way as for the $W$ propagator
in eq.~(\ref{eq:Wtreeampexp}) gives rise to local four-fermion
operators, which can be identified with the (electroweak) penguins
discussed above.

\subsection{Anomalous dimension}
\label{sec:admdf1}

We conclude this Section reporting the LO anomalous dimension for the
full set of four-quark $\Delta F=1$ operators listed above
\cite{Gaillard:1974nj,Altarelli:1974exa,Gilman:1979bc,Guberina:1979ix,Bijnens:1983ye}:
\begin{equation}
  \label{eq:gammadf1LO}
  \scriptstyle
    \gamma_0 =  \left(
    \begin{array}{*{10}{>{\scriptstyle}c}}
      -\frac{6}{N_c} & 6 & -\frac{2}{3 N_c} & \frac{2}{3} &
                                                            -\frac{2}{3
                                                            N_c} &
                                                                   \frac{2}{3}
      & 0 & 0 & 0 & 0 \\
      6 & -\frac{6}{N_c} & 0 & 0 & 0 & 0 & 0 & 0 & 0  & 0\\
      0 & 0 & -\frac{22}{3 N_c} & \frac{22}{3} & -\frac{4}{3N_c} &
                                                                   \frac{4}{3}
      & 0 & 0 & 0 & 0 \\
      0 & 0 & 6 - \frac{2 n_f}{3 N_c} & -\frac{6}{N_c} + \frac{2
                                        n_f}{3} & - \frac{2 n_f}{3 N_c} & \frac{2
                                                                          n_f}{3} & 0 & 0 & 0 & 0\\
      0 & 0 & 0 & 0 & \frac{6}{N_c} & -6 & 0 & 0 & 0 & 0 \\
      0 & 0 & - \frac{2 n_f}{3 N_c} & \frac{2
                                      n_f}{3} & - \frac{2 n_f}{3 N_c}
                                                                 & 6
                                                                   \frac{1-N_c^{2}}{N_c}
                                                                   +\frac{2
                                                                   n_f}{3} & 0 & 0 & 0 & 0 \\
      0 & 0 & 0 & 0 & 0 & 0 & \frac{6}{N_c} & -6 & 0 & 0 \\
      0 & 0 & \frac{-2 (n_u - n_d/2)}{3 N_c} & \frac{2 (n_u -
                                               n_d/2)}{3} &
                                                            \frac{-2
                                                            (n_u
                                                            -
                                                            n_d/2)}{3
                                                            N_c}
                                                                 & \frac{2 (n_u - n_d/2)}{3} & 0 &
                                                                                                   6
                                                                                                   \frac{1-N_c^{2}}{N_c}
              & 0 
                  & 0 \\
      0 & 0 & \frac{2}{3 N_c} & -\frac{2}{3} & \frac{2}{3 N_c} &
                                                                 -\frac{2}{3}
      & 0 & 0 & - \frac{6}{N_c} & 6 \\
      0 & 0 &  \frac{-2 (n_u - n_d/2)}{3 N_c} & \frac{2 (n_u -
                                                n_d/2)}{3} &
                                                             \frac{-2
                                                             (n_u
                                                             -
                                                             n_d/2)}{3
                                                             N_c}
                                                                 & \frac{2 (n_u - n_d/2)}{3} & 0 & 0 &
                                                                                                       6 & -\frac{6}{N_c}
    \end{array}
  \right)\,,
\end{equation}
where $n_{f,u,d}=n_{f,u,d}(\mu)$ is the number of active (up- or
down-type) quarks at the scale $\mu$.

\subsection{The \texorpdfstring{$\Delta I=1/2$}{D12} rule}
\label{sec:di12}

As an example of the applications of the $\Delta S=1$ effective
Hamiltonian, let us consider the $\Delta I=1/2$ rule. We start by
writing down the decay amplitudes for a Kaon (anti-Kaon) to decay in
two pions in terms of final states with different isospin. The
two-pion state in S-wave must have a symmetric isospin wave function,
so it can only be in $I=0$ or $I=2$ states. Denoting by
$\lvert I, I_{3} \rangle$ a state with isospin $I$ and third component
$I_{3}$, we can write
\begin{dgroup*}
  \begin{dmath}[compact,label={eq:pp21}]
    \langle 2,1 \rvert = \frac{1}{\sqrt{2}} 
    \left(
      \langle \pi^{+}\pi^{0} \rvert + \langle \pi^{0}\pi^{+} \rvert
    \right) \,,
  \end{dmath}
  \begin{dmath}[compact,label={eq:pp20}]
    \langle 2,0 \rvert = \frac{1}{\sqrt{6}} 
    \left(
      \langle \pi^{+}\pi^{-} \rvert + \langle \pi^{-}\pi^{+} \rvert +
      2 \langle \pi^{0}\pi^{0} \rvert
    \right) \,,
  \end{dmath}
  \begin{dmath}[compact,label={eq:pp00}]
    \langle 0,0 \rvert = \frac{1}{\sqrt{3}} 
    \left(
      \langle \pi^{+}\pi^{-} \rvert + \langle \pi^{-}\pi^{+} \rvert -
      \langle \pi^{0}\pi^{0} \rvert
    \right) \,.
  \end{dmath}
\end{dgroup*}
The initial state Kaon is a doublet, and the $\Delta S=1$ effective
Hamiltonian has $I=1/2$ and $I=3/2$ components. Coupling the initial
state and the effective Hamiltonian through the Wigner-Eckart theorem,
we have:
\begin{dgroup*}
  \begin{dmath}[compact,label={eq:HKpisospin}]
    \mathcal{H}_{\mathrm{eff}} \lvert K^{+} \rangle =
    (\mathcal{H}_{3/2,1/2} + \mathcal{H}_{1/2,1/2} ) \lvert 1/2,
    1/2 \rangle = \sqrt{\frac{3}{4}} A_{3/2} \lvert 2,1 \rangle -
    \frac{1}{2} A_{3/2} \lvert 1, 1 \rangle +
    A_{1/2} \lvert 1, 1 \rangle\,,
  \end{dmath}
  \begin{dmath}[compact,label={eq:HK0isospin}]
    \mathcal{H}_{\mathrm{eff}} \lvert K^{0} \rangle =
    (\mathcal{H}_{3/2,1/2} + \mathcal{H}_{1/2,1/2} ) \lvert 1/2,
    -1/2 \rangle = \frac{1}{\sqrt{2}} A_{3/2} \lvert 2, 0 \rangle +
    \frac{1}{\sqrt{2}} A_{3/2} \lvert 1,0 \rangle +
    \frac{1}{\sqrt{2}} A_{1/2} \lvert 1,0 \rangle +
    \frac{1}{\sqrt{2}} A_{1/2} \lvert 0,0 \rangle\,.
  \end{dmath}
\end{dgroup*}
Finally, from eqs.~(\ref{eq:pp21})-(\ref{eq:HK0isospin}) we
obtain
\begin{align}
  \label{eq:kpppp0isospin}
   A(K^{+} \to \pi^{+} \pi^{0})  &= \frac{\sqrt{3}}{2\sqrt{2}} A_{3/2}\,,\\
  A(K^{0} \to \pi^{+} \pi^{-}) &= \frac{1}{2\sqrt{3}}
                                 \left(
                                 A_{3/2} + \sqrt{2} A_{1/2}
                                 \right)\,, \nonumber \\
  A(K^{0} \to \pi^{0} \pi^{0}) &= \frac{1}{\sqrt{6}}
                                 \left(
                                  \sqrt{2 }A_{3/2} - A_{1/2}
                                 \right)\,.\nonumber 
\end{align}
It is convenient to define $A_{0,2} = 1/\sqrt{6} A_{1/2,3/2}$ and to
extract, without loss of
generality, a CP-invariant phase $\delta_{0,2}$ from each isospin
amplitude, so that $A_{0,2} \to A_{0,2}^{*}$ under CP:
\begin{align}
    A(K^{0} \to \pi^{+} \pi^{-}) &=  A_{0} e^{i \delta_0} + \frac{A_2}{\sqrt{2}}
    e^{i \delta_2} \,,\nonumber \\
    A(K^{0} \to \pi^{0} \pi^{0}) &= -A_0 e^{i \delta_0} + \sqrt{2} A_2 e^{i \delta_2} \,,\nonumber \\
    A(K^{+} \to \pi^{+} \pi^{0}) &= \frac{3}{2} A_2 e^{i \delta_2}
                                   \,.
    \label{eq:kpipiampli}
\end{align}
We now write the relevant $S$ matrix as:
\begin{equation}
  \label{eq:Smatrix}
  S = 
  \left(
    \begin{array}{ccc}
      K \to K & K \to (\pi\pi)_0& K \to (\pi\pi)_2 \\
      (\pi\pi)_0 \to K & (\pi\pi)_0 \to (\pi\pi)_0 & (\pi\pi)_0 \to
                                                     (\pi\pi)_2 \\
      (\pi\pi)_2 \to K & (\pi\pi)_2 \to (\pi\pi)_0 & (\pi\pi)_2 \to
                                                     (\pi\pi)_2       
    \end{array}
  \right) \simeq \left(
    \begin{array}{ccc}
      1 & -i T_0 & -i T_2 \\
      -i\mathcal{T}(T_0) & e^{i \Delta_0} & 0 \\
      -i\mathcal{T}(T_2)  & 0 & e^{i \Delta_2}
    \end{array}
  \right)\,,
\end{equation}
where we have assumed for simplicity elastic, isospin-conserving
$\pi\pi$ strong interaction scattering, represented by the phases
$\Delta_{0,2}$ for $I=0,2$ $\pi\pi$ states, and we are working at lowest
order in weak interactions. $\mathcal{T}$ denotes time reversal.
Unitarity of the $S$ matrix implies:
\begin{equation}
  \label{eq:SdagS}
  (S^{\dagger} S)_{12} = 0 = -i T_0 + e^{i \Delta} i \mathcal{T}(T_0)^{*}\,.
\end{equation}
Writing, as we did in eq.~(\ref{eq:kpipiampli}),
\begin{equation}
  T_i = A_i e^{i \delta_i}\,, \qquad \mathcal{T}(T_i) =
  \mathcal{CP}(T_i) = A_i^{*} e^{i \delta_i}\,,
  \label{eq:TA}
\end{equation}
we obtain from eq.~(\ref{eq:SdagS})
\begin{equation}
  \label{eq:Watson}
  -i A_0 e^{i \delta_0} + e^{i \Delta_0} i (A_0^{*} e^{i \delta_0})^{*} =
  0 \quad \Rightarrow \delta_0 = \Delta_0/2\,,
\end{equation}
so the CP-even phase of the weak decay amplitude is just half the
phase describing strong-interaction scattering of the final
state. This is known as Watson theorem \cite{Watson:1954uc}, and can
be generalized to the case of multi-channel strong-interactions
unitarity (see ref.~\cite{Franco:2012ck} for an example of application of Watson
theorem to $D$ decays).

Experimentally, $\mathrm{Re}\,A_2 /\mathrm{Re}\,A_0 \sim 1/22$, so $\Delta I=1/2$
transitions happen at a much higher rate than $\Delta I=3/2$. This is
commonly denoted as the $\Delta I=1/2$ rule. One of the most difficult
problems in the study of weak decays, still lacking a complete
solution, is in fact the theoretical prediction of the ratio
$A_2 / A_0$. Let us start from the $\Delta S=1$ Hamiltonian in
eq.~(\ref{eq:heffds1cc_u}). Considering the lowering operator for the
third component of isospin, $I_-$, we have $I_- u = d$, $I_- \overline{d} =
- \overline{u}$, and $I_-d = I_- \overline{u} = 0$. Then the action of
$I_{-}$ on the operator $Q_{-}$ of eq.~(\ref{eq:plusminusbasis}) is
given by
\begin{dmath}[compact,label={eq:ImOm}]
 2 I_{-} Q_{-} =   \overline{s}_{L}\gamma^{\mu} (I_{-} u_{L})
 \overline{u}_{L} \gamma_{\mu} d_{L} +  \overline{s}_{L}\gamma^{\mu} u_{L}
 (I_{-} \overline{u}_{L}) \gamma_{\mu} d_{L} + \overline{s}_{L}\gamma^{\mu} u_{L}
 \overline{u}_{L} \gamma_{\mu} (I_{-} d_{L}) - \overline{s}_{L}^{\alpha}\gamma^{\mu}
 (I_{-} u_{L}^{\beta})
 \overline{u}_{L}^{\beta} \gamma_{\mu} d_{L}^{\alpha} - \overline{s}_{L}^{\alpha}\gamma^{\mu} u_{L}^{\beta}
 (I_{-} \overline{u}_{L}^{\beta}) \gamma_{\mu} d_{L}^{\alpha} - \overline{s}_{L}^{\alpha}\gamma^{\mu} u_{L}^{\beta}
 \overline{u}_{L}^{\beta} \gamma_{\mu} (I_{-} d_{L}^{\alpha}) = \overline{s}_{L}\gamma^{\mu} (d_{L})
 \overline{u}_{L} \gamma_{\mu} d_{L} - \overline{s}_{L}^{\alpha}\gamma^{\mu}
 (d_{L}^{\beta})
 \overline{u}_{L}^{\beta} \gamma_{\mu} d_{L}^{\alpha} = 0\,,
\end{dmath}
where in the last step we have fierzed the Dirac structure. Thus,
$Q_{-}$ is the lower component of an isospin doublet. Doing the same
exercise on $Q_{+}$ shows instead that $Q_{+}$ is an admixture of $I=1/2$
and $I=3/2$. Therefore, the enhancement of $C_{-}$ over $C_{+}$ due to RG
evolution goes in the right direction to explain the $\Delta I=1/2$
rule \cite{Altarelli:1974exa}, although it can only account for about
a factor of two in the amplitude ratio.

Another contribution to the $\Delta I=1/2$ rule comes from QCD penguin
operators $Q_{3...6}$ in eqs.~(\ref{eq:Q3})-(\ref{eq:Q6})
\cite{Vainshtein:1975sv}, since these operators are isospin
doublets. Still, the effect must largely come from the matrix
elements, since perturbative RG effects cannot bring the amplitude
ratio close to the experimental value. Unfortunately, computing the
relevant matrix elements from first principles with Lattice QCD is a
tremendous task. Indeed, this calculation poses all the most difficult
challenges to lattice QCD calculations: final state interactions,
chiral symmetry breaking, power divergences, disconnected diagrams,
etc. Thus, it comes as no surprise that only very recently a
pioneering lattice calculation of the matrix elements relevant for the
$\Delta I=1/2$ rule has been achieved \cite{Boyle:2012ys}. According
to this calculation, there is a large deviation from the Vacuum
Insertion Approximation (VIA) in the matrix elements of
current-current operators, causing a negative interference, and thus a
large cancellation, in $\Delta I=3/2$ matrix elements, which are
therefore suppressed with respect to $\Delta I=1/2$ ones. Such
deviation from the VIA, with the corresponding negative interference,
is also seen in $\Delta S=2$ matrix elements
\cite{Carrasco:2013jda}. However, the same calculation failed to
reproduce the phase of the $\Delta I = 1/2$ amplitude, casting some
doubts on the robustness of the estimate of final state
interactions. Fortunately, with increased statistics and an improved
treatment of the two-pion state, a much better agreement with the
experimental value of $\delta_{0}$ was very recently obtained
\cite{Sachrajda}. We are looking forward to the corresponding update
of the results on the $\Delta I = 1/2$ rule, and hopefully to an
independent confirmation from another lattice collaboration in the
future.

\section{Effective Hamiltonians for \texorpdfstring{$\mathbf{\Delta F=2}$}{DF2} processes}
\label{sec:DF2}

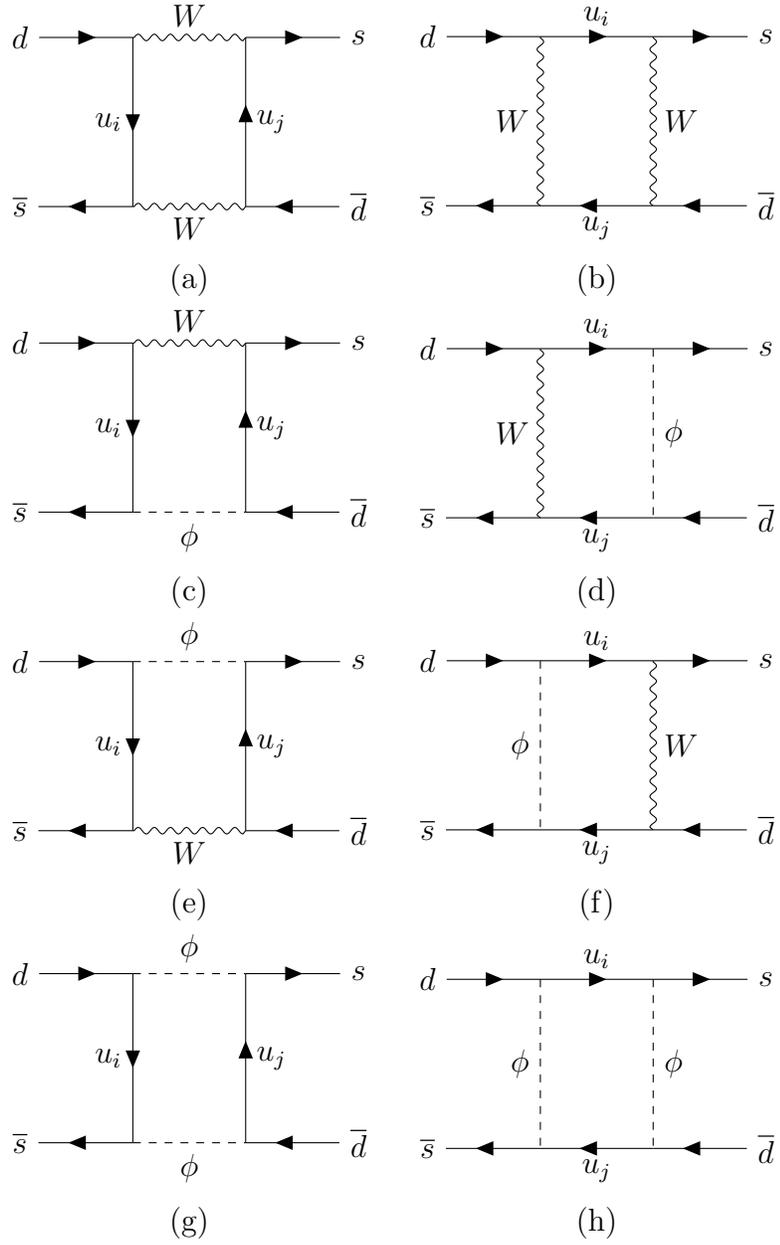
\begin{figure}[!tp]
  \centering
  \begin{tabular}{cc}
    \feynmandiagram [layered layout, horizontal=a to b] {
    i1 [particle=\(d\)]
    -- [fermion] a
    -- [photon, edge label=\(W\)] b
    -- [fermion] f1 [particle=\(s\)],
    i2 [particle=\(\overline s\)]
    -- [anti fermion] c
    -- [photon, edge label'=\(W\)] d
    -- [anti fermion] f2 [particle=\(\overline d\)],
    { [same layer] a -- [fermion,
    edge label'=\(u_{i}\)] c },
    { [same layer] b -- [anti fermion, edge label=\(u_{j}\)] d},
    };
    &
       \feynmandiagram [layered layout, horizontal=a to b] {
    i1 [particle=\(d\)]
    -- [fermion] a
    -- [fermion, edge label=\(u_{i}\)] b
    -- [fermion] f1 [particle=\(s\)],
    i2 [particle=\(\overline s\)]
    -- [anti fermion] c
    -- [anti fermion, edge label'=\(u_{j}\)] d
    -- [anti fermion] f2 [particle=\(\overline d\)],
    { [same layer] a -- [photon, edge label'=\(W\)] c },
    { [same layer] b -- [photon, edge label=\(W\)] d},
      };   \\
    (a) & (b) \\
    \feynmandiagram [layered layout, horizontal=a to b] {
    i1 [particle=\(d\)]
    -- [fermion] a
    -- [photon, edge label=\(W\)] b
    -- [fermion] f1 [particle=\(s\)],
    i2 [particle=\(\overline s\)]
    -- [anti fermion] c
    -- [scalar, edge label'=\(\phi\)] d
    -- [anti fermion] f2 [particle=\(\overline d\)],
    { [same layer] a -- [fermion,
    edge label'=\(u_{i}\)] c },
    { [same layer] b -- [anti fermion, edge label=\(u_{j}\)] d},
    };
    &
       \feynmandiagram [layered layout, horizontal=a to b] {
    i1 [particle=\(d\)]
    -- [fermion] a
    -- [fermion, edge label=\(u_{i}\)] b
    -- [fermion] f1 [particle=\(s\)],
    i2 [particle=\(\overline s\)]
    -- [anti fermion] c
    -- [anti fermion, edge label'=\(u_{j}\)] d
    -- [anti fermion] f2 [particle=\(\overline d\)],
    { [same layer] a -- [photon, edge label'=\(W\)] c },
    { [same layer] b -- [scalar, edge label=\(\phi\)] d},
    };     \\
    (c) & (d) \\
    \feynmandiagram [layered layout, horizontal=a to b] {
    i1 [particle=\(d\)]
    -- [fermion] a
    -- [scalar, edge label=\(\phi\)] b
    -- [fermion] f1 [particle=\(s\)],
    i2 [particle=\(\overline s\)]
    -- [anti fermion] c
    -- [photon, edge label'=\(W\)] d
    -- [anti fermion] f2 [particle=\(\overline d\)],
    { [same layer] a -- [fermion,
    edge label'=\(u_{i}\)] c },
    { [same layer] b -- [anti fermion, edge label=\(u_{j}\)] d},
    };
    &
       \feynmandiagram [layered layout, horizontal=a to b] {
    i1 [particle=\(d\)]
    -- [fermion] a
    -- [fermion, edge label=\(u_{i}\)] b
    -- [fermion] f1 [particle=\(s\)],
    i2 [particle=\(\overline s\)]
    -- [anti fermion] c
    -- [anti fermion, edge label'=\(u_{j}\)] d
    -- [anti fermion] f2 [particle=\(\overline d\)],
    { [same layer] a -- [scalar, edge label'=\(\phi\)] c },
    { [same layer] b -- [photon, edge label=\(W\)] d},
      };     \\
    (e) & (f) \\
    \feynmandiagram [layered layout, horizontal=a to b] {
    i1 [particle=\(d\)]
    -- [fermion] a
    -- [scalar, edge label=\(\phi\)] b
    -- [fermion] f1 [particle=\(s\)],
    i2 [particle=\(\overline s\)]
    -- [anti fermion] c
    -- [scalar, edge label'=\(\phi\)] d
    -- [anti fermion] f2 [particle=\(\overline d\)],
    { [same layer] a -- [fermion,
    edge label'=\(u_{i}\)] c },
    { [same layer] b -- [anti fermion, edge label=\(u_{j}\)] d},
    };
    &
       \feynmandiagram [layered layout, horizontal=a to b] {
    i1 [particle=\(d\)]
    -- [fermion] a
    -- [fermion, edge label=\(u_{i}\)] b
    -- [fermion] f1 [particle=\(s\)],
    i2 [particle=\(\overline s\)]
    -- [anti fermion] c
    -- [anti fermion, edge label'=\(u_{j}\)] d
    -- [anti fermion] f2 [particle=\(\overline d\)],
    { [same layer] a -- [scalar, edge label'=\(\phi\)] c },
    { [same layer] b -- [scalar, edge label=\(\phi\)] d},
      }; \\
    (g) & (h) 
  \end{tabular}
  \caption{Feynman diagrams for $\overline{s}d \to \overline{d}s$ transitions in
    the SM.}
  \label{fig:df2diags}
\end{figure}

Let us now turn to the transitions that give the very stringent
bounds reported in Fig.~\ref{fig:df2bounds}: $\Delta F=2$
processes. In particular, let us consider $\overline{s} d \to \overline{d} s$
transitions. Such FCNC processes cannot arise at the tree level in the
SM, so we must consider one-loop contributions. These contributions
must be finite, since renormalizability of the SM implies that no
counterterm for FCNC amplitudes can arise. In the
't-Hooft-Feynman gauge we have the diagrams in
Fig.~\ref{fig:df2diags}.

Let us start by computing diagram (a). Neglecting external momenta,
the amplitude reads
\begin{align}
  \label{eq:diagaft}
  i\mathcal{A}^{(a)} =& \int \overline{u}_{s} 
  \left(\frac{ig_{2}}{\sqrt{2}} \right) \gamma_{\mu} P_{L} V_{u_{j}s}^{*}
  \frac{i}{\slashed{q} - m_{u_{j}}} 
  \left(\frac{ig_{2}}{\sqrt{2}} \right) \gamma_{\nu} P_{L} V^{\phantom{*}}_{u_{j}d}
                        v_{d} \\
  &\times \overline{v}_{s} 
  \left(\frac{ig_{2}}{\sqrt{2}} \right) \gamma^{\nu} P_{L} V_{u_{i}s}^{*}
  \frac{i}{\slashed{q} - m_{u_{i}}} 
  \left(\frac{ig_{2}}{\sqrt{2}} \right) \gamma^{\mu} P_{L} V^{\phantom{*}}_{u_{i}d}
  u_{d} 
  \left(
    \frac{-i}{q^{2}-M_{W}^{2}}
  \right)^{2} \frac{\mathrm{d}^{4}q}{(2 \pi)^{4}}\,. \nonumber
\end{align}
Left-handed projectors kill the quark mass terms in the numerator of
quark propagators, so we obtain
\begin{equation}
  \label{eq:diagasimpl}
  i\mathcal{A}^{(a)} = \frac{g_2^{4}}{4}V_{u_is}^{*} V^{\phantom{*}}_{u_id}V_{u_js}^{*}V^{\phantom{*}}_{u_jd}
  \overline{u}_s \gamma_\mu \gamma^{\alpha}
  \gamma_\nu P_L v_d  \overline{v}_s \gamma^{\nu} \gamma^{\beta}
  \gamma^{\mu} P_L u_d I_{\alpha\beta}^{ij}\,,
\end{equation}
with
\begin{equation}
  \label{eq:Ialbeij}
  I_{\alpha\beta}^{ij} \equiv \int
    \frac{q_\alpha q_\beta}{(q^{2}-M_W^{2})^{2}(q^{2}-m^{2}_{u_i})(q^{2}-m^{2}_{u_j})}
  \frac{\mathrm{d}^{4}q}{(2 \pi)^{4}}\,.
\end{equation}
We can simplify the integral using partial fractioning in the form
\begin{equation}
  \label{eq:pf}
  \frac{m^{2}_{u_i}-m^{2}_{u_j}}{(q^{2}-m^{2}_{u_i})(q^{2}-m^{2}_{u_j})} =
  \frac{1}{q^{2}-m^{2}_{u_i}} - \frac{1}{q^{2}-m^{2}_{u_j}}\,, 
\end{equation}
obtaining
\begin{equation}
  \label{eq:Ialbeijtoi}
  I_{\alpha\beta}^{ij}=
  \frac{I^{i}_{\alpha\beta} - I^{j}_{\alpha\beta}}{m^{2}_{u_i}-m^{2}_{u_j}}
\end{equation}
with
\begin{dmath}
  I^{i}_{\alpha\beta}= \int
    \frac{q_\alpha q_\beta}{(q^{2}-M_W^{2})^{2}(q^{2}-m^{2}_{u_i})}
  \frac{\mathrm{d}^{4}q}{(2 \pi)^{4}} = \frac{g_{\alpha\beta}}{4}  \int
    \frac{q^{2} + (-m^{2}_{u_i} +m^{2}_{u_i})}{(q^{2}-M_W^{2})^{2}(q^{2}-m^{2}_{u_i})}
                       \frac{\mathrm{d}^{4}q}{(2 \pi)^{4}}=
   \frac{g_{\alpha\beta}}{4} m^{2}_{u_i} \int
    \frac{1}{(q^{2}-M_W^{2})^{2}(q^{2}-m^{2}_{u_i})}
                       \frac{\mathrm{d}^{4}q}{(2 \pi)^{4}} + K\,,
  \label{eq:Ialbei}
\end{dmath}
where $K$ represents terms independent on $m^{2}_{u_i}$ which drop in
$I_{\alpha\beta}^{ij}$. Introducing Feynman parameters as
in eq.~(\ref{eq:Fpars1}), we obtain
\begin{dmath}
  \label{eq:Ialbeint}
  \int
    \frac{1}{(q^{2}-M_W^{2})^{2}(q^{2}-m^{2}_{u_i})}
    \frac{\mathrm{d}^{4}q}{(2 \pi)^{4}}
    = 2 \int_0^{1} \mathrm{d}x
  \int\frac{\mathrm{d}^{4}q}{(2 \pi)^{4}}
                       \frac{x}{[(q^{2}-M_W^{2})x+(q^{2}-m^{2}_{u_i})(1-x)]^{3}}
                       = 2 \int_0^{1} \mathrm{d}x
  \int\frac{\mathrm{d}^{4}q}{(2 \pi)^{4}}
  \frac{1}{[q^{2}-M_W^{2} x-m^{2}_{u_i}(1-x)]^{3}}=
  - \frac{i}{16 \pi^{2}} \int_0^{1} \mathrm{d}x\frac{x}{M_W^{2} x+m^{2}_{u_i}(1-x)}=
  - \frac{i}{16 \pi^{2} M_{W}^{2}} \int_0^{1} \mathrm{d}x\frac{x}{x+x_{i}(1-x)}=
  - \frac{i}{16 \pi^{2} M_{W}^{2}} \int_0^{1} \mathrm{d}x\frac{x}{x_i+x(1-x_i)}=
  - \frac{i}{16 \pi^{2} M_{W}^{2}} \int_0^{1} \frac{\mathrm{d}x}{1-x_i}
  \frac{(1-x_i)x+x_i-x_i}{x_i+x(1-x_i)}=
  - \frac{i}{16 \pi^{2} M_{W}^{2}} \left(\frac{-x_i}{1-x_i}\int_0^{1}
    \frac{\mathrm{d}x}{(1-x_i)x + x_i} +
  \frac{1}{1-x_i}\right)=
  - \frac{i}{16 \pi^{2} M_{W}^{2}} \left(\frac{1}{1-x_i} + \frac{x_i \log x_i}{(1-x_i)^{2}}
  \right)\,,
\end{dmath}
where $x_i = m_{u_i}^{2}/M_W^{2}$.  Thus, up to terms that do not depend
on $m_{u_i}^{2}$, we have
\begin{dmath}[label={eq:Ialbei2}]
  I_{\alpha\beta}^{i}= -\frac{g_{\alpha\beta}}{4}\frac{i}{16 \pi^{2}} J(x_i)\,,
\end{dmath}
with
\begin{dmath}[label={eq:Jxi}]
  J(x_i)=\frac{x_i}{1-x_i} + \frac{x_i^{2} \log x_i}{(1-x_i)^{2}}\,,
\end{dmath}
and therefore
\begin{dmath}[label={eq:Ialbeij2}]
  I_{\alpha\beta}^{ij}= -\frac{g_{\alpha\beta}}{4M_W^{2}}\frac{i}{16 \pi^{2}} A(x_i,x_j)\,,
\end{dmath}
with
\begin{dmath}[label={eq:Axixj}]
  A(x_i,x_j)=\frac{J(x_i)-J(x_j)}{x_i-x_j}\,.
\end{dmath}
We now turn to the Dirac structure
\begin{dmath}[label={eq:dstringdf21}]
\overline{u}_s \gamma_\mu \gamma_\alpha
  \gamma_\nu P_L v_d  \overline{v}_s \gamma^{\nu} \gamma^{\beta}
  \gamma^{\mu} P_L u_d  
\end{dmath}
and use again the Fierz identity in eq.~(\ref{eq:Fierz1}) to obtain 
\begin{dmath}[label={eq:dstringdf2f}]
- \frac{1}{2} \overline{v}_s \gamma^{\rho} P_L
  v_d \overline{u}_s \gamma_\mu \gamma_\alpha
  \gamma_\nu \gamma_\rho P_R \gamma^{\nu} \gamma^{\alpha}
  \gamma^{\mu}  u_d  = 4 \overline{v}_s \gamma^{\mu} P_L
  v_d \overline{u}_s \gamma_\mu P_L u_d\,.
\end{dmath}
Putting everything together we obtain the amplitude generated by
diagram (a) as
\begin{dmath}[label={eq:ampda}]
i \mathcal{A}^{(a)} = - \frac{i}{16 \pi^{2}} \frac{4 g^{4}}{16 M_W^{2}}
\sum_{i,j=u,c,t} v^{*}_{is}V^{\phantom{*}}_{id} V^{*}_{js}V^{\phantom{*}}_{jd} A(x_i,x_j) \overline{v}_s \gamma^{\mu} P_L
  v_d \overline{u}_s \gamma_\mu P_L u_d = - \frac{i G_F^{2} M_W^{2}}{2 \pi^{2}}
  \sum_{i,j} \lambda_{sd}^{i} \lambda_{sd}^{j} A(x_i,x_j) \overline{v}_s \gamma^{\mu} P_L
  v_d \overline{u}_s \gamma_\mu P_L u_d\,,
\end{dmath}
with $\lambda_{sd}^{i}=V^{*}_{is}V^{\phantom{*}}_{id}$.

Diagram (b) in Fig.~\ref{fig:df2diags} is identical to diagram (a)
if we exchange an incoming $s$ antiquark with an outgoing $s$ quark
and viceversa:
\begin{dmath}[label={eq:ampdb}]
i \mathcal{A}^{(b)} = \frac{i G_F^{2} M_W^{2}}{2 \pi^{2}}
  \sum_{i,j} \lambda_{sd}^{i} \lambda_{sd}^{j} A(x_i,x_j) \overline{u}_s \gamma^{\mu} P_L
  u_d \overline{v}_s \gamma_\mu P_L v_d\,.
\end{dmath}
We now notice that the amplitudes generated by diagrams (a) and (b)
can be written as the matrix element of a local operator, so we can
introduce the following $\Delta S=2$ effective Hamiltonian:
\begin{equation}
  \label{eq:Heffdf2}
  \mathcal{H}_{\mathrm{eff}}^{\Delta S = 2} = C \overline{s}\gamma^{\mu} P_L
  d \overline{s} \gamma_\mu P_L d\,,
\end{equation}
with $C$ a Wilson coefficient with mass dimension $-2$. This effective
Hamiltonian generates the following amplitude:
\begin{dmath}[label={eq:Adf2eff}]
  i T^{\mathcal{H}} (\overline{s}d \to \overline{d}s) = -i C \langle \overline{d}s
  \lvert
  \overline{s} \gamma^{\mu} P_L
  d \overline{s} \gamma_\mu P_L d \rvert \overline{s}d \rangle = -2 i C 
  \left(
    \overline{u}_s \gamma^{\mu} P_L
  v_d \overline{v}_s \gamma_\mu P_L u_d - \overline{u}_s \gamma^{\mu} P_L
  u_d \overline{v}_s \gamma_\mu P_L v_d  
  \right)\,.
\end{dmath}
Matching it with the amplitude in the full theory $i
\mathcal{A}^{(a)+(b)}$ we obtain
\begin{dmath}[label={eq:Capb}]
  C^{(a)+(b)} = \frac{G_F^{2} M_W^{2}}{4 \pi^{2}} \sum_{i,j} \lambda_{sd}^{i} \lambda_{sd}^{j} A(x_i,x_j)\,. 
\end{dmath}
Evaluating diagrams (c) to (h) in Fig.~\ref{fig:df2diags} we obtain
\begin{dgroup*}
  \begin{dmath}
    C^{(c)+(d)} = C^{(e)+(f)} = - \frac{G_F^{2} M_W^{2}}{4 \pi^{2}}
    \sum_{i,j} \lambda_{sd}^{i} \lambda_{sd}^{j} A^{\prime}(x_i,x_j) x_i
    x_j\,,
  \end{dmath}
  \begin{dmath}
    C^{(g)+(h)} = \frac{1}{4} \frac{G_F^{2} M_W^{2}}{4 \pi^{2}} \sum_{i,j}
    \lambda_{sd}^{i} \lambda_{sd}^{j} A(x_i,x_j) x_i x_j\,,
  \end{dmath}
\end{dgroup*}
with
\begin{equation}
  A^{\prime}(x_i,x_j) =
  \frac{J^{\prime}(x_i)-J^{\prime}(x_j)}{x_i-x_j}\,,\qquad
  J^{\prime}(x) = \frac{1}{1-x} + \frac{x \log x}{(1-x)^{2}}\,.
  \label{eq:aprimej}
\end{equation}
Putting everything together we obtain
\begin{dmath}[label={eq:Cds2}]
  C = \frac{G_F^{2} M_W^{2}}{4 \pi^{2}} \sum_{i,j} \lambda_{sd}^{i}
  \lambda_{sd}^{j} \overline{A}(x_i,x_j)\,,
\end{dmath}
where
\begin{dmath}[label={eq:Abarij}]
  \overline{A}(x_i,x_j) = A(x_i,x_j) - x_i x_j A^{\prime}(x_i,x_j) +
  \frac{1}{4} x_i x_j A(x_i,x_j)\,.
\end{dmath}
Next, we use CKM unitarity in the form
\begin{equation}
  \label{eq:CKMunids2}
  \lambda_{sd}^{u} = - \lambda_{sd}^{c} - \lambda_{sd}^{t}
\end{equation}
to eliminate $\lambda_{sd}^{u}$, and we obtain
\begin{align}
  \sum_{i,j} \lambda_{sd}^{i} \lambda_{sd}^{j} \overline{A}(x_i,x_j) =& \left(
    \lambda_{sd}^{c} + \lambda_{sd}^{t} \right)^{2} \overline{A}(x_u,x_u) + 2
  \lambda_{sd}^{c} \lambda_{sd}^{t} \overline{A}(x_c,x_t) \nonumber \\
  &+\left(\lambda_{sd}^{c} \right)^{2} \overline{A}(x_c,x_c) +
  \left(\lambda_{sd}^{t} \right)^{2} \overline{A}(x_t,x_t) \nonumber
  \\
  &- 2 \lambda_{sd}^{t}
  \left( \lambda_{sd}^{c} + \lambda_{sd}^{t} \right) \overline{A}(x_u,x_t) - 2
  \lambda_{sd}^{c} \left( \lambda_{sd}^{c} + \lambda_{sd}^{t} \right)
    \overline{A}(x_c,x_u) \nonumber \\
  =& \left( \lambda_{sd}^{t} \right)^{2} S_0(x_t) + \left(
    \lambda_{sd}^{c} \right)^{2} S_0(x_c) + 2 \lambda_{sd}^{t}
  \lambda_{sd}^{c} S_0(x_c,x_t)\,,
  \label{eq:ds2sum}
\end{align}
with
\begin{dgroup*}
  \begin{dmath}
    S_0(x_t) = \overline{A}(x_t,x_t) + \overline{A}(x_u,x_u) - 2
    \overline{A}(x_u,x_t)\,,
  \end{dmath}
  \begin{dmath}
    S_0(x_c) = \overline{A}(x_c,x_c) + \overline{A}(x_u,x_u) - 2
    \overline{A}(x_u,x_c)\,,
  \end{dmath}
  \begin{dmath}
    S_0(x_c,x_t) = \overline{A}(x_c,x_t) + \overline{A}(x_u,x_u) -
    \overline{A}(x_u,x_c) - \overline{A}(x_u,x_t) \,.
  \end{dmath}
\end{dgroup*}
We finally obtain
\begin{dmath}[label={eq:Heffds2}]
  \mathcal{H}_{\mathrm{eff}}^{\Delta S=2} = \frac{G_{F}^{2}M_{W}^{2}}{4 \pi^{2}} 
  \left[
    \left(
    \lambda_{sd}^{t} 
  \right)^{2} S_{0}(x_{t}) +  \left(
    \lambda_{sd}^{c} 
  \right)^{2} S_{0}(x_{c}) + 2 \lambda_{sd}^{t} \lambda_{sd}^{c} S_{0}(x_{c},x_{t})
  \right] \overline{s}\gamma^{\mu} P_{L}
  d \overline{s} \gamma_{\mu} P_{L} d\,.
\end{dmath}
Notice that the $S_0$ functions are differences of $\overline{A}$
functions with different arguments. If we Taylor-expand $\overline{A}$ in
powers of quark masses, the zeroth-order term cancels in $S_0$. Thus,
for massless quarks no FCNC vertices arise, and the latter are
suppressed by the GIM cancellation mechanism \cite{Glashow:1970gm}.
For small $x$ the loop function $S_0$ vanishes linearly. Neglecting
the contribution of the third family, the FCNC coupling in
eq.~(\ref{eq:Heffdf2}) is proportional to
\begin{equation}
  \label{eq:smallx}
  G_F^{2} M_W^{2} \lambda^{2} x_c = G_F^{2} m_c^{2} \lambda^{2} \,, 
\end{equation}
where $\lambda$ is the Wolfenstein parameter introduced in
eq.~(\ref{eq:Wolfenstein-Buras}).  In other words, the (hard) GIM
mechanism converts the effective $\Delta S=2$ coupling from an
$\mathcal{O}(1/M_W^{2})$ effect to an $\mathcal{O}(m_c^{2}/M_W^{4})$
one. Notice also that SM fermions do not decouple, since $S_0$ grows
linearly for large $x$; this non-decoupling explains the relevance of
the top quark even in low-energy FCNC processes, and can be easily
understood since the coupling to the would-be Goldstone bosons is
proportional to fermion masses (or, more precisely, to Yukawa
couplings).

\subsection{Locality and higher dimensional operators}
\label{sec:local}

The matching calculation we performed above might look as an academic
exercise: what can we say on $K$-$\overline{K}$ mixing from a matrix
element between zero-momentum quarks with no strong interactions?
Indeed, neglecting external momenta a local operator is generated by
construction, but this is a reasonable approximation only if the
dependence on external momenta is negligible. Let us now discuss this
problem in some detail. First of all, we notice that diagrams
containing up quarks only are in fact non-local contributions, which
cannot be estimated by matching onto a local effective Hamiltonian.
Indeed, up-quark contributions cancel in the matching against the
matrix element of two $\Delta S=1$ effective Hamiltonians. The latter
represents the long-distance contribution to $K$-$\overline{K}$ mixing
which must be evaluated using some non-perturbative method. This point
can be explicitly checked using the same argument discussed in
Sec.~\ref{sec:penguins}. The diagrams in Fig.~\ref{fig:df2diags}
become divergent when substituting the $W$ propagator with a local
interaction, which corresponds to the matrix element of two
$\Delta S=1$ effective Hamiltonians. However, differentiating thrice
with respect to quark masses and/or momenta, the diagram becomes
convergent even when the $W$ propagator becomes local, allowing us to
identify the diagrams in Fig.~\ref{fig:df2diags} with the matrix
element of two $\Delta S=1$ effective Hamiltonians up to a constant
term, to a term proportional to $p^{2}/M_{W}^{2}$ and to a term
proportional to $m_{i}^{2}/M_{W}^{2}$, where $p$ represents external
momenta and $m_{i}$ the mass of the quark running in the loop.

However, if we look at the CP-odd part of the effective
Hamiltonian we can drop the up-quark contribution since we can always
choose a phase convention such that Im$\lambda^{u}_{sd}=0$. In this
convention, we have Im$\lambda^{c}_{sd}= - $  Im$\lambda^{t}_{sd}$, so
that
\begin{dgroup*}
  \begin{dmath}
    \mathrm{Im} \left( \lambda^{t}_{sd} \right)^{2} = 2 \mathrm{Im}\,
    \lambda^{t}_{sd} \mathrm{Re}\, \lambda^{t}_{sd}\,,
  \end{dmath}
  \begin{dmath}
    \mathrm{Im} \left( \lambda^{c}_{sd} \right)^{2} = - 2 \mathrm{Im}\,
    \lambda^{t}_{sd} \mathrm{Re} \, \lambda^{c}_{sd} \,,
  \end{dmath}
  \begin{dmath}
    \mathrm{Im} \, \lambda^{c}_{sd} \lambda^{t}_{sd} = \mathrm{Im}\,
    \lambda^{t}_{sd} \left( \mathrm{Re} \, \lambda^{c}_{sd} - \mathrm{Re}
      \, \lambda^{t}_{sd} \right)\,,
  \end{dmath}
\end{dgroup*}
leading to the following Wilson coefficient:
\begin{dmath}[label={eq:imheff}]
  \frac{G_F^{2}M_W^{2}}{2 \pi^{2}} 
    \mathrm{Im}\,
    \lambda_{sd}^{t}
    \left[
       \mathrm{Re}\, \lambda_{sd}^{t} 
       \left(
         S_0(x_t) - S_0(x_t,x_c) 
       \right) - \mathrm{Re}\, \lambda_{sd}^{c} \left(
         S_0(x_c) - S_0(x_t,x_c) 
       \right) 
    \right]\,.
\end{dmath}
We can indeed check that loops of up quarks drop in the differences of
$S_0$ functions in eq.~(\ref{eq:imheff}), leaving us with the
following expressions:
\begin{dgroup*}
  \begin{dmath}
    S_0(x_t) - S_0(x_c,x_t) = \overline{A}(x_t,x_t) - \overline{A}(x_t,x_c) -
    \overline{A}(x_t,x_u) + \overline{A}(x_c,x_u)\,,
  \end{dmath}
  \begin{dmath}
    S_0(x_c) - S_0(x_c,x_t) = \overline{A}(x_t,x_u) - \overline{A}(x_t,x_c) -
    \overline{A}(x_c,x_u) + \overline{A}(x_c,x_c)\,.
  \end{dmath}
\end{dgroup*}
Now,
\begin{dgroup*}
  \begin{dmath}
    S_0(x_t) - S_0(x_c,x_t) = \frac{m_t^{2}}{M_W^{2}} S_t(x_t,x_c)\,,
  \label{eq:S0GIMt}
  \end{dmath}
  \begin{dmath}
    S_0(x_c) - S_0(x_c,x_t) = \frac{m_c^{2}}{M_W^{2}} S_c(x_t,x_c)\,,
  \label{eq:S0GIMc}  
  \end{dmath}
\end{dgroup*}
with $S_{c,t}(x_t,x_c)$ non-vanishing in the limit $x_c \to 0$.
Had we kept the dependence on external momenta $p$ in the evaluation of
the loop functions, we would have obtained terms of
$\mathcal{O}(p^{2}/M_W^{2})$ (or higher) in $S_0$, corresponding to a correction of
$\mathcal{O}(p^{2}/m_t^{2})$ to eq.~(\ref{eq:S0GIMt}) and to an
$\mathcal{O}(p^{2}/m_c^{2})$ correction to eq.~(\ref{eq:S0GIMc}). The
first one is fully negligible, but the second one is potentially
relevant, since $m_K^{2}/m_c^{2} \sim 10\%$. Fortunately, we have a systematic
way to keep these corrections into account, since a contribution of
$\mathcal{O}(p^{2}/M_W^{2})$ to the amplitude can be described by the
matrix element of a local operator of dimension eight. To perform the
matching of the full amplitude onto the effective theory including
dimension eight operators, we need to expand the diagrams we computed
above at $\mathcal{O}(p^{2}/M_W^{2})$. However, this is not enough since
at dimension eight the operator basis includes an operator involving
the commutator of two covariant derivatives,
\begin{equation}
  \label{eq:d8op}
  g_s \overline{s} \gamma_\mu P_L \tilde{G}^{\mu\nu}d \overline{s} \gamma_\nu d\,,
\end{equation}
which has vanishing matrix element on four-quark states. Therefore, we
need to consider external states with four quarks and a gluon to
complete the matching at dimension eight. In this way we can estimate
the corrections of $\mathcal{O}(p^{2}/m_c^{2})$, which turn out to be at
the few percent level \cite{Cata:2003mn,khighdim}.

To summarize, the expansion in local operators is safe and
systematically improvable by going to dimension eight operators for
the CP violating part of the Hamiltonian, while the CP conserving one
is dominated by long distance contributions, which must be evaluated
as a long distance matrix element of two $\Delta S=1$ effective
Hamiltonians.

\subsection{QCD corrections}
\label{sec:QCDcorrDF2}

The inclusion of LO QCD corrections goes exactly along the lines of
Sec. \ref{sec:ccLO}, the only difference being that in the $\Delta
F=2$ case we do not need to introduce the second operator with a
different colour structure since we can Fierz Dirac
indices to fall back on the original operator in
eq.~(\ref{eq:Heffds2}):
\begin{dmath}[label={eq:ds2Fierz}]
  \overline{s}^{\alpha}\gamma^{\mu} P_Ld^{\beta} \overline{s}^{\beta} \gamma_\mu
  P_L d^{\alpha} = \overline{s}\gamma^{\mu} P_L  d \overline{s} \gamma_\mu P_L
  d\,.
\end{dmath}
The relevant anomalous dimension can then be obtained by a
straightforward combination of the results in Sec.~\ref{sec:ccLO},
yielding the same result as for $Q_+$ in
eq.~(\ref{eq:plusminusbasis}), namely
\begin{dmath}[label={eq:gammadf2LO}]
  \gamma_0 = 6 \frac{N_c-1}{N_c}\,,
\end{dmath}
leading to a suppression of $C(\mu_h)$ with respect to $C(M_W)$.

The calculation of NLO (and of NNLO) QCD corrections is more involved
and goes beyond the scope of these lectures; the interested reader can
find all the details in refs.~\cite{Buras:1990fn,Herrlich:1993yv,Herrlich:1995hh,Nierste:1995fr,Herrlich:1996vf,Buras:1998raa,Brod:2010mj,Brod:2011ty}.

\subsection{\texorpdfstring{$\Delta B=2$}{DB2} effective Hamiltonian}
\label{sec:HeffDB2}

In the previous paragraphs we introduced the $\Delta S=2$ effective
Hamiltonian. If we consider instead $\overline{b}q \to \overline{q}b$
transitions, with $q=d,s$, we see that in this case at the scale
$\mu_h \sim m_b$ up and charm quarks remain dynamical and thus their
contribution cancels in the matching, leaving us with the top-quark
contribution only. Recalling the relative size of the relevant CKM
factors,
\begin{equation}
  \label{eq:bCKMl}
  \lambda^{t}_{bs} \sim \lambda^{c}_{bs} \gg \lambda^{u}_{bs}\,, \qquad
  \lambda^{t}_{bd} \sim \lambda^{c}_{bd} \sim \lambda^{u}_{bd}\,,
\end{equation}
and the relative size of the loop functions $S_0(x_t)$ and
$S_0(x_c,x_t)$, we immediately realize that the top-charm contribution
enters at $\mathcal{O}(m_c^{2}/m_t^{2})$ and is therefore fully
negligible, leaving us with the effective Hamiltonian
\begin{dmath}[label={eq:HeffDB2}]
  \mathcal{H}_{\mathrm{eff}}^{\Delta B=2} =  \frac{G_F^{2}M_W^{2}}{4 \pi^{2}} 
    \left(
    \lambda_{bq}^{t} 
  \right)^{2} S_0(x_t) \overline{b}\gamma^{\mu} P_L
  q \overline{b} \gamma_\mu P_L q\,.
\end{dmath}
QCD corrections can be included up to NLO following the same line as
for the top-top contribution to $\Delta S=2$. Electroweak corrections
have been computed in ref.~\cite{Gambino:1998rt}.

\subsection{\texorpdfstring{$\Delta C=2$}{DC2} effective Hamiltonian}
\label{sec:HeffDC2}

One could think of applying the same procedure as for $\Delta S=2$
processes to obtain an effective Hamiltonian for $\Delta C=2$
transitions. However, in this case the role played by the charm in
$\Delta S=2$ goes to the strange quark, which is still dynamical at
the charm scale. One would then be in a situation similar to $\Delta
B=2$, except that for $\Delta C=2$ one has
\begin{equation}
  \label{eq:cCKMl}
  \lambda^{d}_{cu} \sim \lambda^{s}_{cu} \gg \lambda^{b}_{cu}\,,
\end{equation}
and $\lambda^{b}_{cu}/\lambda^{s}_{cu} \ll m_b/m_s$,\footnote{The strange
  quark mass in the denominator should actually be replaced by a suitable
  hadronic scale, making the ratio even smaller.} so that the process
is dominated by the matrix element of two $\Delta C=1$ effective
Hamiltonians. Indeed, to an excellent approximation GIM cancellation
in $\Delta C=2$ processes coincides with flavour $SU(3)$.

\subsection{\texorpdfstring{$\Delta F = 2$}{DF2} Hamiltonians beyond the SM}
\label{sec:HeffDF2NP}

While generalizing $\Delta F=1$ Hamiltonians beyond the SM,
\emph{i.e.} writing down the most general $\Delta F=1$ Hamiltonian
including all dimension six, $SU(3) \otimes U(1)_{\mathrm{em}}$
gauge-invariant operators, increases the number of operators up to
$\sim 120$ \cite{Buras:2000if,Aebischer:2017gaw}, the number of
independent operators that may arise is much smaller for $\Delta F=2$
transitions, so let us discuss this as an illustrative example of
going beyond the SM.

There is a large degree of arbitrariness in the choice of the operator
basis, since Fierz transformations can be used to get rid of a Dirac
and colour structure in favour of a different one. As an example, let
us choose the basis in ref.~\cite{Gabbiani:1996hi}:
\begin{dmath}
  \begin{array}{lcl}
    Q_1^{sdsd} = \overline{s}_L \gamma^{\mu} d_L \overline{s}_L \gamma^{\mu} d_L
       &\qquad &    \tilde{Q}_1^{sdsd} = \overline{s}_R \gamma^{\mu} d_R \overline{s}_R
    \gamma^{\mu} d_R
  \\
    Q_2^{sdsd} = \overline{s}_L d_R \overline{s}_L d_R    &\qquad &    \tilde{Q}_2^{sdsd}
    = \overline{s}_R d_L \overline{s}_R d_L
  \\
    Q_3^{sdsd} = \overline{s}_L^{\alpha} d_R^{\beta} \overline{s}_L^{\beta} d_R^{\alpha}
       &\qquad&    \tilde{Q}_3^{sdsd} = \overline{s}_R^{\alpha} d_L^{\beta}
    \overline{s}_R^{\beta} d_L^{\alpha}
  \\
    Q_4^{sdsd} = \overline{s}_L d_R \overline{s}_R d_L    &&\\    Q_5^{sdsd} =
    \overline{s}_L^{\alpha} d_R^{\beta} \overline{s}_R^{\beta} d_L^{\alpha} &&
  \end{array}
  \label{eq:df2genbasis}
\end{dmath}
With respect to the SM, where only $Q_1^{sdsd}$ is present, we need to
add two new Dirac structures, each one with two different colour
structures, plus the operators obtained by the $L \leftrightarrow R$
transformation. As originally pointed out in ref.~\cite{Bagger:1997gg}
and confirmed at NLO in
refs.~\cite{Ciuchini:1997bw,Buras:2000if,Ciuchini:2006dw}, the
additional operators have large anomalous dimensions (especially
$Q_4^{sdsd}$) which strongly
enhance their coefficients at the hadronic scale with respect to the
high scale, making them very important in phenomenological studies of
$\Delta F=2$ processes beyond the SM.

\subsection{\texorpdfstring{$\Delta F=2$}{DF2} matrix elements in the VIA}
\label{sec:DF2VIA}

Before closing this Section on $\Delta F=2$ effective Hamiltonians,
let us briefly discuss how their matrix elements can be estimated in
the VIA. VIA matrix elements are useful not only because they give a
first (and not too rough) estimate of the matrix elements, but also
because it is often easier and more accurate to compute the ratio of
the full matrix element normalized to the VIA than the absolute matrix
element. For this reason, matrix elements are often expressed
in terms of VIA results times the so-called $B$-parameters, which in
fact parameterize the ratio of the full matrix element with respect to
the VIA one.

For the sake of concreteness, let us consider $\Delta S=2$
processes. The SM effective Hamiltonian only contains $Q_1^{sdsd}$,
whose VIA matrix element is given by
\begin{dmath}[label={eq:DS2SMVIA}]
  \langle \overline{K}^{0} \lvert \overline{s} \gamma^{\mu} P_L d \overline{s} \gamma_\mu
  P_L d \rvert K^{0} \rangle_{\mathrm{VIA}} = 2 (\langle \overline{K}^{0} \lvert \overline{s}
  \gamma^{\mu} P_L d \rvert 0 \rangle \langle 0 \lvert \overline{s} \gamma_\mu
  P_L d \rvert K^{0} \rangle + \langle \overline{K}^{0} \lvert \overline{s}^{\alpha}
  \gamma^{\mu} P_L d^{\beta} \rvert 0 \rangle \langle 0 \lvert \overline{s}^{\beta} \gamma_\mu
  P_L d^{\alpha} \rvert K^{0} \rangle)\,,
\end{dmath}
where the second term corresponds to Fierzed contractions with
respect to the first term. Using
\begin{equation}
  \label{eq:FK}
  \langle 0 \lvert \overline{s} \gamma^{\mu} \gamma_5 d \rvert K^{0} \rangle = -i
  p_K^{\mu} \frac{F_K}{\sqrt{2 m_K}}
\end{equation}
we obtain for the first term
\begin{equation}
  \label{eq:DS21stterm}
  \frac{1}{4}\frac{F_K^{2} m_K^{2}}{2 m_K} = \frac{1}{8} m_K F_K^{2} \,.
\end{equation}
For the second term we perform a colour Fierz transformation:
\begin{equation}
  \label{eq:cfierz}
  \delta_{\alpha\beta} \delta_{\gamma\delta} = 2 T^{a}_{\alpha\delta}
  T^{a}_{\gamma\beta} + \frac{1}{3} \delta_{\alpha\delta} \delta_{\gamma\beta}
\end{equation}
getting
\begin{dmath}[label={eq:DS22ndterm}]
  \langle \overline{K}^{0} \lvert \overline{s}^{\alpha}
  \gamma^{\mu} P_L d^{\beta} \rvert 0 \rangle \langle 0 \lvert \overline{s}^{\beta} \gamma_\mu
  P_L d^{\alpha} \rvert K^{0} \rangle = \frac{1}{3} \langle \overline{K}^{0} \lvert \overline{s}
  \gamma^{\mu} P_L d \rvert 0 \rangle \langle 0 \lvert \overline{s} \gamma_\mu
  P_L d \rvert K^{0} \rangle + 2 \langle \overline{K}^{0} \lvert \overline{s}
  T^{a} \gamma^{\mu} P_L d^{\beta} \rvert 0 \rangle \langle 0
  \lvert \overline{s} T^{a} \gamma_\mu
  P_L d \rvert K^{0} \rangle\,.
\end{dmath}
The second term vanishes and the first one reduces to
eq.~(\ref{eq:DS21stterm}), so that in the end we obtain
\begin{equation}
  \label{eq:DS2SMVIAME}
   \langle \overline{K}^{0} \lvert \overline{s} \gamma^{\mu} P_L d \overline{s} \gamma_\mu
  P_L d \rvert K^{0} \rangle_{\mathrm{VIA}} = 2 (1+\frac{1}{3}) \frac{1}{8} F_K^{2} m_K = \frac{1}{3} F_K^{2} m_K\,.
\end{equation}
In general we can therefore write
\begin{dmath}[label={eq:ds2smfull}]
  \langle \overline{K}^{0} \lvert \overline{s} \gamma^{\mu} P_L d \overline{s} \gamma_\mu
  P_L d \rvert K^{0} \rangle = \langle \overline{K}^{0} \lvert \overline{s} \gamma^{\mu} P_L d \overline{s} \gamma_\mu
  P_L d \rvert K^{0} \rangle_{\mathrm{VIA}} B_K = \frac{1}{3} F_K^{2} m_K B_K\,.
\end{dmath}

It is now interesting to look at the VIA matrix elements of the
additional operators that arise beyond the SM. From
eq.~(\ref{eq:df2genbasis}) we see that the operators $Q_{i}^{sdsd}$
for $i>1$ are built by products of scalar/pseudoscalar
densities. We have
\begin{dmath}[label={eq:pdensity}]
  \partial_\mu \langle 0 \lvert \overline{s} \gamma^{\mu} \gamma_5 d \rvert K^{0}
  \rangle =
  \begin{array}{l}
    \nearrow \\
    \searrow
  \end{array}
  \setlength{\arraylinesep}{10pt}
  \begin{array}{l}
     -i (m_s + m_d) \langle0 \lvert \overline{s} \gamma_5 d \rvert K^{0}\rangle
    \\
    \frac{m_K^{2} F_K}{\sqrt{2 m_K}}
  \end{array}
  \setlength{\arraylinesep}{0pt}
\end{dmath}
where in the first case we applied the derivative to the quark
bilinear while in the second case we applied it to the whole matrix
element. Using eq.~(\ref{eq:pdensity}) we obtain
\begin{dmath}[label={eq:psdensityvia}]
  \langle \overline{K}^{0} \lvert \overline{s} \gamma_5 d \lvert 0 \rangle \langle 0
  \lvert\overline{s} \gamma_5 d \rvert K^{0} \rangle = - 
  \left(
    \frac{m_K}{m_s + m_d}
\right)^{2} \frac{m_K F_K^{2}}{2}\,,
\end{dmath}
which for Kaons is chirally enhanced by one order of magnitude since
$m_K \sim 3 (m_s + m_d)$. Combining this chiral enhancement with the
RG enhancement one sees that these operators play a crucial role in $\Delta
S=2$ processes beyond the SM. We will return to this point later. For
completeness, we write down the matrix elements for all the operators
in eq.~(\ref{eq:df2genbasis}) in the VIA:
\begin{dgroup*}
  \begin{dmath}[compact]
    \langle \overline{K}^{0} \vert Q_{1} 
    \vert K^{0} \rangle = \frac{1}{3}m_{K}F_{K}^{2}\; ,
  \end{dmath}
  \begin{dmath}[compact]
    \langle \overline{K}^{0} \vert Q_{2} 
    \vert K^{0} \rangle = -\frac{5}{24}  
    \left(\frac{m_{K}}{m_{s}+m_{d}}\right)^{2}m_{K}F_{K}^{2}\; ,
  \end{dmath}
  \begin{dmath}[compact]
    \langle \overline{K}^{0} \vert Q_{3} 
    \vert K^{0} \rangle  =  \frac{1}{24} 
    \left(\frac{m_{K}}{m_{s}+m_{d}}\right)^{2}m_{K}F_{K}^{2}\; ,
  \end{dmath}
  \begin{dmath}[compact]
    \langle \overline{K}^{0} \vert Q_{4} 
    \vert K^{0} \rangle  =  \left[\frac{1}{24} +
      \frac{1}{4} 
      \left(\frac{m_{K}}{m_{s}+m_{d}}\right)^{2}\right]m_{K}F_{K}^{2}\; ,
  \end{dmath}
  \begin{dmath}[compact]
    \langle \overline{K}^{0} \vert Q_{5}
    \vert K^{0} \rangle  = \left[\frac{1}{8} +
      \frac{1}{12} 
      \left(\frac{m_{K}}{m_{s}+m_{d}}\right)^{2}\right]
    m_{K}F_{K}^{2}\; .
  \end{dmath}  
\end{dgroup*}
A word of caution is necessary at this point. Operators $Q_{4,5}$ have
VIA matrix elements that contain two contributions, one from pseudoscalar density
matrix elements and one from axial vector currents. To define the
corresponding $B$-parameters it is convenient to choose as
normalization just the pseudoscalar density contributions. However,
this corresponds to having $B \neq 1$ in the VIA \cite{Allton:1998sm}.

\section{Effective Hamiltonians at work: meson-antimeson mixing and CP
  violation}
\label{sec:mixCP}

Having discussed the basics of $\Delta F=1$ and $\Delta F=2$ effective
Hamiltonians, let us now use them to study meson-antimeson mixing and
CP violation.

\subsection{Meson-antimeson mixing}
\label{sec:mixing}

There are four neutral mesons which differ from their antiparticles
just because of their flavour quantum numbers: $K^{0}$, $D^{0}$, $B_d$ and
$B_s$ mesons. While strong and electromagnetic interactions preserve
flavour, the full Hamiltonian does not, due to flavour-changing weak
interactions. Therefore, its eigenstates will be superpositions of
mesons and antimesons, giving rise to the phenomenon of
meson-antimeson oscillations, which entails a difference of mass and
width of the two eigenstates \cite{GellMann:1955jx}. Let us first
write down the formalism for a generic neutral meson, which we denote
by $M^{0}$, and then specialize to the four cases above, in which
different simplifying assumptions can be made.

Notice that a CP transformation takes a neutral meson into its
antiparticle with an arbitrary phase shift $\xi$:
\begin{eqnarray}
  \label{eq:cpm}
  \mathcal{CP} \vert M^{0} \rangle = e^{i \xi} \vert \overline{M}^{0} \rangle\,, \\
  \mathcal{CP} \vert \overline{M}^{0} \rangle = e^{-i \xi} \vert M^{0} \rangle\,. \nonumber
\end{eqnarray}

The matrix elements of the full Hamiltonian between $M^{0}$ and
$\overline{M}^{0}$ states can be written as a two-by-two complex matrix:
\begin{equation}
  \label{eq:Hij}
\hat{H} = \left(
    \begin{array}{cc}
      H_{11}& H_{12} \\
      H_{21}& H_{22}
    \end{array}
  \right) \equiv \left(
    \begin{array}{cc}
      \langle M^{0} \lvert H \rvert M^{0} \rangle&  \langle M^{0} \lvert H \rvert \overline{M}^{0} \rangle\\
     \langle \overline{M}^{0} \lvert H \rvert M^{0} \rangle & \langle \overline{M}^{0}
                                                   \lvert H \rvert \overline{M}^{0} \rangle
    \end{array}
  \right) \equiv \hat{M} - \frac{i}{2} \hat{\Gamma} \,,
\end{equation}
where in the last equality we have split the complex matrix $\hat{H}$
in its Hermitian ($\hat{M}$) and anti-Hermitian ($-i/2 \hat{\Gamma}$)
parts.

CPT invariance requires $M_{11}=M_{22}$ and $\Gamma_{11} =
\Gamma_{22}$, while it does not constrain the off-diagonal matrix
elements. CP invariance instead connects off-diagonal elements among
themselves:
\begin{dmath}[label={eq:CP}]
  H_{21} = \langle \overline{M}^{0} \lvert H \rvert M^{0} \rangle
  \xrightarrow{\mathcal{CP}} e^{i \xi} \langle M^{0} \lvert
  H^{\mathcal{CP}} \rvert \overline{M}^{0} \rangle e^{i \xi}\,,
\end{dmath}
so that CP conservation ($H^{\mathcal{CP}} = H$) implies
\begin{equation}
  H_{21} =
e^{2 i \xi} H_{12} \Rightarrow \lvert H_{21} \rvert = \lvert H_{12}
\rvert \Rightarrow \mathrm{Im} 
\left(
  M_{12}^{*} \Gamma_{12}
\right) = 0
\Rightarrow \mathrm{Im} 
\left(
  \frac{\Gamma_{12}}{M_{12}}
\right) = 0\,.
\label{eq:CPcons}
\end{equation}

The eigenvalue equation reads, assuming CPT invariance,
\begin{dmath}[label={eq:Heig},compact]
  \det 
  \left(
    \hat{H} - \lambda  \mathbb{1} 
  \right) = 0 = (H_{11} - \lambda)^{2} - H_{12}H_{21} 
  \Rightarrow \lambda = H_{11} \pm \sqrt{H_{12}H_{21}}\,.
\end{dmath}

Defining 
\begin{eqnarray}
  \label{eq:dmdefs}
  &&\lambda_{1,2} = m_{1,2} - i/2 \Gamma_{1,2}\,,\quad m=\frac{m_1 +
  m_2}{2}\,,\quad \Gamma=\frac{\Gamma_1 + \Gamma_2}{2}\,, \\
  &&\Delta m = m_1 -
  m_2\,,\quad \Delta \Gamma = \Gamma_1 - \Gamma_2\,,\quad x = \frac{\Delta
    m}{\Gamma}\,, \quad y = \frac{\Delta \Gamma}{2 \Gamma}\,, \nonumber
\end{eqnarray}
one can
alternatively label the two eigenstates by their mass, \textit{i.e.}
defining $\Delta m = m_H-m_L $ to be positive (here $H$ stands for
heavy and $L$ for light), or by their width, \textit{i.e.}
defining $\Delta \Gamma = \Gamma_S-\Gamma_L$ to be
positive (here $S$ stands for short-lived and $L$ for long-lived).

We have
\begin{dgroup*}[noalign]
  \begin{dmath}[label={eq:eigenv1},compact]
    \Delta \lambda = \lambda_1 - \lambda_2 = \Delta m - \frac{i}{2}
    \Delta\Gamma = 2 \sqrt{H_{12}H_{21}}\,, 
  \end{dmath}
  \begin{dmath}[label={eq:eigenv2},compact]
    (\Delta m)^{2} -
    \frac{1}{4} (\Delta\Gamma)^{2} -i \Delta m \Delta \Gamma = 4 H_{12}
    H_{21} =
  \end{dmath}
  \begin{dmath*}[compact]
    \phantom{~~~~~~~~}=4 
    \left(
      \lvert M_{12}^{2} \rvert - \frac{1}{4} \lvert \Gamma_{12}
      \rvert^{2} 
    \right) - 4 i \mathrm{Re}
    \left(
      M_{12}\Gamma_{12}^{*}
    \right)\,.
  \end{dmath*}
\end{dgroup*}
Taking real and imaginary parts we obtain
\begin{dgroup*}
  \begin{dmath}
    (\Delta m)^{2} - \frac{1}{4} (\Delta \Gamma)^{2} = 4 \left(
      \lvert M_{12}\rvert^{2} - \frac{1}{4} \lvert \Gamma_{12} \rvert^{2}
    \right)\,,
  \end{dmath}
  \begin{dmath}
    \Delta m \Delta \Gamma = 4 \mathrm{Re} \left(
      M_{12} \Gamma_{12}^{*}
    \right)\,.
  \end{dmath}
\end{dgroup*}
Notice that
\begin{dmath}[label={eq:dm2dg2},compact]
  (\Delta m)^{2} = 4 (\mathrm{Re} \sqrt{H_{12}H_{21}})^{2} = 2 \lvert H_{12}
  H_{21} \rvert + 2 \mathrm{Re} H_{12}H_{21}\,,
\end{dmath}
so that
\begin{dmath}[label={eq:h12ph21sq}]
  \left(
    \lvert H_{12}\rvert + \lvert H_{21} \rvert
  \right)^{2} =  \lvert H_{12}\rvert^{2} + \lvert H_{21} \rvert^{2} +
  2 \lvert H_{12}  H_{21}\rvert = \lvert H_{12}\rvert^{2} + \lvert
  H_{21} \rvert^{2}
  - 2 \mathrm{Re} (H_{12}  H_{21}) + (\Delta m)^{2} = \lvert
  H_{12}-H_{21}^{*}\rvert^{2}+ (\Delta m)^{2} = \lvert \Gamma_{12} \rvert^{2}
  +  (\Delta m)^{2}\,.
\end{dmath}

Let us write the eigenstates as
\begin{dmath}[label={eq:eigdef},compact]
  \vert M_{1,2} \rangle = p \vert M^{0} \rangle \pm q \vert
  \overline{M}^{0} \rangle\,,~\mathrm{with}~
  \lvert p \rvert^{2} + \lvert q \rvert^{2} = 1\,.
\end{dmath}
Then we have
\begin{dmath}[label={eq:df2evec},compact]
  H_{11} p \pm H_{12} q = \lambda_{1,2} p \Rightarrow H_{11} \pm
  \frac{q}{p} H_{12} = H_{11} \pm \sqrt{H_{12} H_{21}} \Rightarrow
  \frac{q}{p} = \sqrt{\frac{H_{21}}{H_{12}}} = \frac{2 M_{12}^{*} - i
    \Gamma_{12}^{*}}{\Delta m - \frac{i}{2}\Delta \Gamma} = \frac{\Delta
  m - \frac{i}{2} \Delta \Gamma}{2 M_{12} - i
    \Gamma_{12}}\,.
\end{dmath}
Using eq.~(\ref{eq:CPcons}) we see that CP conservation implies
\begin{dmath}[label={eq:qopCPcons},compact]
    \frac{q}{p} = \sqrt{\frac{H_{21}}{H_{12}}} =
    \sqrt{\frac{H_{12}e^{2 i \xi}}{H_{12}}} = e^{i \xi}\,,\qquad
  \mathrm{so~that} \qquad
    \left\vert
      \frac{q}{p}
    \right\vert = 1\,.
\end{dmath}
Thus,
\begin{dmath}[label={eq:CPVmix}]
  \left\vert \frac{q}{p} \right\vert \neq 1 \mathrm{~implies~CP~violation,}
\end{dmath}
usually denoted as \emph{CP violation
in mixing}.

It is useful to define
\begin{dseries}[label={eq:x12y12phi12},compact]
  \begin{math}
    x_{12} = \frac{2\lvert M_{12} \rvert}{\Gamma}
  \end{math},
  \begin{math}
    y_{12} = \frac{\lvert \Gamma_{12} \rvert}{\Gamma}
  \end{math} and 
  \begin{math}
    \Phi_{12} = \mathrm{arg} 
    \left(
      \frac{\Gamma_{12}}{M_{12}}
    \right)
  \end{math},
\end{dseries}
so that 
\begin{dmath}[label={eq:qopPhi12},compact]
    \left\vert \frac{q}{p} \right\vert = \frac{\lvert 2 M_{12}^{*} - i
      \Gamma_{12}^{*}\rvert}{\Gamma \lvert x - i y \rvert} = 
    \frac{\sqrt{x_{12}^{2} + y_{12}^{2} - 2 x_{12} y_{12} \sin
        \Phi_{12}}}{\sqrt{x^{2} + y^{2}}}\,,
\end{dmath}
implying that
\begin{dmath}[label={eq:Phi12cpv},compact]
  \left\vert \frac{q}{p} \right\vert \neq 1 \Leftrightarrow \sin \Phi_{12} \neq 0\,.
\end{dmath}
Finally, CP violation in meson-antimeson mixing can also be expressed
in terms of the parameter $\delta$ defined as
\begin{dmath}[label={eq:deltaCPV},compact]
  \delta \equiv \frac{\lvert H_{12}\rvert- \lvert H_{21}\rvert}{\lvert H_{12}\rvert+ \lvert H_{21}\rvert}=\langle M_1 \vert M_2 \rangle = \lvert p
    \rvert^{2} - \lvert q
    \rvert^{2} = \frac{1 - \left\vert \frac{q}{p}
    \right\vert^{2}}{1 + \left\vert \frac{q}{p}
    \right\vert^{2}}\,.
\end{dmath}
One has
\begin{dmath}[label={eq:upd2},compact]
  1+\delta^{2} = 1 + \frac{\lvert H_{12} \rvert^{2} + \lvert H_{21}
    \rvert^{2} - 2 \lvert H_{12} \rvert \lvert H_{21} \rvert}{\lvert
    H_{12} \rvert^{2} + \lvert H_{21} \rvert^{2} + 2 \lvert H_{12}
    \rvert \lvert H_{21} \rvert} = 2 \frac{\lvert H_{12} \rvert^{2} +
    \lvert H_{21} \rvert^{2}}{ \left( \lvert H_{12} \rvert^{ }+ \lvert
      H_{21} \rvert \right)^{2}}\,, 
\end{dmath}
\begin{dmath}[label={eq:h12sqph21sq},compact]
  \lvert H_{12} \rvert^{2} + \lvert
  H_{21} \rvert^{2} = \frac{4 \lvert M_{12} \rvert^{2} + \lvert
    \Gamma_{12} \rvert^{2}}{2}\,,
\end{dmath}
so that
\begin{dmath}[label={eq:doupd2},compact]
  \frac{\delta}{1+\delta^{2}} = \frac{1}{2} \frac{\lvert H_{12} \rvert^{2} - \lvert H_{21}
    \rvert^{2}}{\lvert H_{12} \rvert^{2} + \lvert H_{21}
    \rvert^{2}} = \frac{2 \lvert M_{12} \Gamma_{12} \rvert
  \sin \Phi_{12}}{4 \lvert M_{12} \rvert^{2} + \lvert \Gamma_{12} \rvert^{2}}
\end{dmath}
and
\begin{dmath}[label={eq:delta},compact]
  \delta = \frac{2 \lvert M_{12} \Gamma_{12} \rvert
  \sin \Phi_{12}}{(\Delta m)^{2} + \lvert \Gamma_{12} \rvert^{2}}\,.
\end{dmath}

Let us also write down the expressions for
  \begin{math}
    \lvert M_{12}\rvert
  \end{math},
  \begin{math}
    \lvert \Gamma_{12} \rvert
  \end{math}
  and
  \begin{math}
    \Phi_{12}
  \end{math}
in terms of
  \begin{math}
\Delta m
  \end{math},
  \begin{math}
\Delta \Gamma
  \end{math}
  and
  \begin{math}
\delta
  \end{math}:
\begin{dgroup*}
  \begin{dmath}[label={eq:m12exp},compact]
    \lvert M_{12} \rvert = \sqrt{4 \frac{(\Delta m)^{2}+\delta^{2}
      (\Delta \Gamma)^{2}}{16(1-\delta^{2})}}\sim \frac{\Delta
    m}{2}+\mathcal{O}(\delta^{2})\,,  
  \end{dmath}
  \begin{dmath}[label={eq:G12exp}]
    \vert \Gamma_{12} \vert= \sqrt{\frac{(\Delta \Gamma)^{2}+ 4 \delta^{2}
        (\Delta m)^{2}}{4(1-\delta^{2})}}\sim \frac{\Delta \Gamma}{2}+\mathcal{O}(\delta^{2})\,,
  \end{dmath}
  \begin{dmath}[label={eq:Phi12exp}]
    \sin \Phi_{12} = \frac{4 \vert \Gamma_{12}\vert^{2} + 16 \vert
      M_{12}\vert^{2} - (4 (\Delta m)^{2}+(\Delta\Gamma)^{2})\vert q/p\vert^{2}}{16 \vert
      M_{12} \Gamma_{12}\vert}\sim
    \frac{4(\Delta m)^{2}+(\Delta\Gamma)^{2}}{2\Delta m \Delta\Gamma} \delta+\mathcal{O}(\delta^{2})
  \end{dmath}
\end{dgroup*}
and viceversa:
\begin{dgroup*}
  \begin{dmath}[compact]
    (\Delta m)^{2} = \frac{4 \lvert M_{12}\rvert^{2} - \delta^{2} \lvert
      \Gamma_{12}\rvert^{2}}{1+\delta^{2}}
    \sim 4 \lvert M_{12}\rvert^{2} + \mathcal{O}(\delta^{2}) \,,
  \end{dmath}
  \begin{dmath}[compact]
    (\Delta \Gamma)^{2} = \frac{4 \lvert \Gamma_{12}\rvert^{2} - 16
      \delta^{2} \lvert M_{12}\rvert^{2}}{1+\delta^{2}} \sim 4 \lvert
    \Gamma_{12}\rvert^{2} + \mathcal{O}(\delta^{2})\,.
  \end{dmath}
\end{dgroup*}

Notice that the transformation in eq.~(\ref{eq:eigdef}) is unitary
only if $\lvert p \rvert^2 - \lvert q \rvert^2 = \delta = 0$, i.e.\ if
CP is conserved in the mixing. Therefore, if CP is violated, one must
be careful in defining outgoing $M_{1,2}$ states using the so-called
reciprocal basis~\cite{Sachs:1963zz,Sachs:1964zz,Enz:1965tr,Wolfenstein:1970wb,Beuthe:1997fu,AlvarezGaume:1998yr,Branco:1999fs,Silva:2004gz}:
\begin{dmath}[label={eq:eigdefr},compact]
  \langle \tilde{M}_{1,2} \vert = \frac{q \langle M^{0} \vert  \pm p \langle
  \overline{M}^{0} \vert}{2 p q}\,.
\end{dmath}

\subsection{Time evolution of mixed meson states}
\label{sec:timemix}

Having obtained the eigenstates of the Hamiltonian in
eqs.~(\ref{eq:Heig}) and (\ref{eq:dmdefs}), we can write down the time
evolution of a state initially produced as an $M^{0}$ or as an 
$\overline{M}^{0}$. We start from the time evolution of the
eigenstates,
\begin{dmath}[label={eq:te-eig}]
  \vert M_{1,2}(t) \rangle = e^{-i\lambda_{1,2}t} \vert M_{1,2}(0) \rangle\,, 
\end{dmath}
and use eq.~(\ref{eq:eigdef}) to rotate back to the \stackon[.3pt]{$M$}{\brabar}$\phantom{}^{0}$:
\begin{dgroup*}
  \begin{dmath}[compact,label={eq:M0oft}]
    \vert M^{0}(t) \rangle = \frac{1}{2p} 
    \left(
      M_{1}(t) + M_{2}(t)
    \right) = g_{+}(t) \vert M^{0} \rangle +
    \frac{q}{p} g_{-}(t) \vert \overline{M}^{0} \rangle\,,
  \end{dmath}
  \begin{dmath}[compact,label={eq:M0boft}]
    \vert \overline{M}^{0}(t) \rangle = \frac{1}{2q} 
    \left(
      M_{1}(t) - M_{2}(t)
    \right) = \frac{p}{q }g_{-}(t) \vert M^{0} \rangle +
    g_{+}(t) \vert \overline{M}^{0} \rangle\,,
  \end{dmath}
\end{dgroup*}
with
\begin{dmath}[label={eq:gpm},compact]
  g_{\pm}(t) = \frac{e^{-i\lambda_{1}t} \pm e^{-i\lambda_{2}t}}{2} \,.
\end{dmath}
The probability that a meson initially produced as a
\stackon[.3pt]{$M$}{\brabar}$\phantom{}^{0}$ remains such at time $t$
is given by $\lvert g_{+}(t)\rvert^{2}$, while the probabilities of an
$M^{0}$ becoming an $\overline{M}^{0}$ and viceversa are not equal to
each other if CP is violated:
\begin{dgroup*}
  \begin{dmath}[label={eq:P00b},compact]
    \mathcal{P}(M^{0}(0)\rightarrow \overline{M}^{0}(t))) = 
    \left\vert
      \frac{q}{p}
    \right\vert^{2} \lvert g_{-}(t)\rvert^{2}\,,
  \end{dmath}
   \begin{dmath}[label={eq:P0b0},compact]
    \mathcal{P}(\overline{M}^{0}(0)\rightarrow M^{0}(t))) = 
    \left\vert
      \frac{p}{q}
    \right\vert^{2} \lvert g_{-}(t)\rvert^{2}\,.
  \end{dmath} 
\end{dgroup*}

We have
\begin{dgroup*}
  \begin{dmath}[label={eq:gpm2}]
    \left\vert
      g_{\pm}(t)
    \right\vert^{2} = \frac{
      e^{-\Gamma_{1}t} + e^{-\Gamma_{2}t} \pm 2 e^{-\Gamma t}\cos(\Delta m\,t)
    }{4} = \frac{e^{-\Gamma t}}{2} 
    \left(
      \cosh (\Delta\Gamma\,t/2) \pm \cos(\Delta m\,t)
    \right)\,,
  \end{dmath}
  \begin{dmath}[label={eq:gpgmst}]
      g_{+}(t) g_{-}^{*}(t) = \frac{
      e^{-\Gamma_{1}t} - e^{-\Gamma_{2}t} - 2 e^{-\Gamma t}\sin (\Delta m\,t)
    }{4} = - \frac{e^{-\Gamma t}}{2} 
    \left(
      \sinh (\Delta\Gamma\,t/2) - i \sin(\Delta m\,t)
    \right)\,.
  \end{dmath}
\end{dgroup*}

\subsection{Observables for meson-antimeson mixing}
\label{sec:mixobs}

Since the $M^{0}$ mesons are unstable, we must consider their weak
decays in building observables related to meson mixing. The
probabilities in eqs.(\ref{eq:P00b}) and (\ref{eq:P0b0}) can be
directly accessed using semileptonic decays, since for decays of
down-type quarks one has $M^{0} \not \to \ell \overline{\nu}_{\ell} X$
and $\overline{M}^{0} \not \to \overline{\ell} \nu_{\ell} X$, where
$X$ represents an unspecified hadronic final state. Therefore, those
decays can only happen through mixing, and one can define the
semileptonic CP asymmetry as the difference of the number of
semileptonic decays to wrong sign leptons in
\stackon[.3pt]{$M$}{\brabar}$\phantom{}^{0}$ decays normalized to the
total number of such decays:
\begin{dmath}[label={eq:asl}]
  a_{\mathrm{SL}} \equiv \frac{N(\overline{M}^{0}\to \overline{\ell}
    \nu_{\ell} X) -N(M^{0} \to \ell \overline{\nu}_{\ell} X) }{N(\overline{M}^{0}\to \overline{\ell}
    \nu_{\ell} X) +N(M^{0} \to \ell \overline{\nu}_{\ell} X)}\,.
\end{dmath}
Assuming that $M^{0}$ and $\overline{M}^{0}$ are produced in equal
number $N_{0}$ and CP invariance of the semileptonic decay amplitude $A$, one
has
\begin{dgroup*}
  \begin{dmath}[label={eq:N0},compact]
    N(M^{0} \to \ell \overline{\nu}_{\ell} X) = N_{0} \lvert A \rvert^{2}
    \left\vert
      \frac{q}{p}
    \right\vert^{2} \int_{0}^{\infty} \lvert g_{-}(t)\rvert^{2} \mathrm{d}t\,,
  \end{dmath}
   \begin{dmath}[label={eq:N0b},compact]
   N(\overline{M}^{0}\to \overline{\ell}
    \nu_{\ell} X)  = N_{0} \lvert A \rvert^{2}
    \left\vert
      \frac{p}{q}
    \right\vert^{2} \int_{0}^{\infty} \lvert g_{-}(t)\rvert^{2} \mathrm{d}t\,.
  \end{dmath} 
\end{dgroup*}
All factors except for the mixing parameters drop in the ratio,
leading to
\begin{dmath}[label={eq:aslqop},compact]
  a_{\mathrm{SL}} = \frac{ \left\vert
      \frac{p}{q}
    \right\vert^{2} - \left\vert
      \frac{q}{p}
    \right\vert^{2}}{ \left\vert
      \frac{p}{q}
    \right\vert^{2} + \left\vert
      \frac{q}{p}
    \right\vert^{2}} = \frac{ 1 - \left\vert
      \frac{q}{p}
    \right\vert^{4}}{ 1 + \left\vert
      \frac{q}{p}
    \right\vert^{4}}\,.
\end{dmath}

In general, the decay amplitude into a given final state $\vert f\rangle$ and its
CP-conjugate $\vert \overline{f}\rangle =
e^{-i\xi_{f}}\mathcal{CP}\vert f \rangle$ in the SM can be always written in the
following form:\footnote{We prefer to keep the equations in a
  symmetric form, keeping in mind that an overall phase could be
  dropped since it is physically irrelevant.}
\begin{dgroup*}
  \begin{dmath}[label={eq:Af},compact]
    A(M \to f) \equiv \mathcal{A}_{f} = A_{f} e^{i \phi_{f}} e^{i \delta_{f}} (1 + r_{f}
    e^{i \phi_{r_{f}}} e^{i \delta_{r_{f}}})\,, 
  \end{dmath}
  \begin{dmath}[label={eq:Abfb},compact]
    A(\overline{M} \to \overline{f}) \equiv \overline{\mathcal{A}}_{\overline{f}}
    = e^{-i \xi} e^{i \xi_{f}} A_{f} e^{-i \phi_{f}} e^{i
      \delta_{f}} (1 + r_{f} e^{-i \phi_{r_{f}}} e^{i \delta_{r_{f}}})\,, 
  \end{dmath}
  \begin{dmath}[label={eq:Afb},compact]
    A(M \to \overline{f}) \equiv \mathcal{A}_{\overline{f}}
    =A_{\overline{f}} e^{i
      \phi_{\overline{f}}} e^{i \delta_{\overline{f}}} (1 +
    r_{\overline{f}} e^{i \phi_{r_{\overline{f}}}} e^{i \delta_{r_{\overline{f}}}})\,, 
  \end{dmath}
  \begin{dmath}[label={eq:Abf},compact]
    A(\overline{M} \to f) \equiv \overline{\mathcal{A}}_{f} = e^{-i \xi} e^{-i \xi_{f}} A_{\overline{f}} e^{-i
      \phi_{\overline{f}}} e^{i \delta_{\overline{f}}} (1 +
    r_{\overline{f}} e^{-i \phi_{r_{\overline{f}}}}  e^{i \delta_{r_{\overline{f}}}})\,,
  \end{dmath}
\end{dgroup*}
with $A_{f}$, $A_{\overline{f}}$, $r_{f}$ and $r_{\overline{f}}$ real,
and $\mathcal{CP}\vert M \rangle = e^{i \xi}\vert \overline{M} \rangle $.
Indeed, using CKM unitarity where needed one can have at most two
independent CKM factors in each decay amplitude, corresponding for
example in eq.~(\ref{eq:Af}) to $A_{f}$ and $A_{f} r_{f}$. For later
convenience, we have written the second amplitude as a multiplicative
factor, since in several cases one has $\lvert r_{f} \rvert \ll 1$ and
an expansion in $r_{f}$ can be performed. We have written explicitly
the CP-odd weak phases $\phi_{i}$ and the CP-even strong phases $\delta_{i}$.

We see that
\begin{dmath}[label={eq:CPVdir},compact]
  \phi_{r_{f}}\neq 0~\mathrm{and}~\delta_{r_{f}}\neq 0 \Leftrightarrow
  \mathcal{A}_{f} \neq \overline{\mathcal{A}}_{\overline{f}}\,.
\end{dmath}
This is usually denoted as \emph{``direct'' CP violation or CP
  violation in the decay}. The corresponding CP asymmetry can be
written as
\begin{dmath}[compact,label={eq:ACPdir}]
  A_{\mathrm{CP}}^{\mathrm{dir}}(f) \equiv \frac{\lvert A(\overline{M} \to
  \overline{f}) \rvert^{2} - \lvert A(M \to f) \rvert^{2}}{\lvert A(\overline{M} \to
  \overline{f}) \rvert^{2} + \lvert A(M \to f) \rvert^{2}}
= \frac{2 r_{f} \sin \phi_{r_{f}} \sin \delta_{r_{f}}}{1 + r_{f}^{2} +
2 r_{f} \cos \phi_{r_{f}} \cos \delta_{{r_{f}}}}\,.
\end{dmath}

For neutral meson decays, one can consider the case of a final state
which is a CP eingenstate with eigenvalue $\eta_{f}$:
\begin{dgroup*}
  \begin{dmath}[label={eq:Afcp},compact]
    A(M \to f_{\mathrm{CP}}) = A_{f} e^{i \phi_{f}} e^{i \delta_{f}} (1 + r_{f}
    e^{i \phi_{r_{f}}} e^{i \delta_{r_{f}}})\,, 
  \end{dmath}
  \begin{dmath}[label={eq:Abfcp},compact]
    A(\overline{M} \to f_{\mathrm{CP}}) = e^{-i \xi} \eta_{f} A_{f} e^{-i \phi_{f}} e^{i
      \delta_{f}} (1 + r_{f} e^{-i \phi_{r_{f}}} e^{i \delta_{r_{f}}})\,.
  \end{dmath}
\end{dgroup*}

It is useful to introduce
\begin{dmath}[compact,label={eq:lambdaf}]
  \lambda_{f} = \frac{q}{p} \frac{\overline{\mathcal{A}}_{f}}{\mathcal{A}_{f}}
  = \frac{q}{p} \frac{ e^{-i \xi} e^{-i \xi_{f}} A_{\overline{f}} e^{-i
      \phi_{\overline{f}}} e^{i \delta_{\overline{f}}} (1 +
    r_{\overline{f}} e^{-i \phi_{r_{\overline{f}}}}  e^{i \delta_{r_{\overline{f}}}})}{A_{f} e^{i \phi_{f}} e^{i \delta_{f}} (1 + r_{f}
    e^{i \phi_{r_{f}}} e^{i
      \delta_{r_{f}}})}\,,
\end{dmath}
which is manifestly rephasing invariant since for $\vert M^{0} \rangle
\to e^{i \Xi} \vert M^{0} \rangle$ and $\vert \overline{M}^{0} \rangle
\to e^{i \overline{\Xi}} \vert \overline{M}^{0} \rangle$ we have
\begin{dmath}[compact,label={eq:lambdainv}]
  \frac{q}{p} \to e^{i (\Xi - \overline{\Xi})} \frac{q}{p}\,,\quad
  \frac{\overline{\mathcal{A}}_{f}}{\mathcal{A}_{f}} \to e^{i (\overline{\Xi} - \Xi
    )} \frac{\overline{\mathcal{A}}_{f}}{\mathcal{A}_{f}}
\end{dmath}
(or equivalently $\xi \to \xi  + \Xi - \overline{\Xi}$), so that
$\lambda_{f} \to \lambda_{f}$.

For decays to a CP eigenstate this simplifies to
\begin{dmath}[label={eq:lambdafcp}]
  \lambda_{f_{\mathrm{CP}}} = \frac{q}{p}e^{-i \xi} \eta_{f}
    e^{-2 i \phi_{f}}\frac{ (1 + r_{f} e^{-i \phi_{r_{f}}} e^{i \delta_{r_{f}}})}{(1 + r_{f}
    e^{i \phi_{r_{f}}} e^{i \delta_{r_{f}}})}=\frac{q}{p}e^{-i \xi} \eta_{f}
    e^{-2 i \phi_{f}} (1 - 2 i r_{f} e^{i \delta_{r_{f}}} \sin
    \phi_{r_{f}} + \mathcal{O}(r_{f}^{2}))\,.
\end{dmath}
CP conservation implies $q/p=e^{i \xi}$ (see eq.~(\ref{eq:CPcons}))
and $\phi_{f,r_{f}} = 0$, \emph{i.e.} Im
$\lambda_{f_{\mathrm{CP}}}=0$. Therefore,
\begin{dmath}[label={eq:CPVint}]
  \mathrm{Im}
  \lambda_{f_{\mathrm{CP}}}\neq 0~\mathrm{implies~CP~violation.}
\end{dmath}
This form of CP violation requires the \emph{interference
between mixing and decay}. Indeed, $\lambda_{f_{\mathrm{CP}}}$ determines the
time evolution of \stackon[.3pt]{$M$}{\brabar}$\phantom{}^{0}$ decays
into $f_{\mathrm{CP}}$. Using eqs.~(\ref{eq:Af}), (\ref{eq:Abf}) and
(\ref{eq:M0oft})-(\ref{eq:M0boft}) we can write
\begin{eqnarray}
  \label{eq:GM0oft}
    \Gamma( M^{0}(t) \to f_{\mathrm{CP}}) &=& 
    \left\vert
      g_{+}(t)
    \right\vert^{2} \left\vert \mathcal{A}_{f_{\mathrm{CP}}}\right\vert^{2}+
    \left\vert\frac{q}{p} \right\vert^{2} \left\vert g_{-}(t)
    \right\vert^{2} \left\vert \overline{\mathcal{A}}_{
                                              f_{\mathrm{CP}}}\right\vert^{2} \\
  &&+ 2 \mathrm{Re} \left( g_{+}^{*}(t)g_{-}(t)
      \frac{q}{p}\mathcal{A}_{f_{\mathrm{CP}}}^{*}\overline{\mathcal{A}}_{
     f_{\mathrm{CP}}}\right)\nonumber\\
  &=& \left\vert
    \mathcal{A}_{f_{\mathrm{CP}}}\right\vert^{2} 
  \left[
      \left\vert
      g_{+}(t)
    \right\vert^{2} +\left\vert \lambda_{f_{\mathrm{CP}}}
    \right\vert^{2}  \left\vert g_{-}(t)
      \right\vert^{2} + 2 \mathrm{Re} \left( g_{+}^{*}(t)g_{-}(t)
      \lambda_{f_{\mathrm{CP}}}\right)
      \right]\nonumber\\
  &=& \left\vert
    \mathcal{A}_{f_{\mathrm{CP}}}\right\vert^{2} \frac{e^{-\Gamma t}}{2} \left[
      \left(1 + \left\vert \lambda_{f_{\mathrm{CP}}}
    \right\vert^{2}
      \right)  \cosh (\Delta\Gamma\,t/2) \right.\nonumber\\
  && \left.
     +\left(1 - \left\vert \lambda_{f_{\mathrm{CP}}}
    \right\vert^{2}\right)
      \cos(\Delta m\,t)\right.\nonumber\\
  && \left.   - 2 
      \mathrm{Re }\lambda_{f_{\mathrm{CP}}} \sinh (\Delta\Gamma\,t/2) +2
      \mathrm{Im}\lambda_{f_{\mathrm{CP}}}\sin(\Delta m\,t)
    \right]
    \,,\nonumber\\
    \Gamma(\overline{M}^{0}(t) \to f_{\mathrm{CP}}) &=& \left\vert\frac{p}{q
      } \right\vert^{2} \left\vert g_{-}(t)  \right\vert^{2}
    \left\vert \mathcal{A}_{f_{\mathrm{CP}}} \right\vert^{2} +
    \left\vert g_{+}(t)   \right\vert^{2} \left\vert
      \overline{\mathcal{A}}_{f_{\mathrm{CP}}}\right\vert^{2}\nonumber\\
 && + 2 \mathrm{Re}\left( g_{+}^{*}(t)g_{-}(t)
      \frac{p}{q}\mathcal{A}_{f_{\mathrm{CP}}}\overline{\mathcal{A}}_{
    f_{\mathrm{CP}}}^{*}\right)\nonumber\\
  &=& \left\vert
    \overline{\mathcal{A}}_{f_{\mathrm{CP}}}\right\vert^{2} 
  \left[
      \left\vert
      g_{+}(t)
    \right\vert^{2} +\left\vert \lambda_{f_{\mathrm{CP}}}
    \right\vert^{-2}  \left\vert g_{-}(t)
    \right\vert^{2} + 2 \mathrm{Re} \left( g_{+}(t)^{*}g_{-}(t)
      \lambda_{f_{\mathrm{CP}}}^{-1}\right)
      \right]\nonumber\\
  &=& \left\vert
    \overline{\mathcal{A}}_{f_{\mathrm{CP}}}\right\vert^{2}
  \frac{e^{-\Gamma t}}{2}
  \left[
      \left(1 + \left\vert \lambda_{f_{\mathrm{CP}}}
    \right\vert^{-2}
    \right)  \cosh (\Delta\Gamma\,t/2) \right. \nonumber\\
 &&  \left.   +\left(1 - \left\vert \lambda_{f_{\mathrm{CP}}}
    \right\vert^{-2}\right)
      \cos(\Delta m\,t)\right.\nonumber\\
   &&\left.   - 2 
      \mathrm{Re}\lambda_{f_{\mathrm{CP}}}^{-1} \sinh (\Delta\Gamma\,t/2) +2
      \mathrm{Im}\lambda_{f_{\mathrm{CP}}}^{-1}\sin(\Delta m\,t)
    \right]\,,\nonumber
\end{eqnarray}
where in the last step we have used eqs.~(\ref{eq:gpm2}) and
(\ref{eq:gpgmst}).

Using the expressions above we can find an explicit form for the
so-called ``time-dependent CP asymmetry'', defined as follows:
\begin{dmath}[label={eq:acpt}]
  \mathcal{A}_{\mathrm{CP}}(t) \equiv \frac{\Gamma(M^{0}(t)\to
    f_{\mathrm{CP}})-\Gamma(\overline{M}^{0}(t)\to f_{\mathrm{CP}})}{\Gamma(M^{0}(t)\to
    f_{\mathrm{CP}})+\Gamma(\overline{M}^{0}(t)\to f_{\mathrm{CP}})} = 
   \frac{
    \left(
      1-\lvert \lambda_{f_{\mathrm{CP}}}\rvert^{2}
    \right) \cos(\Delta m\,t) - 2 \mathrm{Im}\lambda_{f_{\mathrm{CP}}}
  \sin(\Delta m\,t)}{ \left(
      1+\lvert \lambda_{f_{\mathrm{CP}}}\rvert^{2}
    \right) \cosh(\Delta \Gamma\,t/2) - 2 \mathrm{Re}\lambda_{f_{\mathrm{CP}}}
  \sinh(\Delta \Gamma\,t/2)} + \mathcal{O}
\left(1-\left\vert\frac{p}{q
      } \right\vert^{2}\right)\,.~~~~~~~~~~~~~~~~~
\end{dmath}
Neglecting CP violation in mixing, \emph{i.e.} assuming, according to
eq.~(\ref{eq:CPVmix}), $\lvert q/p \rvert =1$, the coefficient of the
$\cos(\Delta m\,t)$ term is nonvanishing in the presence of direct CP
violation in the $M \to f_{\mathrm{CP}}$ decay (see
eq.~(\ref{eq:CPVdir})), while the coefficient of the
$\sin(\Delta m\,t)$ term signals CP violation in the interference
between mixing and decay (see eq.~(\ref{eq:CPVint})).

Let us now discuss in turn $K$, $D$, $B_{d}$ and $B_{s}$ mixing. As we
shall see, different simplifying assumptions can be made in each sector.

\subsection{Kaon mixing and \texorpdfstring{$\epsilon_K$}{epsK}}
\label{sec:kaondf2}

If CP were conserved, the CP-odd eigenstate would not decay in a
two-pion final state, resulting in a much longer lifetime. Allowing
for small CP violation, it remains true that one eigenstate has a much
longer lifetime, so it is convenient to label the eigenstate by the
lifetime as Long- and Short-lived. Thus, we have
\begin{dmath}[label={eq:keig}]
  K_{S,L}= p_{K} \vert K^{0} \rangle \pm q_{K} \vert \overline{K}^{0}  \rangle\,.
\end{dmath}
We can simplify the general expressions in
eqs.~(\ref{eq:eigenv1})-(\ref{eq:df2evec}) using two peculiarities of
the Kaon system. First of all, the $\Delta I=1/2$ rule implies that
\begin{equation}
  \label{eq:G12K}
  \Gamma_{12} \approx A_0^{*} \overline{A}_0\,.
\end{equation}
Furthermore, one has
\begin{dmath}[label={eq:DGDMK}]
  \Delta \Gamma_{K} \sim -2 \Delta m_{K}\,.
\end{dmath}

From eq.~(\ref{eq:delta}), using eqs.~(\ref{eq:G12exp}),
(\ref{eq:DGDMK}) and (\ref{eq:G12K}), we obtain
\begin{dmath}[compact,label={eq:delta_K}]
  \delta = \frac{2 \mathrm{Im} 
    \left(
      M_{12}^{*} \Gamma_{12}
    \right)}{(\Delta m)^{2}
    + \lvert \Gamma_{12} \rvert^{2}} \simeq  \frac{2 \mathrm{Im}
    \left(
      M_{12}^{*} \Gamma_{12}
    \right)}{-2(\Delta m) 
    \lvert \Gamma_{12} \rvert} \simeq \frac{\mathrm{Im}
    \left(
      M_{12} A_0 \overline{A}_0^{*}
    \right)}{(\Delta m) 
    \lvert A_0 \overline{A}_0^{*} \rvert} = \frac{1}{\Delta
    m}\mathrm{Im}
  \left(
    M_{12} 
  \left[
    1 + 2 i \frac{\mathrm{Im} A_{0}}{\mathrm{Re} A_{0}} +
    \mathcal{O} 
    \left(
      \frac{\mathrm{Im} A_{0}}{\mathrm{Re} A_{0}}
    \right)^{2}
  \right]
  \right)
   \qquad\qquad \simeq \frac{1}{\Delta m} 
  \left(
    \mathrm{Im} M_{12} +  2 \frac{\mathrm{Im}
      A_{0}}{\mathrm{Re} A_{0}} \mathrm{Re} M_{12}
  \right)\simeq \frac{1}{\Delta m} \mathrm{Im} M_{12}
     + \frac{\mathrm{Im}
      A_{0}}{\mathrm{Re} A_{0}} \,,
\end{dmath}
where we have taken into account that in the standard phase convention
\begin{dmath}[label={eq:ImA0ReA0}]
  \mathrm{Im} A_{0} \ll \mathrm{Re} A_{0}\,.
\end{dmath}
Neglecting Im$A_{0}$, dimension eight operators in Im$M_{12}$ and
nonlocal matrix elements of two insertions of $\Delta S=1$
Hamiltonians, one has
\begin{equation}
  \delta \approx \frac{\mathrm{Im} M_{12}^{\mathrm{SD;}D=6}}{\Delta m_{K}}\,.
  \label{eq:deltaKsimp}
\end{equation}
This
approximation has an accuracy of $\sim 5\%$; going beyond it requires
evaluating Im$A_{0}$, long-distance contributions to Im$M_{12}$ and
the contribution of dimension eight operators to Im$M_{12}$, a
formidable task \cite{Cata:2003mn,Buras:2010pza,khighdim}.

To evaluate eq.~(\ref{eq:deltaKsimp}) we make use of the results
obtained in Sec.~\ref{sec:DF2}, and in particular of
eqs.~(\ref{eq:imheff}) and (\ref{eq:ds2smfull}):
\begin{dmath}[label={eq:imm12sdd6}]
  \mathrm{Im} M_{12}^{\mathrm{SD;}D=6} = \frac{G_{F}^{2}M_{W}^{2}}{6
    \pi^{2}}F_{K}^{2}m_{K}B_{K}(\mu) \mathrm{Im}\,
    \lambda_{sd}^{t}
    \left[\mathrm{Re}\, \lambda_{sd}^{c} \left(
         \eta_{c}(\mu) S_0(x_c) - \eta_{tc}(\mu) S_0(x_t,x_c) 
       \right) 
        - \mathrm{Re}\, \lambda_{sd}^{t} 
       \left(
         \eta_{t}(\mu) S_0(x_t) - \eta_{tc}(\mu) S_0(x_t,x_c) 
       \right)
    \right]\,,
\end{dmath}
where the QCD corrections from the matching and from the RG evolution
have been lumped in the factors $\eta_{t,c,tc}(\mu)$. Notice that in the
literature it is customary to define the scale-invariant parameters
$\hat{B}_{K}$ and $\eta_{1,2,3}$ by combining the $\mu$-dependent part of
$\eta_{t,c,tc}(\mu)$ with $B_{K}(\mu)$, thereby cancelling explicitly
the $\mu$ dependence at the given order, see for example
ref.~\cite{Buchalla:1995vs}.  

\subsubsection{Phenomenology of CP violation in the Kaon system}
\label{sec:epsk}

In the CP-invariant case, $K_{S,L}$ would correspond to CP
eigenstates and $K_L$ decays to two pions would be forbidden. To test
CP conservation, one can therefore measure
\begin{dmath}[label={eq:etaK},compact]
    \eta_{00} = \frac{\langle \pi^{0} \pi^{0} \lvert H \rvert K_L
      \rangle}{\langle \pi^{0} \pi^{0} \lvert H \rvert K_S \rangle} \qquad
  \textrm{and} \qquad
    \eta_{+-} = \frac{\langle \pi^{+} \pi^{-} \lvert H \rvert K_L
      \rangle}{\langle \pi^{+} \pi^{-} \lvert H \rvert K_S \rangle}\,.
\end{dmath}
Defining
\begin{dmath}[label={eq:AAbarK},compact]
    A_{f} = \langle f \lvert H \rvert K^{0} \rangle\,,~
    \overline{A}_{f} = \langle f \lvert H \rvert \overline{K}^{0} \rangle~
  \mathrm{and}~
    \lambda_f = 
    \left(
      \frac{q}{p}
    \right)_{K} \frac{\overline{A}_f}{A_f}\,,
\end{dmath}
we have
\begin{dmath}[label={eq:etalambda}]
    \eta_f = \frac{1 - \lambda_f}{1 + \lambda_f}\,.
\end{dmath}
Writing $\pi\pi$ decay amplitudes in terms of final states with
fixed isospin as in eq.~(\ref{eq:kpipiampli}), we take the
combination
\begin{dmath}[label={eq:epsetas},compact]
  \epsilon_K = \frac{1}{3} 
  \left(
    \eta_{00} + 2 \eta_{+-}
  \right) = \frac{1- \lambda_0}{1 + \lambda_0} + \mathcal{O}
  \left(
    \frac{A_2^{2}}{A_0^{2}}
  \right)\,,
\end{dmath}
selecting a pure $I=0$ state up to $2$\textperthousand. Since there is
only one final state strong phase, the conditions of
eq.~(\ref{eq:CPVdir}) are not met and there is no direct CP
violation in $\epsilon_{K}$. We have
\begin{dmath}[label={eq:Reepsk}]
  \mathrm{Re}\,\epsilon_{K} = \frac{1 -
    \lvert\lambda_{0}\rvert^{2}}{1+ 2\mathrm{Re}\,\lambda_{0} + \lvert\lambda_{0}\rvert^{2}}\,,
\end{dmath}
so that
\begin{dmath}[label={eq:Reepsn0},compact]
  \mathrm{Re}\,\epsilon_{K} \neq 0 \qquad \Rightarrow \qquad \lvert
  \lambda_{0}\rvert \neq 1 \qquad \Rightarrow \qquad 
  \left\vert
    \frac{q_{K}}{p_{K}}\right\vert \neq 1
\end{dmath}
implies CP violation in $K-\bar{K}$ mixing (see eq.(\ref{eq:CPVmix})), while
\begin{dmath}[label={eq:Imepsk}]
  \mathrm{Im}\,\epsilon_{K} = \frac{-2
    \mathrm{Im}\,\lambda_{0}}{1+ 2\mathrm{Re}\,\lambda_{0} +
    \lvert\lambda_{0}\rvert^{2}}\,, 
\end{dmath}
so that
\begin{dmath}[label={eq:Imepsn0},compact]
  \mathrm{Im}\,\epsilon_{K} \neq 0 \qquad \Rightarrow \qquad \mathrm{Im}\,
  \lambda_{0} \neq 0
\end{dmath}
implies CP violation in the interference between mixing and decay (see
eq.(\ref{eq:CPVint})). Experimentally, arg\,$\epsilon_{K}\approx
\pi/4$, so the two CP-violating effects are comparable. 

From eqs.~(\ref{eq:Reepsk}), (\ref{eq:AAbarK}), (\ref{eq:deltaCPV}),
(\ref{eq:deltaKsimp}) and (\ref{eq:imm12sdd6}) we obtain
\begin{dmath}[compact,label={eq:ReepsKphen}]
  \mathrm{Re}\,\epsilon_{K} \simeq \frac{1}{2}\frac{1-\lvert
    \lambda_{0}\rvert^{2}}{1+\lvert \lambda_{0}\rvert^{2}} =
  \frac{1}{2}\frac{1-\lvert
    \frac{q_{K}}{p_{K}}
    \rvert^{2}}{1+\lvert \frac{q_{K}}{p_{K}}\rvert^{2}} = \frac{\delta}{2} =
  \frac{G_{F}^{2}M_{W}^{2}}{12 \Delta m_{K}
    \pi^{2}}F_{K}^{2}m_{K}B_{K}(\mu) \mathrm{Im}\,
    \lambda_{sd}^{t}
    \left[\mathrm{Re}\, \lambda_{sd}^{c} \left(
         \eta_{c}(\mu) S_0(x_c) - \eta_{tc}(\mu) S_0(x_t,x_c) 
       \right)
        -  \mathrm{Re}\, \lambda_{sd}^{t} 
       \left(
         \eta_{t}(\mu) S_0(x_t) - \eta_{tc}(\mu) S_0(x_t,x_c) 
       \right)
    \right]\,.
\end{dmath}
The expression above is valid up to corrections from dimension eight
operators, from nonlocal matrix elements of two $\Delta S=1$
effective Hamiltonians and from the deviation from $\pi/4$ of the
phase of $\epsilon_{K}$. These effects have been partially estimated
in ref.~\cite{Buras:2010pza}, leading to a correction factor of $0.94
\pm 0.02$. At NLO, the SM prediction from ref.~\cite{Bona:2005vz}
\begin{dmath}[label={eq:epsKSM}]
  \lvert\epsilon_{K}\rvert = (1.97 \pm 0.18)\cdot 10^{-3}
\end{dmath}
compares very well with the experimental value
\begin{dmath}[label={eq:epsKexp}]
  \lvert\epsilon_{K}\rvert = (2.228 \pm 0.011)\cdot 10^{-3}\,.
\end{dmath}
We will come back again to $\epsilon_{K}$ when discussing the UTA in
the SM and beyond in Section \ref{sec:UTA}.

We can form another interesting combination of $\eta_{+-}$ and
$\eta_{00}$:
\begin{dmath}[compact,label={eq:epsp}]
  \epsilon^{\prime} \equiv \frac{1}{3} 
  \left(
    \eta_{{+-}}-\eta_{00}
  \right) \simeq \frac{\langle (\pi \pi)_{I=0}
    \lvert H \rvert K_L\rangle\langle (\pi \pi)_{I=2}
    \lvert H \rvert K_S\rangle - \langle (\pi \pi)_{I=0}
    \lvert H \rvert K_S\rangle\langle (\pi \pi)_{I=2}
    \lvert H \rvert K_L\rangle }{\sqrt{2}\langle (\pi \pi)_{I=0}
    \lvert H \rvert K_S\rangle^{2}} 
  \simeq \frac{i e^{i
    (\delta_{2}- \delta_{0})}}{\sqrt{2}}  \mathrm{Im}\,
  \left(
    \frac{A_{2}}{A_{0}}
  \right)\simeq  \frac{i e^{i
    (\delta_{2}- \delta_{0})}}{\sqrt{2}} 
\left(
  \frac{\mathrm{Im}\,A_{2}}{\mathrm{Re}\,A_{0}} - \omega
  \frac{\mathrm{Im}\,A_{0}}{\mathrm{Re}\,A_{0}}
\right)\,,
\end{dmath}
where $\omega=\mathrm{Re}\,A_{2}/\mathrm{Re}\,A_{0}$ and the
equalities are valid up to corrections of relative order
$\mathcal{O}(\omega,\epsilon_{K},\mathrm{Im}\,A_{0}/\mathrm{Re}\,A_{0})$.
For $\delta_{2} \neq \delta_{0}$ and Im $(A_{2}/A_{0}) \neq 0$ the
conditions for CP violation in the decay are satisfied and we have Re
$\epsilon^{\prime}\neq 0$.

Obtaining a solid estimate of $\epsilon^{\prime}$ is an extremely
difficult task: it contains all the difficulties of the $\Delta I=1/2$
rule and it is also affected by the cancellation between the two terms
in the right-hand side of eq.~(\ref{eq:epsp}). Indeed, in the SM the
CP-violating effects from QCD penguins in $A_{0}$ and from electroweak
penguins in $A_{2}$ cancel to a large extent, leading typically to
predictions for Re $\epsilon^{\prime}/\epsilon$ in the $10^{-4}$ range
\cite{Buras:1993dy,Ciuchini:1995cd,Bosch:1999wr}, below the world average of Re
$\epsilon^{\prime}/\epsilon = (16.6 \pm 2.3) 10^{-4}$
\cite{Batley:2002gn,AlaviHarati:2002ye,Abouzaid:2010ny}. Very
recently, a first estimate of Re $\epsilon^{\prime}/\epsilon$ in
Lattice QCD has been obtained in the same framework of the first
estimate of the $\Delta I=1/2$ rule, pointing to a value in the low
$10^{-4}$ range, but with a large uncertainty
\cite{Bai:2015nea,Blum:2015ywa}. This result has triggered a
reanalysis of the SM prediction combining lattice QCD results with
phenomenological considerations and/or arguments based on Dual QCD
\cite{Buras:2015xba,Buras:2015yba,Buras:2016fys,Kitahara:2016nld,Buras:2018ozh}, leading to a claimed
discrepancy of $\sim 3\,\sigma$ with the experimental value. On the
other hand, the lattice calculation underestimates the $I=0$ strong
interaction phase, and underestimating final state interactions could
bring to an underestimate of $\epsilon^{\prime}/\epsilon$, as noted in
\cite{Antonelli:1995gw,Bertolini:1995tp,Pallante:1999qf,Pallante:2000hk,Buchler:2001np,Buchler:2001nm,Pallante:2001he}
and more recently stressed in
\cite{Gisbert:2017vvj,Gisbert:2018tuf,Gisbert:2018niu}. Further
progress in the evaluation of the relevant matrix elements is needed
to assess the compatibility of the SM prediction with the experimental
value, keeping in mind that $\epsilon^{\prime}/\epsilon$ is one of the
observables with higher sensitivity to NP.

\subsection{\texorpdfstring{$\mathbf{D-\bar{D}}$}{DDbar} mixing and CP violation}
\label{sec:DDbar}

\begin{figure}[htbp]
\begin{center}      
  \includegraphics[width=0.4\textwidth]{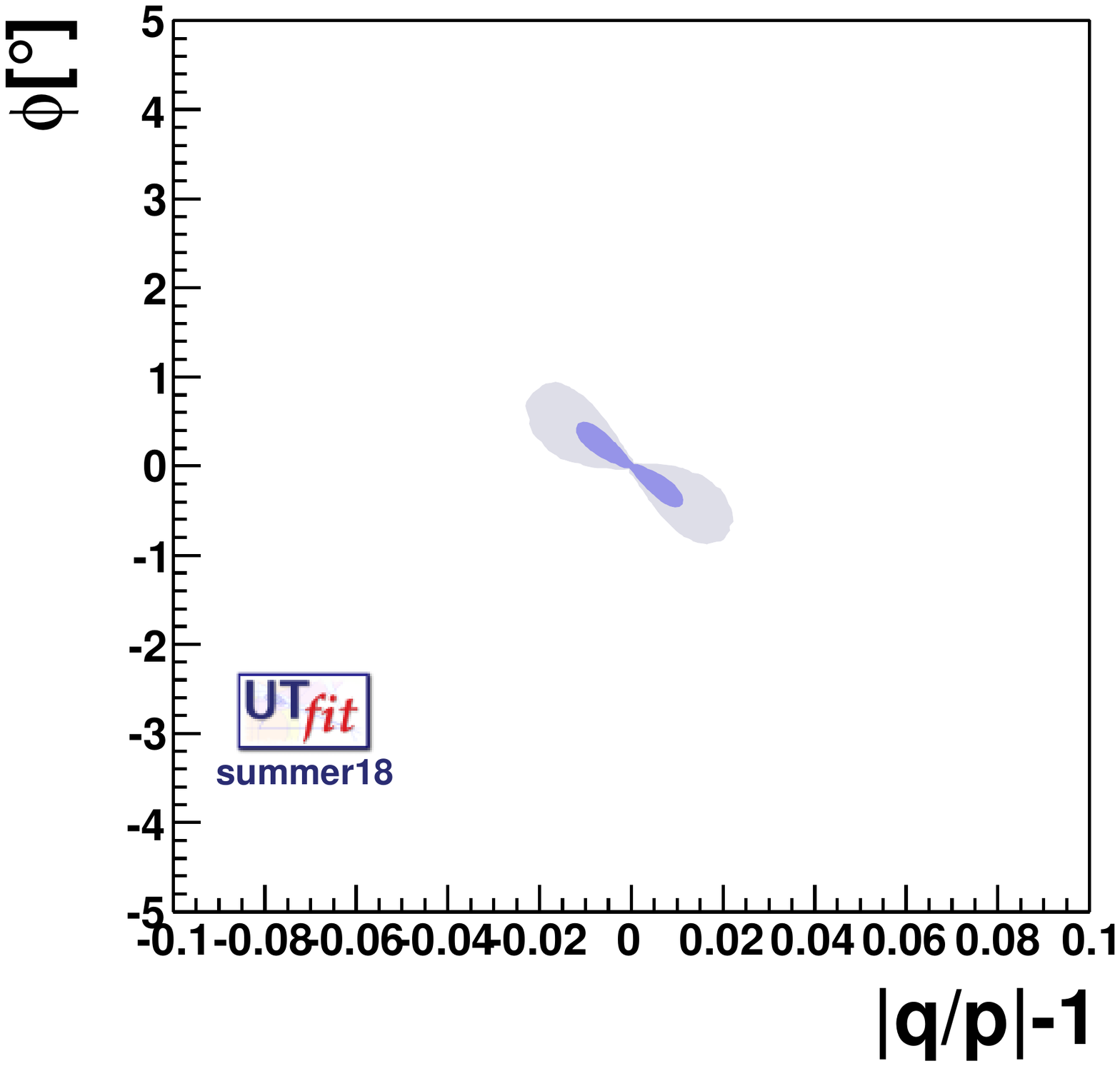}
  \includegraphics[width=0.4\textwidth]{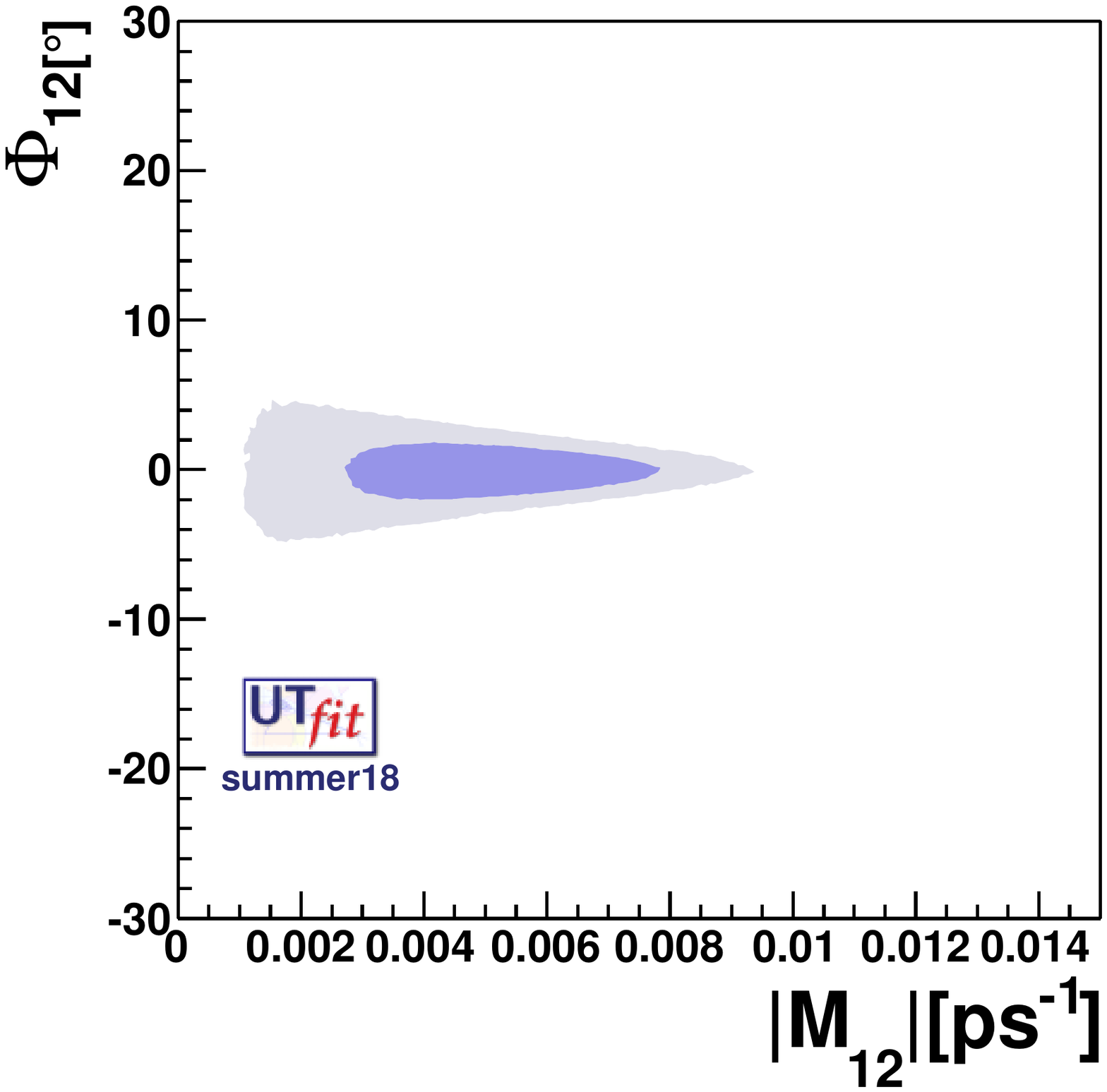}
\end{center}
\caption{Probability density functions for $\phi$ vs
  $\lvert q_{D}/p_{D} \rvert-1$ (left panel) and for $\Phi_{12}$ vs
  $\lvert M_{12} \rvert$ (right panel). The darker (lighter) regions
  correspond to $68\%$ ($95\%$) probability.}
\label{fig:DDbar} 
\end{figure}

In complete analogy with $\Delta S=2$ transitions, $M_{12}$ and
$\Gamma_{12}$ for $D-\bar{D}$ mixing have the following structure:
\begin{dmath}[compact,label={eq:ddbarm12g12}]
  (\lambda_{cu}^{s})^{2} (f_{dd}+ f_{ss}-2 f_{ds}) + 2 \lambda_{cu}^{s}\lambda_{cu}^{b}
  (f_{dd}-f_{ds}-f_{db}+f_{sb}) + (\lambda_{cu}^{b})^{2} (f_{dd}+ f_{bb}-2 f_{db})\,,
\end{dmath}
where $\lambda_{cu}^{q} = V^{\phantom{*}}_{cq}V_{uq}^{*}$, $f_{q_{i}q_{j}}$ represents an
intermediate state with flavours $q_{i}$ and $q_{j}$, and intermediate
states containing a $b$ quark only appear in $M_{12}$. We see that the
third generation here plays a very minor role with respect to
$K - \bar{K}$ mixing, since its contribution is suppressed by
$m_{b}^{2}/m_{t}^{2}$ with respect to $\Delta S=2$ amplitudes. Indeed,
we can safely neglect the term proportional to
$(\lambda_{cu}^{b})^{2}$. Then, the GIM mechanism essentially coincides with
the U-spin subgroup of the flavour $SU(3)$ symmetry of strong
interactions. Repeating the arguments of Sec.~\ref{sec:local} we see
that in this case the mixing amplitudes are dominated by non-local
contributions, making even a rough estimate of $M_{12}$ and
$\Gamma_{12}$ a tremendous task. While we may hope that in the future
the pioneering studies of $\Delta m_{K}$ on the lattice
\cite{Christ:2014qwa} may be extended to $D-\bar{D}$ mixing, it turns
out that CP violation in $D-\bar{D}$ mixing is already today a very
powerful probe of NP. Indeed, the approximate decoupling of the third
generation implies a strong suppression of CP-violating effects. We
can quantify this suppression by looking at the relevant combination
of CKM elements:
\begin{dmath}[compact,label={eq:RCKMD}]
  r = \mathrm{Im}\, \frac{\lambda_{cu}^{b}}{\lambda_{cu}^{s}} \simeq
  6.5 \cdot 10^{-4}\,.
\end{dmath}
The long-distance contributions to $M_{12}$ and $\Gamma_{12}$ can be
parameterized in terms of their U-spin quantum numbers:
\begin{dmath}[compact,label={eq:DC2Usp}]
  (\lambda_{cu}^{s})^{2} 
  \left(
    \Delta U=2
  \right) + 2 \lambda_{cu}^{s}\lambda_{cu}^{b} (\Delta U=1 +
  \Delta U=2) + \mathcal{O}((\lambda_{cu}^{b})^{2}) \approx (\lambda_{cu}^{s})^{2}
  \epsilon^{2} + 2 \lambda_{cu}^{s}\lambda_{cu}^{b} \epsilon\,,
\end{dmath}
so that we expect CP violation to arise at the level of
$r\epsilon/\epsilon^{2} \approx 2\cdot 10^{-3} \approx 0.1^{\circ}$
for an U-spin breaking of the order of $30\%$. Given the current
experimental errors, it is therefore adequate to assume all SM
amplitudes to be real, and interpret the (non)-observation of CP
violation in $D-\overline{D}$ mixing as an effect of (a constraint on)
NP. In fact, heavy NP could generate a short-distance contribution to
Im $M_{12}$, which could be observable either via
$\lvert q_{D}/p_{D} \rvert \neq 1$ or equivalently via
$\phi \equiv \mathrm{arg}(q/p)_{D} \neq 0$ (the two are not independent if
all decay amplitudes are real \cite{Ciuchini:2007cw}). Allowing
for NP-induced CP violation in $M_{12}$ only, and keeping all decay
amplitudes real, a global combination of $D$-mixing related decays can
be performed, leading to stringent constraints on NP. For example, the
Summer 2018 update of the analysis of refs.~\cite{Bevan:2014tha} finds
the distributions for $\lvert q_{D}/p_{D} \rvert$,
$\phi$, $\lvert M_{12} \rvert$ and its phase $\Phi_{12}$ reported in
Fig. \ref{fig:DDbar}, corresponding to a bound on
$\lvert\Phi_{12}\rvert < 3.5^{\circ} @ 95\%$ probability.

\subsection{\texorpdfstring{$\mathbf{B_{d}-\bar{B}_{d}}$}{BBar}
  mixing} 
\label{sec:Bdmix}

Let us consider the structure of $M_{12}$ and
$\Gamma_{12}$ for $B_{d}-\bar{B}_{d}$ mixing:
\begin{dmath}[label={eq:bbbarm12g12}]
  (\lambda_{bd}^{c*})^{2} (f_{uu}+ f_{cc}-2 f_{uc}) + 2 \lambda_{bd}^{c*}\lambda_{bd}^{t*}
  (f_{uu}-f_{uc}-f_{ut}+f_{ct}) + (\lambda_{bd}^{t*})^{2} (f_{uu}+ f_{tt}-2 f_{ut})\,,
\end{dmath}
where $\lambda_{bd}^{q} = V_{qb}^{*}V^{\phantom{*}}_{qd}$, $f_{q_{i}q_{j}}$ represents
an intermediate state with flavours $q_{i}$ and $q_{j}$, and again
intermediate states containing a $t$ quark only appear in $M_{12}$. A
few remarks are in order:
\begin{itemize}
\item contrary to the case of $\Delta S=2$ transitions,
  $\lambda_{bd}^{c} \sim \lambda_{bd}^{t}$, so there is no CKM enhancement of
  light quark contributions and $M_{12}$ is dominated by top quark
  exchange, \emph{i.e.}
  \begin{dmath}[compact,label={eq:m12bd}]
    M_{12} \simeq (\lambda^{t*}_{bd})^{2} f_{tt}\,.
  \end{dmath}
  Following the reasoning in Sec.~\ref{sec:local} we see
  that corrections to the leading contribution from nonlocal matrix
  elements of two $\Delta B=1$ effective Hamiltonians and from higher
  dimensional operators arise at
  $\mathcal{O}(m_{b}^{2}/M_{W}^{2},m_{b}^{2}/m_{t}^{2})$ and are thus
fully negligible;
\item also at variance with $K - \bar{K}$ mixing, $\Gamma_{12}$ is
  suppressed with respect to $M_{12}$ since the top quark does not
  contribute there, so that
  \begin{dmath}[compact,label={eq:G12oM12}]
    \left\vert
      \frac{\Gamma_{12}}{M_{12}}
    \right\vert \sim \mathcal{O}
    \left(
      \frac{m_{b}^{2}}{m_{t}^{2}} 
    \right) \ll 1\,;
  \end{dmath}
\item CP violation in mixing is even further suppressed, since the
  dominant contribution to both $M_{12}$ and $\Gamma_{12}$ is
  proportional to $(\lambda_{bd}^{t*})^{2}$, so that the CKM phase drops in the
  ratio $\Gamma_{12}/M_{12}$. CP violation is then induced solely by the other
  GIM-suppressed contributions to $\Gamma_{12}$;
\item last but not least, since the number of channels contributing to
  $\Gamma_{12}$ is large, and the momentum of intermediate
  states is of $\mathcal{O}(m_{b})$, we can advocate quark-hadron
  duality and perform an operator product expansion for $\Gamma_{12}$
  as well. While a detailed discussion of this subject goes well
  beyond the scope of these lectures, the interested reader will find
  all the details in
  refs.~\cite{Beneke:1996gn,Beneke:1998sy,Ciuchini:2003ww}.
\end{itemize}

Let us now work out the expressions of Sec.~\ref{sec:mixing} with the
approximation $\lvert\Gamma_{12}\rvert \ll \lvert M_{12}\rvert$:
\begin{dgroup*}
  \begin{dmath}[compact]
    \Delta m_{B_{d}} = 2 \lvert M_{12}\rvert\,,
  \end{dmath}
  \begin{dmath}[compact]
    \frac{\Delta\Gamma_{B_{d}}}{\Delta m_{B_{d}}} = \mathrm{Re}
    \frac{\Gamma_{12}}{M_{12}}\,,
  \end{dmath}
  \begin{dmath}[compact]
    \left(
      \frac{q}{p}
    \right)_{B_{d}} = \frac{M_{12}^{*}}{\lvert M_{12}\rvert} 
    \left(
      1 - \frac{1}{2} \mathrm{Im} \frac{\Gamma_{12}}{M_{12}}
    \right)\,,
  \end{dmath}
  \begin{dmath}[compact]
    \left\vert
      \frac{q}{p}
    \right\vert_{B_{d}} - 1 = -\frac{1}{2} \mathrm{Im} \frac{\Gamma_{12}}{M_{12}}\,.
  \end{dmath}
\end{dgroup*}

The mass difference is obtained taking the matrix element of the
$\Delta B=2$ effective Hamiltonian as
\begin{dmath}[compact,label={eq:deltambd}]
  \Delta m_{B_{d}} = \frac{G_{F}^{2} M_{W}^{2}}{2 \pi^{2}}
  \left\vert
    V^{\phantom{*}}_{tb}V^{\phantom{*}}_{td}
  \right\vert^{2} S_{0}(x_{t}) \eta_{b} m_{B_{d}} f_{B_{d}}^{2}
  B_{B_{d}}\,,
\end{dmath}
where the QCD corrections
\cite{Buras:1992tc,Buras:1992zv,Ciuchini:1993vr} have been absorbed in
$\eta_{b}$ and $B_{B_{d}}$ is the $B$-parameter computed in the same
scheme and at the same scale as $\eta_{b}$. The Summer 2018 SM
prediction by the UTfit collaboration is
\begin{dmath}[label={eq:dmbdsm}]
  \Delta m_{B_{d}}^{\mathrm{SM}} = (0.54 \pm 0.03)~\mathrm{ps}^{-1}
\end{dmath}
which compares very well with the experimental average
\begin{dmath}[label={eq:dmbdexp}]
  \Delta m_{B_{d}}^{\mathrm{exp}} = (0.5064 \pm 0.0019) \mathrm{ps}^{-1}\,.
\end{dmath}
The experimental sensitivity to $\Delta \Gamma_{B_{d}}$ is still well
above the SM prediction, and the same is true for the semileptonic
asymmetry $A^{\mathrm{SL}}_{B_{d}}$ defined in eq.~(\ref{eq:asl}),
which measures CP violation in mixing.

From the phenomenological point of view, $B_{d}$ mesons have three
peculiarities that make them a golden system to study meson-antimeson
oscillations and CP violation  \cite{Carter:1980tk,Bigi:1981qs}:
\begin{itemize}
\item since CKM angles involving the third generation are small, the
  $B_{d}$ lifetime is of $\mathcal{O}($ps$^{-1})$, so that a
  relatively small boost is enough to allow for a $B_{d}$ meson to
  fly a measurable distance before it decays;
\item the $B_{d}-\bar{B}_{d}$ mass difference is comparable to the
  $B_{d}$ lifetime, opening the possibility to measure the time
  dependence of the oscillations;
\item the time-dependent CP asymmetry defined in eq.~(\ref{eq:acpt})
  allows to measure the CP-violating Im$\lambda_{f}$ for a variety of
  final states $f$, allowing for an extensive test of the CKM
  mechanism and of possible NP contributions.
\end{itemize}
For these reasons, the idea of an asymmetric $B$-factory, where
entangled pairs of $B_{d}-\bar{B}_{d}$ mesons could be produced with a
boost sufficient to observe the time oscillation, was put
forward \cite{Oddone:1987up} and developed, leading to the
extraordinary success of the BaBar and Belle experiments at SLAC and
KEK \cite{Bevan:2014iga}. 

\subsubsection{Time-dependent CP asymmetry in
  \texorpdfstring{$B_{d} \to J/\Psi K_{S}$}{JPsiKS}}
\label{sec:JPsiKS}

\begin{figure}[!htbp]   
    \begin{flushleft}
      \scalebox{.8}{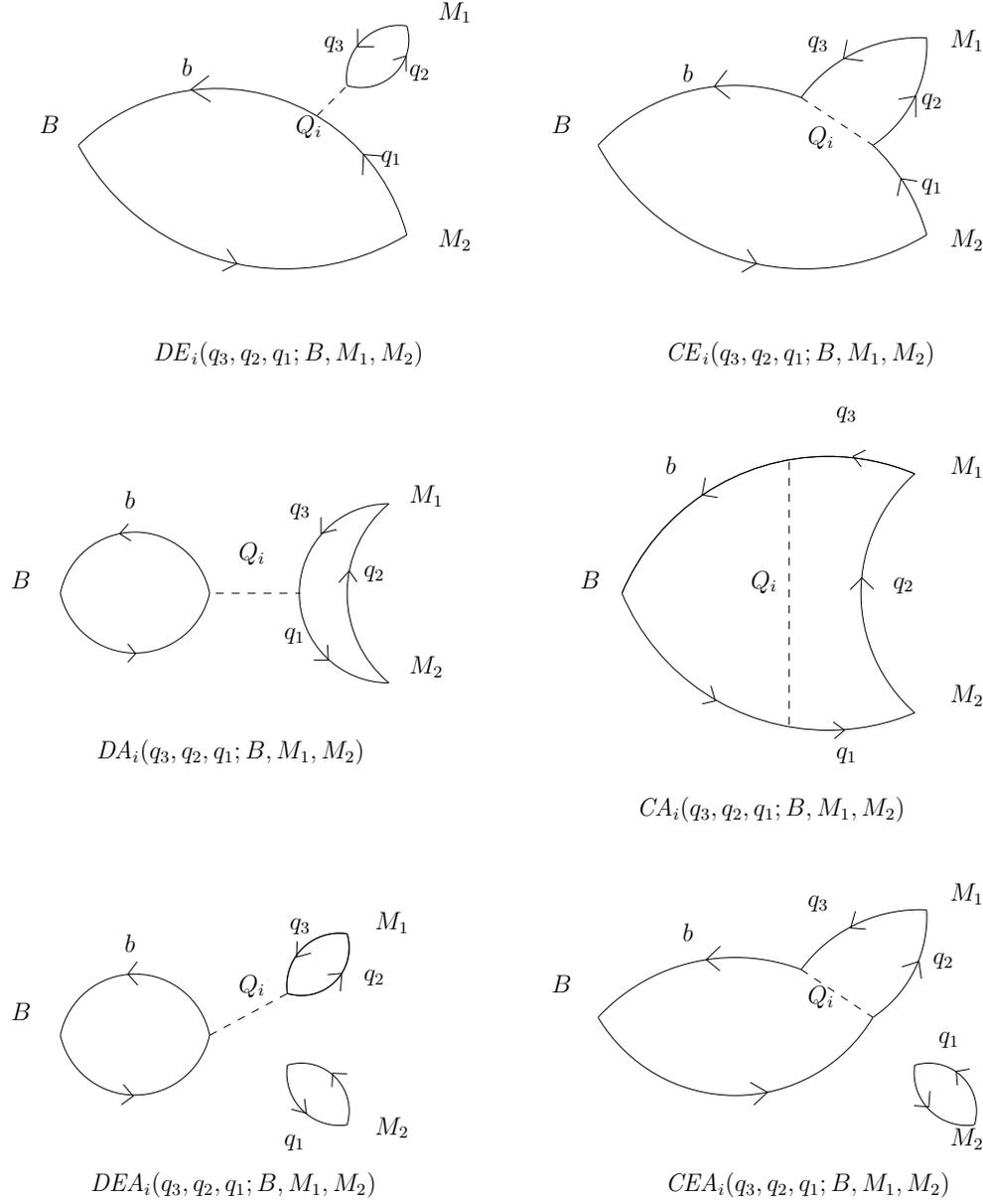}
    \end{flushleft}
    \caption[]{Emission, annihilation and emission-annihilation
      topologies of Wick contractions in the matrix elements of
      operators $Q_i$. From ref.~\protect\cite{Buras:1998ra}.}
    \label{fig:emiss-ann}
\end{figure}

\begin{figure}[!htbp]   
    \begin{flushleft}
      \scalebox{.8}{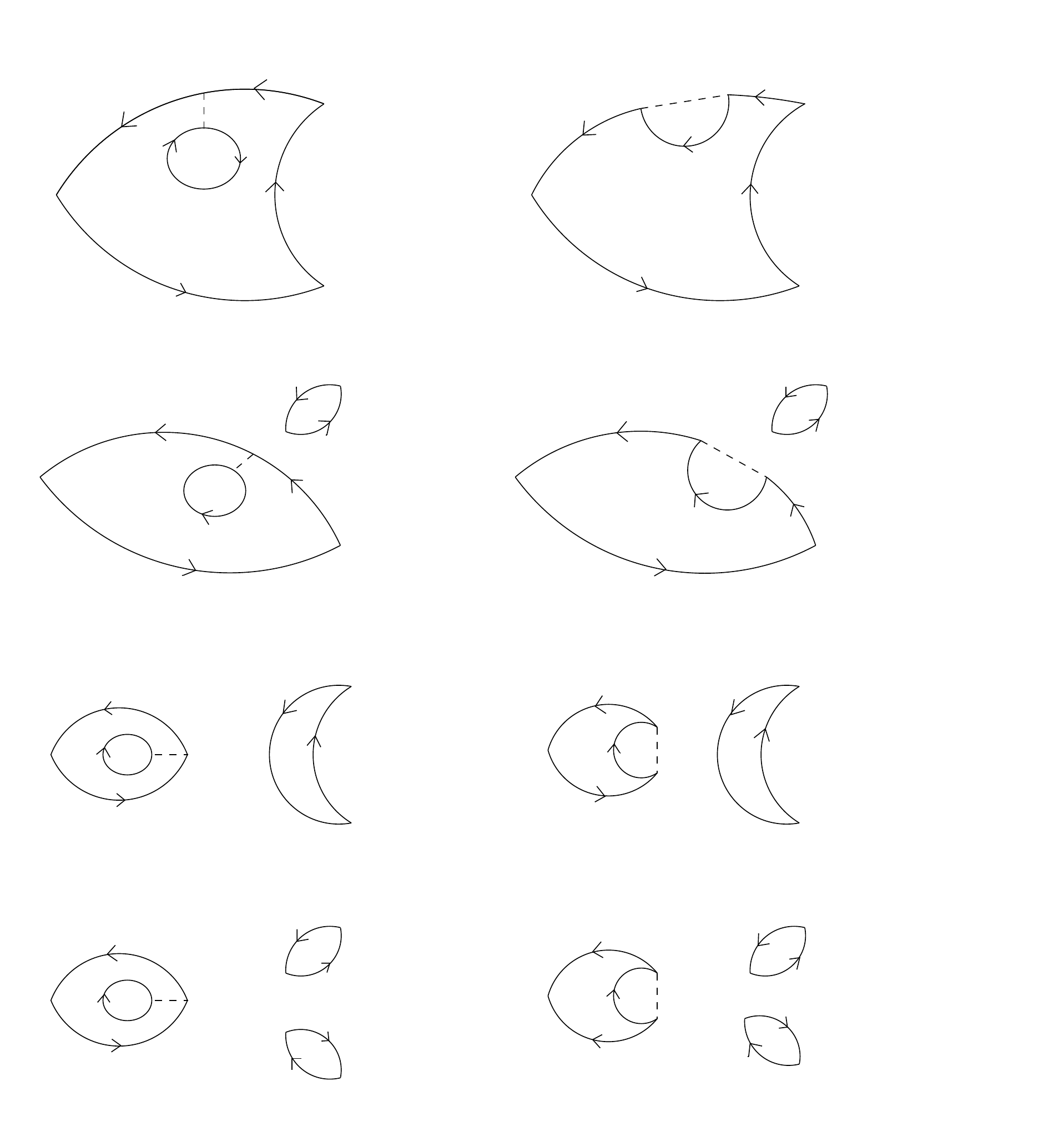}
    \end{flushleft}
    \caption[]{Penguin, penguin-emission, penguin-annihilation and
      double-penguin-annihilation topologies of Wick contractions in
      the matrix elements of operators $Q_i$. From
      ref.~\protect\cite{Buras:1998ra}.} 
    \label{fig:penguinst}
\end{figure}

Let us now discuss the time-dependent CP asymmetries for a series of
final states, starting from the famous ``golden channel''
$B_{d} \to J/\Psi K_{S}$. The underlying weak decay is
$\bar{b} \to \bar{c} c \bar{s}$, which is generated by the following
piece of the $\Delta B=1$ effective Hamiltonian (see
eqs.~(\ref{eq:heffds1}) and (\ref{eq:Q7})-(\ref{eq:Q10})):
\begin{dmath}[compact,label={eq:DB1bccbars}]
  \mathcal{H}_{\mathrm{eff}}^{\bar{b}\to \bar{c} c \bar{s}} = \frac{4 G_F}{\sqrt{2}}
  \left\{
   \lambda_{bs}^{c} 
    \left(
      C_1 Q_1^{\overline{b}c\overline{c} s} + C_2
      Q_2^{\overline{b}c\overline{c} s} +
      \sum_{i=3}^{10} C_i Q_i^{\overline{b}s} \right) +
      \lambda_{bs}^{u} \left(
      C_1 Q_1^{\overline{b}u\overline{u} s} + C_2
     Q_2^{\overline{b}u\overline{u} s} + \sum_{i=3}^{10} C_i Q_i^{\overline{b}s} \right) 
  \right\}\,.
\end{dmath}
To obtain the $\langle J/\Psi K^{0} \lvert {H}_{\mathrm{eff}}^{\bar{b}\to \bar{c}
  c \bar{s}} \rvert B_{d} \rangle$ matrix element we need to consider
all possible Wick contractions of the fields in
${H}_{\mathrm{eff}}^{\bar{b}\to \bar{c}c \bar{s}}$ with the initial
and final states. Following refs.~\cite{hep-ph/9703353,Buras:1998ra},
where the interested reader can find all details, we can classify the
different Wick contraction topologies as in
Figures \ref{fig:emiss-ann} and \ref{fig:penguinst}, where the left
(right) panels contain ``disconnected'' (``connected'')
topologies. In the infinite $m_{b}$ limit, the case in which the
``emitted'' meson (\emph{i.e.} $M_{1}$ in Figure \ref{fig:emiss-ann})
is light becomes computable in terms of form factors, decay constants
and perturbative QCD corrections, as argued in
ref.~\cite{Bjorken:1988kk} and carefully demonstrated in
refs.~\cite{hep-ph/9905312,Beneke:2000ry}. The basic idea is that the
emitted light meson flies away too fast for soft gluons to be
exchanged with the $B$ meson and with the other final state meson, the
so-called ``colour transparency'' argument. In spite of this
tremendous theoretical progress, however, a full-fledged computation
of the diagrams in Figures \ref{fig:emiss-ann} and \ref{fig:penguinst}
for realistic values of the $b$-quark mass
remains well beyond our capabilities. Indeed, long-distance
contributions and rescattering effects arising at $O(\Lambda/m_{b})$
are not systematically computable and have a strong phenomenological
impact in two-body nonleptonic $B$ decays, as emphasized in
refs.~\cite{hep-ph/9703353,hep-ph/0104126,Beneke:2001ev}. A
particularly dangerous class of long-distance contributions are the
so-called ``charming penguins'', namely penguin matrix elements as in
Figure \ref{fig:penguinst} with a charm quark running in the loop,
which are affected by $D_{(s)}^{(*)}-\bar{D}_{(s)}^{(*)}$ rescattering
into light mesons. 

To be able to obtain robust phenomenological results, one must
therefore seek observables where the dangerous long-distance
contributions are either absent or strongly suppressed. To this aim,
it is convenient to consider renormalization-group invariant
combinations of Wilson coefficients times Wick contractions, as
detailed in ref.~\cite{Buras:1998ra}, and to express the decay
amplitudes in terms of these parameters. In the case of
$B_{d} \to J/\Psi K_{S}$ we obtain, using eq.~(\ref{eq:eigdefr}) for
the $K_{S}$ in the final state:
\begin{dmath}[compact,label={eq:AJPsiKS}]
  A_{B_{d} \to J/\Psi K_{S}} = 
  \left[
     \lambda^{c}_{bs} (E_{2} + P_{2}) +
  \lambda^{u}_{bs} (P_{2} - P_{2}^{\mathrm{GIM}})
  \right]/(2 p_{K})\,,
\end{dmath}
where $E_{2}$ contains emission matrix elements of
$Q_{1,2}^{\overline{b}c\overline{c} d}$ in the colour-suppressed
combination $C_{1} CE + C_{2} DE$; $P_{2}$ contains penguin-emission
matrix elements of $Q_{1,2}^{\overline{b}c\overline{c} d}$ together
with emission, emission-annihilation and penguin emission matrix
elements of $Q_{3-10}^{\overline{b}d}$; $P_{2}^{\mathrm{GIM}}$
contains penguin-emission matrix elements of the GIM-suppressed
combinations $Q_{1,2}^{\overline{b}c\overline{c} d} -
Q_{1,2}^{\overline{b}u\overline{u} d}$; $p_{K}$ appears to project the
$K^{0}$ onto the $K_{S}$ final state. We expect the dominant
contribution to come from $E_{2}$, since $P_{2}$ is suppressed either
by small Wilson coefficients or by penguin matrix elements, and
$P_{2}^{\mathrm{GIM}}$ is suppressed by penguin matrix elements and by
the GIM mechanism. 

Thus, we can use the expansion of
eqs.~(\ref{eq:Afcp})-(\ref{eq:Abfcp}) with
\begin{dmath}[compact,label={eq:rJpsiKs}]
  r_{J/\Psi K_{S}} = 
  \left\vert
    \frac{\lambda^{u}_{bs}}{\lambda^{c}_{bs}}
  \right\vert  \left\vert
    \frac{P_{2} - P_{2}^{\mathrm{GIM}}}{E_{2} + P_{2}}
  \right\vert \lesssim 
  \left\vert
    \frac{V^{\phantom{*}}_{ub}V^{\phantom{*}}_{us}}{V^{\phantom{*}}_{cb}}
  \right\vert \sim \mathcal{O}(10^{-2})\,.
\end{dmath}
Let us first be bold and put $r_{J/\Psi K_{S}}$ to zero. Then we
obtain from eqs. (\ref{eq:lambdafcp}) and (\ref{eq:AJPsiKS})
\begin{align}
    \lambda_{J/\Psi K_{s}} &= \left( \frac{q}{p} \right)_{B_{d}}
    \frac{\lambda^{c*}_{bs}}{\lambda^{c}_{bs}}\left(\frac{p}{q}
    \right)_{K} =
    \frac{(\lambda^{t}_{bd})^{2}}{(\lambda^{t*}_{bd})^{2}}
    \frac{\lambda^{c*}_{bs}}{\lambda^{c}_{bs}}
                             \frac{\lambda^{c*}_{sd}}{\lambda^{c}_{sd}}
                             \nonumber \\
  &=
    \frac{V_{tb}^{*}V^{\phantom{*}}_{td}}{V^{\phantom{*}}_{tb}V_{td}^{*}}
    \frac{V^{\phantom{*}}_{cb}V_{cs}^{*}}{V_{cb}^{*}V^{\phantom{*}}_{cs}}
    \frac{V^{\phantom{*}}_{cs}V_{cd}^{*}}{V_{cs}^{*}V^{\phantom{*}}_{cd}} =
    \frac{V^{\phantom{*}}_{cb}V_{cd}^{*}}{V^{\phantom{*}}_{tb}V_{td}^{*}}
    \frac{V_{tb}^{*}V^{\phantom{*}}_{td}}{V_{cb}^{*}V^{\phantom{*}}_{cd}} = e^{- 2 i \beta}\,,
  \label{eq:ljpsiksr0}
\end{align}
where the angle $\beta$ of the Unitarity Triangle is defined in
eq.~(\ref{eq:UTdef}). Plugging eq.~(\ref{eq:ljpsiksr0}) in
eq.~(\ref{eq:acpt}) we obtain
\begin{dmath}[compact]
  \label{eq:acpjpsiksr0}
  \mathcal{A}_{\mathrm{CP}}^{B_{d} \to J/\Psi K_{S}}(t) = - \sin 2
  \beta \sin(\Delta m_{B_{d}} t)\,,
\end{dmath}
taking into account that the final state is CP odd.

Thus, in the approximation $r_{J/\Psi K_{S}}=0$, measuring the
time-dependent asymmetry in this channel we should find a vanishing
coefficient of the $\cos (\Delta m_{B_{d}}t)$ term, and the
coefficient of the $\sin (\Delta m_{B_{d}}t)$ measures $\sin
2\beta$. The current world average of
$\mathcal{A}_{\mathrm{CP}}^{B_{d} \to J/\Psi K_{S,L}}(t)$ gives
\cite{Aubert:2009aw,Adachi:2012et,Aaij:2015vza,Aaij:2017yld,HFLAV16}
\begin{dmath}[compact,label={eq:jpsiksexp}]
  \sin 2 \beta = 0.690 \pm 0.018\,,
\end{dmath}
corresponding to $\beta \approx 21.8^{\circ}$.

Let us now go back to the assumption $r_{J/\Psi K_{S}}=0$, under which
we obtained eq.~(\ref{eq:acpjpsiksr0}), and investigate if we can get
any theoretical or experimental handle on the actual value of
$r_{J/\Psi K_{S}}$, or at least an upper bound on its value. A
theoretical calculation of $r_{J/\Psi K_{S}}$ from first principles is
currently impossible even in the infinite $m_{b}$ limit, since the
emitted meson is heavy. The direct CP asymmetry, \emph{i.e.} the
coefficient of the $\cos (\Delta m_{B_{d}} t)$ term in the
time-dependent asymmetry, according to eq.~(\ref{eq:ACPdir})
is sensitive to
\begin{dmath}[compact,label={eq:ACPdirJPsiKS}]
  r_{J/\Psi K_{S}} \sin\phi_{r_{J/\Psi K_{S}}} \sin\delta_{r_{J/\Psi
      K_{S}}} \approx \lambda^{2} R_{b}\left\vert
    \frac{P_{2} - P_{2}^{\mathrm{GIM}}}{E_{2} + P_{2}}
  \right\vert  \sin\gamma \sin \mathrm{arg} 
  \left(
    \frac{P_{2} - P_{2}^{\mathrm{GIM}}}{E_{2} + P_{2}}
  \right)\,,
\end{dmath}
with the UT parameters $\lambda$, $R_{b}$ and $\gamma$ defined in
eqs.~(\ref{eq:Wolfenstein-Buras}), (\ref{eq:UTdef}) and
(\ref{eq:angdef}). The last term in eq.~(\ref{eq:ACPdirJPsiKS}),
\emph{i.e.} the sine of the strong phase difference between the two
amplitudes, prevents us from using directly the direct CP asymmetry as
a bound on $r_{J/\Psi K_{S}}$. Even if we ignore this problem,
bounding $r_{J/\Psi K_{S}}$ at the level of the direct CP asymmetry
would anyway give a theoretical error comparable to the experimental
one. A way out can be found using the SU(3)-related decay channel
$B_{d} \to J/\Psi \pi^{0}$ \cite{Ciuchini:2005mg,Ciuchini:2011kd}. The effective
Hamiltonian governing this decay is given by
\begin{dmath}[compact,label={eq:DB1bccbard}]
  \mathcal{H}_{\mathrm{eff}}^{\bar{b}\to \bar{c} c \bar{d}} = \frac{4 G_F}{\sqrt{2}}
  \left\{
   \lambda_{bd}^{c} 
    \left(
      C_1 Q_1^{\overline{b}c\overline{c} d} + C_2
      Q_2^{\overline{b}c\overline{c} d} +
      \sum_{i=3}^{10} C_i Q_i^{\overline{b}d} \right) +
      \lambda_{bd}^{u} \left(
      C_1 Q_1^{\overline{b}u\overline{u} d} + C_2
     Q_2^{\overline{b}u\overline{u} d} + \sum_{i=3}^{10} C_i Q_i^{\overline{b}d} \right) 
  \right\}\,,
\end{dmath}
and the decay amplitude is given in the SU(3) limit by
\begin{dmath}[compact,label={eq:AJPsipi0}]
  A_{B_{d} \to J/\Psi \pi^{0}} =
     \lambda^{c}_{bd} (E_{2} + P_{2}) +
  \lambda^{u}_{bd} (P_{2} - P_{2}^{\mathrm{GIM}})\,,
\end{dmath}
neglecting a small, colour suppressed emission-annihilation
contribution. In eq.~(\ref{eq:AJPsipi0}) the second term is not
doubly-Cabibbo-suppressed anymore, so that this channel is much more
sensitive to $P_{2}^{\mathrm{GIM}} - P_{2}$. Using the information
from $B_{d} \to J/\Psi \pi^{0}$ one can constrain the theoretical
error in the extraction of $\sin 2 \beta$ from $B_{d} \to J/\Psi
K_{S}$ to be subdominant even allowing for an SU(3)
breaking of $100\%$. It is however crucial that in the future the
experimental progress on $B_{d} \to J/\Psi \pi^{0}$ parallels the one
on $B_{d} \to J/\Psi K_{S}$, so that the theory uncertainty remains
subdominant. 

\subsubsection{Time-dependent CP asymmetry in
  \texorpdfstring{$B \to \pi\pi$}{pipi}}
\label{sec:pipi}

Thanks to isospin symmetry, $B \to \pi\pi$ decays have the unique
property that all decay amplitudes can be determined experimentally,
allowing for a measurement of the CKM angle $\alpha$ with essentially
no theoretical input other than isospin
\cite{Gronau:1990ka}. Effects of isospin breaking due to
electromagnetic interactions and to quark masses are negligible with
respect to current experimental uncertainties, so we will not discuss
them here \cite{Gronau:2005pq}.

Using the isospin decomposition of eq.~(\ref{eq:kpipiampli}) for $B$
decays, we see that the independent parameters are the relative strong
phase of $I=0$ and $I=2$ amplitudes, the weak phases of $I=0$ and
$I=2$ amplitudes, and their absolute values, so five independent
parameters. If we consider time-dependent CP asymmetries, we should
add $(q/p)_{B_{d}}$; neglecting CP violation in the mixing, this
amounts to another parameter, arg$(q/p)_{B_{d}}$. The observables are
three CP-averaged branching ratios ($\mathcal{B}_{+-}$,
$\mathcal{B}_{+0}$ and $\mathcal{B}_{00}$) and four CP asymmetries (the
coefficients $\mathcal{S}_{+-,00}$ of $\sin \Delta m_{B_{d}} t$ and
$\mathcal{C}_{+-,00}$ of $\cos \Delta m_{B_{d}} t$ terms in
$\mathcal{A}_{CP}^{B_{d} \to \pi^{+,0}\pi^{-,0}}$), so the system is
overdetermined. In practice, however, the measurement of the
time-dependent CP asymmetry in $B_{d} \to \pi^{0} \pi^{0}$ is very
difficult, but there are enough observables to determine all
parameters even if only $\mathcal{C}_{00}$ is used.

It is convenient to write the decay amplitudes separating terms with
different weak phases rather than with different strong phases as in
eq.~(\ref{eq:kpipiampli}). In particular, using CKM unitarity, we
separate the amplitudes in terms
proportional to $\lambda^{u}_{bd}$ and $\lambda^{t}_{bd}$. Taking
into account that $(q/p)_{B_{d}} \simeq
(\lambda^{t}_{bd}/\lambda^{t*}_{bd})^{2}$ and that $\lambda_{f} = q/p
\bar{A}_{f}/A_{f}$, it is convenient to absorb a factor of
$\lambda^{t*}_{bd}$ ($\lambda^{t}_{bd}$) in $A_{f}$
($\bar{A}_{f}$). In this way we obtain
\begin{dgroup*}
  \begin{dmath}[compact,label={eq:ABdpppm}]
    A(B_{d} \to \pi^{+} \pi^{-}) = e^{-i \alpha} T^{+-} + P\,,
  \end{dmath}
  \begin{dmath}[compact,label={eq:ABdp0p0}]
    A(B_{d} \to \pi^{0} \pi^{0}) = 
    \left(
      e^{-i \alpha} T^{00} - P
    \right)\,,
  \end{dmath}
  \begin{dmath}[compact,label={eq:ABmpmp0}]
    A(B^{-} \to \pi^{-} \pi^{0}) = \frac{1}{\sqrt{2}} 
    e^{-i \alpha} \left(
      T^{+-} + T^{00} 
    \right)\,.
  \end{dmath}
\end{dgroup*}
One can then extract $\alpha$, together with $T^{+-}$, $T^{00}$, $P$
and their relative phases, up to an eight-fold ambiguity (explicit
formul{\ae} can be found in
refs.~\cite{Charles:1998qx,Bona:2007qta}). It is however clear that
the degeneracies in $\alpha$ correspond to different values of the
parameters $T^{+-}$, $T^{00}$ and $P$. One can then follow the same
argument used in Sec.~\ref{sec:JPsiKS} and relate
$B \to \pi^{+}\pi^{-}$ decays to $B_{s} \to K^{+}K^{-}$ via a U-spin
transformation. Since $B_{s} \to K^{+}K^{-}$ is a
$\bar{b} \to \bar{s} u \bar{u}$ transition, the $T$ and $P$ terms in
the amplitudes are weighted by a different CKM factor, breaking the
degeneracy between different solutions of the $B \to \pi\pi$
system. Thus, the isospin analysis of $B \to \pi\pi$ supplemented by
$B_{s} \to K^{+}K^{-}$ is more efficient \cite{Ciuchini:2012gd}.

Notice that the isospin analysis of $B \to \pi\pi$ can be generalized
beyond the SM as long as new physics does not enhance electroweak
penguins by orders of magnitude, and as long as it does not contribute
sizeably to current-current operators. Then, one can still extract
$\alpha$ even allowing for a NP weak phase to be present in $P$
\cite{Lipkin:1991st,Bona:2007vi}, although with a slightly larger
uncertainty.

Finally, the same analysis presented for $B \to \pi\pi$ can be carried
out for each polarization of the $B \to \rho\rho$ decays; it turns out
that the latter profits from larger branching ratios, making it more
sensitive than the $\pi\pi$ channel.

\subsubsection{Extracting \texorpdfstring{$\alpha$}{alpha} from 
  \texorpdfstring{$B \to \rho \pi$}{rhopi} decays}
\label{sec:rhopi}

In general, decays to final states including vector mesons can be
analyzed with a very powerful tool, the Dalitz plot, which allows in
principle to extract the absolute values of all amplitudes
contributing to a given final state, and all their relative phases,
provided that they interfere among each other in a non-negligible
region of phase space. Although the isospin structure of
$B \to \rho \pi$ decays is richer than the one of $\pi \pi$, since
the final state can also have isospin one, this just turns the
triangular relation for $B \to \pi \pi$,
$A(B_{d} \to \pi^{+} \pi^{-}) + A(B_{d} \to \pi^{0}\pi^{0}) = \sqrt{2}
A(B^{+} \to \pi^{+}\pi^{0})$, into a pentagonal relation,
$A(B_{d} \to \pi^{+} \rho^{-}) + A(B_{d} \to \pi^{-} \rho^{+}) + 2
A(B_{d} \to \pi^{0} \rho^{0}) = \sqrt{2} \left( A(B^{+} \to
  \rho^{+}\pi^{0}) + A(B^{+} \to \rho^{0}\pi^{+}) \right)$. Again,
this allows to determine the relative phase of the $I=3/2$ amplitudes
for $B$ and $\bar{B}$ decays, which corresponds to $2\alpha$
\cite{Lipkin:1991st,Snyder:1993mx,Quinn:2000by}. While a detailed
discussion of Dalitz analyses of three-body heavy meson decays goes
beyond the scope of these lectures, we refer the interested reader to
chapter 13 of ref.~\cite{Bevan:2014iga} for a review of several Dalitz
analysis techniques.

\subsection{\texorpdfstring{$\mathbf{B_{s}-\bar{B}_{s}}$}{BsBsbar}
  mixing} 
\label{sec:Bsmix}

The structure of $M_{12}$ and
$\Gamma_{12}$ for $B_{s}-\bar{B}_{s}$ mixing is analogous to the one
for $B_{d}-\bar{B}_{d}$ mixing given in eq.~(\ref{eq:bbbarm12g12}),
with the substitution $\lambda_{bd}^{f} \to
\lambda_{bs}^{f}$. However, while in the case of $B_{d}-\bar{B}_{d}$
mixing one has $\lvert \lambda_{bd}^{u,c,t} \rvert \sim \lambda^{3}$,
so all three factors arise at third order in the CKM parameter
$\lambda$, for $b \to s$ transitions the
relative weight of the three CKM factors is instead hierarchical:
\begin{dmath}[compact,label={eq:lambdabs}]
  \lvert \lambda_{bs}^{t,c}\rvert \simeq \lambda^{2} \gg
  \lambda_{bs}^{u} \simeq \lambda^{4}\,.
\end{dmath}
This has three very important phenomenological consequences:
\begin{enumerate}
\item CP violation in $B_{s}-\overline{B}_{s}$ mixing is tiny, since
  the $\mathcal{O}(\lambda^{2})$ decoupling of the first generation is
  reflected in the smallness of the angle $\beta_{s} \sim
  \mathcal{O}(\lambda^{2})$ defined in eq.~(\ref{eq:angdef}). This
  suppression acts on top of the mechanism already discussed for
  $B_{d}-\bar{B}_{d}$ mixing, leading to Im$(\Gamma_{12}/M_{12})\sim
  \mathcal{O}(10^{{-5}})$;
\item since $\Delta m_{B_{s}}/\Delta m_{B_{d}}$ goes approximately
  like the ratio $V^{\phantom{*}}_{ts}/V^{\phantom{*}}_{td} \sim 1/\lambda$
  while $\Gamma_{B_{s}} \sim \Gamma_{B_{d}}$, one has $\Delta
  m_{B_{s}}/\Gamma_{B_{s}} \sim 25$, making it much more difficult to
  resolve experimentally the time-dependence of the mixing;
\item the enhancement factor $\Delta m_{B_{s}}/\Gamma_{B_{s}}$ brings
  $\Delta \Gamma_{B_{s}}/\Gamma_{B_{s}} \sim 25 \Delta
  \Gamma_{B_{s}}/\Delta m_{B_{s}}$ to the observable level of
  $\mathcal{O}(10\%)$.
\end{enumerate}

Therefore, in studying $B_{s}-\bar{B}_{s}$ mixing we should keep the
terms proportional to $\Delta \Gamma_{B_{s}}$ in the expressions of
Sec.~\ref{sec:mixobs}, in particular in eq.~(\ref{eq:acpt}).

The Summer 2018 prediction for $\Delta m_{B_{s}}$ in the SM by the UTfit
collaboration is
\begin{dmath}[label={eq:dmbssm}]
  \Delta m_{B_{s}}^{\mathrm{SM}} = (17.25 \pm 0.85) \mathrm{ps}^{-1}\,,
\end{dmath}
which compares very well with the experimental average
\begin{dmath}[label={eq:dmbsexp}]
  \Delta m_{B_{s}}^{\mathrm{exp}} = (17.757 \pm 0.0021) \mathrm{ps}^{-1}\,,
\end{dmath}
while the prediction for $\Delta \Gamma_{B_{s}}$ yields
\begin{dmath}[label={eq:dgbssm}]
  (\Delta \Gamma_{B_{s}}/\Gamma_{B_{s}})^{\mathrm{SM}} = 0.15 \pm 0.01 \,,
\end{dmath}
well compatible with the experimental average
\begin{dmath}[label={eq:dgbsexp}]
  (\Delta \Gamma_{B_{s}}/\Gamma_{B_{s}})^{\mathrm{exp}} = 0.132 \pm
  0.008\,. 
\end{dmath}

\subsubsection{Time-dependent CP asymmetry in
  \texorpdfstring{$B_{s} \to J/\Psi \phi$}{JPsiphi}}
\label{sec:JPsiphi}

If we apply the same arguments presented in Sec.~\ref{sec:JPsiKS} and consider a
$\bar{b} \to \bar{c} c \bar{s}$ transition for $B_{s}$ decays, we are
led to $B_{s} \to J/\Psi \phi$ as the golden channel for the
measurement of the CKM angle $\beta_{s}$:
\begin{align}
    \lambda_{J/\Psi \phi} &= \left( \frac{q}{p} \right)_{B_{s}}
    \frac{\lambda^{c*}_{bs}}{\lambda^{c}_{bs}} =
    \frac{(\lambda^{t}_{bs})^{2}}{(\lambda^{t*}_{bs})^{2}}
    \frac{\lambda^{c*}_{bs}}{\lambda^{c}_{bs}}
                             \nonumber \\
  &=
    \frac{V_{tb}^{*}V^{\phantom{*}}_{ts}}{V^{\phantom{*}}_{tb}V_{ts}^{*}}
    \frac{V^{\phantom{*}}_{cb}V_{cs}^{*}}{V_{cb}^{*}V^{\phantom{*}}_{cs}}
     =
    \frac{V_{tb}^{*}V^{\phantom{*}}_{ts}}{V_{cb}^{*}V^{\phantom{*}}_{cs}}
    \frac{V^{\phantom{*}}_{cb}V_{cs}^{*}}{V^{\phantom{*}}_{tb}V_{ts}^{*}} = e^{2 i \beta_{s}}\,,
  \label{eq:ljpsiphir0}
\end{align}
where we have assumed $r_{J/\Psi \phi}=0$ and for simplicity we have
omitted the CP parity of the final state, to be determined with an
angular analysis of the decay products of the $J/\Psi \phi$
intermediate state. In the case of the $B_{s}$ meson, one cannot
neglect the terms proportional to $\Delta \Gamma_{B_{s}}$ in
eq.~(\ref{eq:acpt}), so the result of the measurement is a combined
fit of $\Delta \Gamma_{B_{s}}$ and Im$\lambda_{J/\Psi \phi}$.

However, if we now allow for a nonvanishing $r_{J/\Psi \phi}$, which
again we can estimate, following eq.~(\ref{eq:rJpsiKs}), as
\begin{dmath}[compact,label={eq:rJpsiphi}]
  r_{J/\Psi \phi} = 
  \left\vert
    \frac{\lambda^{u}_{bs}}{\lambda^{c}_{bs}}
  \right\vert  \left\vert
    \frac{P_{2} - P_{2}^{\mathrm{GIM}}}{E_{2} + P_{2}}
  \right\vert \lesssim 
  \left\vert
    \frac{V^{\phantom{*}}_{ub}V^{\phantom{*}}_{us}}{V^{\phantom{*}}_{cb}}
  \right\vert \sim \mathcal{O}(10^{-2})\,,
\end{dmath}
we immediately see that the correction to Im$\lambda_{J/\Psi \phi}$ is
of the same order of $\sin 2 \beta_{s}$:
\begin{dmath}[compact,label={eq:imljpsiphir0}]
  \mathrm{Im}\,\lambda_{J/\Psi \phi} = \sin 2 \beta_{s} -2 r_{J/\Psi \phi}\sin\gamma
  \cos \delta_{r_{J/\Psi \phi}} + \mathcal{O}(r_{J/\Psi \phi}^{2},r_{J/\Psi \phi}
    \lambda^{2})\,.
\end{dmath}
In other words, both $B_{d} \to J/\Psi K_{S}$ and $B_{s} \to J/\Psi
\phi$ suffer from doubly-Cabibbo suppressed corrections, but the
leading term is of $\mathcal{O}(1)$ for $B_{d}$ and doubly Cabibbo
suppressed for $B_{s}$. Still, the time-dependent CP asymmetry in $B_{s} \to J/\Psi
\phi$ remains a most precious tool to constrain possible NP
contributions to CP violation in $B_{s}$ mixing, at least down to the
level of $r_{J/\Psi \phi}$. One could of course envisage a strategy to
keep the corrections due to $r_{J/\Psi \phi}$ under control, using
$SU(3)$ as was discussed in Sec.~\ref{sec:JPsiKS}. However, this
approach is complicated by the mixed singlet-octet flavour structure
of the $\phi$ meson, requiring a detailed analysis of several final
states. We refer the interested reader to the discussion in
ref.~\cite{DeBruyn:2014oga}. 

\section{The Unitarity Triangle Analysis in the SM and beyond}
\label{sec:UTA}

Let us now very quickly review how we can combine a large amount of
theoretical and experimental information using the Unitarity Triangle
introduced in Sec.~\ref{sec:ckm}. Since the CKM matrix is governing
all flavour and CP violation in weak interactions, we can translate
virtually any flavour- or CP-violating process into a constraint on
the UT. Let us start from charged-current processes arising at the
tree level in the SM, before turning to FCNC transitions.

\subsection{The UT from tree-level decays}
\label{sec:treeUT}

The CKM matrix elements $|V^{\phantom{*}}_{ud}|$ and
$|V^{\phantom{*}}_{us}|$ can be measured from super-allowed $\beta$
decays~\cite{Towner:2010zz,Hardy:2016vhg} and from
semileptonic/leptonic kaon
decays~\cite{Moulson:2017ive,Tanabashi:2018oca,Aoki:2019cca}
respectively, providing an accurate determination of the sine of the
Cabibbo angle. Similarly, $|V^{\phantom{*}}_{cb}|$ and
$|V^{\phantom{*}}_{ub}|$ can be determined using (semi-)leptonic $B$
decays. In this case, one can use either exclusive or inclusive
decays, which have different theoretical and experimental systematic
errors. For $b \to c$ transitions, the analysis of inclusive
semileptonic decays relies on heavy quark symmetry and on global
quark-hadron duality, while the study of inclusive semileptonic
$b \to u$ transitions requires local quark-hadron duality, as well as
some model-dependent regularization of singularities that are absent
in $b \to c$ decays. In exclusive decays, an estimate of the relevant
form factors, as well as of their momentum dependence, is needed to
extract CKM factors. Unfortunately, determinations of
$|V^{\phantom{*}}_{cb}|$ and $|V^{\phantom{*}}_{ub}|$ from inclusive
and exclusive semileptonic $B$ decays have been displaying a
$\sim 3\,\sigma$ discrepancy for quite a while
\cite{Tanabashi:2018oca}, although it was recently noticed that for
$|V^{\phantom{*}}_{cb}|$ the situation improves considerably if one
relaxes some assumptions on the momentum dependence of the form
factors based on the heavy quark limit
\cite{Grinstein:2017nlq,Bigi:2017njr,Bernlochner:2019ldg}.  Hopefully
more precise data and improved lattice calculations will bring to a
resolution of this long-standing puzzle.

The measurements discussed above provide us with the normalization of
the UT and with the length of one of the non-unit sides,
$R_{b}$. Fortunately, we can complete the determination of the UT
using only tree-level decays by measuring the angle $\gamma$, defined
in eq.~(\ref{eq:UTdef}). The measurement of $\gamma$ can be achieved
by exploiting the interference between
$\bar{b} \to \bar{c} u \bar{q} \to f$ and
$\bar{b} \to \bar{u} c \bar{q} \to f$ transitions, where $q=d,s$ and
$f$ is a generic final state accessible through both decay chains
\cite{Gronau:1990ra,Atwood:1996ci,Atwood:2000ck,Giri:2003ty}. The
theoretical uncertainty in the extraction of $\gamma$ can be always
kept subdominant \cite{Brod:2013sga}, so future experimental progress
will have a strong impact on the UT analysis.

\begin{figure}[!tb]
  \centering
  \includegraphics[width=0.7\textwidth]{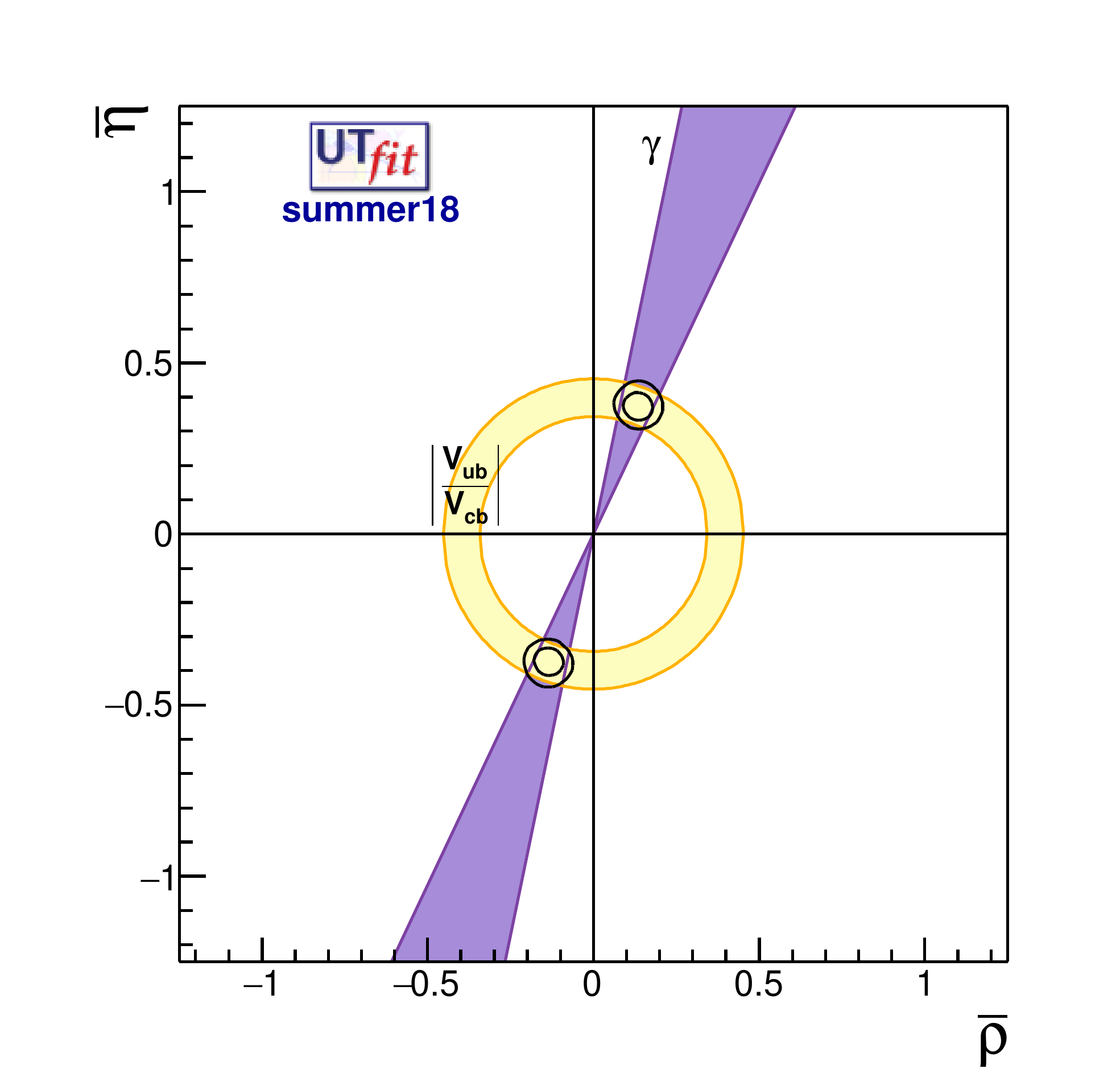}
  \caption{Current status of the UT determination from tree-level
    decays, from the UTfit Collaboration.}
  \label{fig:treeUT}
\end{figure}

Figure \ref{fig:treeUT} shows the current status of the UT determined
through tree level decays only. Notice that two regions in the
$\bar{\rho}-\bar{\eta}$ planes are selected, since we can determine
$\gamma$ only up to $\pm 180^{\circ}$. We will discuss below how this
ambiguity can be lifted using measurements of CP violation in
$B-\bar{B}$ mixing \cite{Laplace:2002ik}.

\subsection{Adding FCNC to the UT Analysis in the Standard Model and
  Beyond} 
\label{sec:loopUT}

\begin{figure}[!tb]
  \centering
  \includegraphics[width=0.7\textwidth]{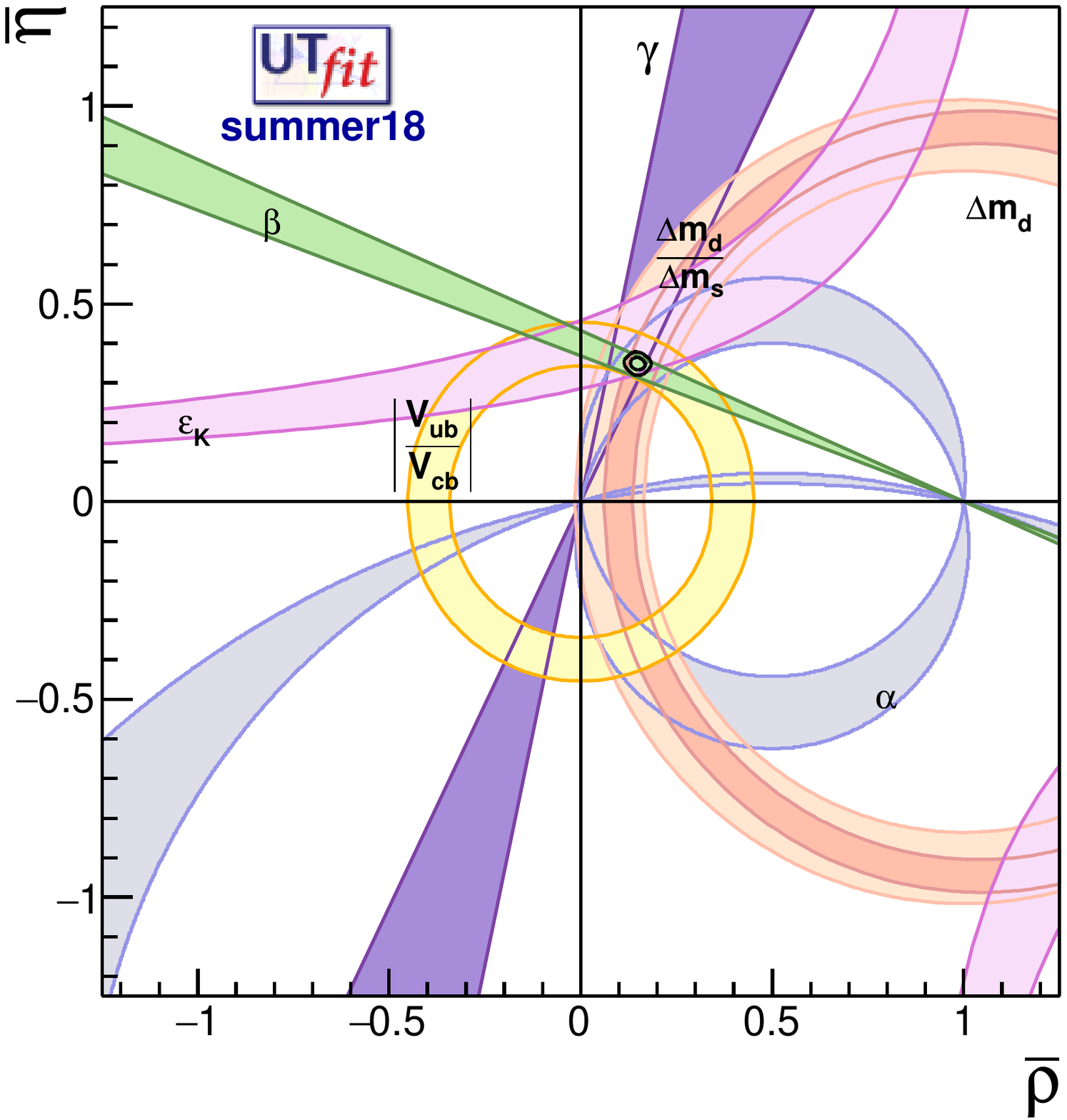}
  \caption{Current status of the UT determination in the SM, from the
    UTfit Collaboration.} 
  \label{fig:fullUT}
\end{figure}

We are now ready to add to the processes used in Sec.~\ref{sec:treeUT}
meson-antimeson mixing in $K$, $B_{d}$ and $B_{s}$ sectors, using
eq.~(\ref{eq:ReepsKphen}) for $\epsilon_{K}$, eq.~(\ref{eq:deltambd})
for $\Delta m_{B_{d,s}}$, eq.~(\ref{eq:acpjpsiksr0}) for $\sin 2
\beta$ and the results of Sec.~\ref{sec:pipi} for $\alpha$. This
allows to break the degeneracy between the first and third
quadrant. The global fit displays a very good consistency of all
observables within the SM, as can be seen from Fig.~\ref{fig:fullUT}. 

The consistency of the SM fit can be translated into constraints on NP
contributions to meson-antimeson mixing. Let us proceed in two
steps. First, we generalize the UT analysis by parameterizing the
relevant NP contributions. Second, we translate the constraints on NP
contributions into bounds on the scale of NP.

Following refs.~\cite{Bona:2005eu,Bona:2007vi}, we introduce the following
parameters to account for possible NP contributions to meson-antimeson
mixing:
\begin{dgroup*}
  \begin{dmath}[compact,label={eq:CBqphiBq}]
   C_{B_q}
  \, e^{2 i \phi_{B_q}} = \frac{M_{12}^{B_{q},\mathrm{full}}} {M_{12}^{B_{q},\mathrm{SM}}}\,, \qquad
  (q=d,s)
  \end{dmath}
  \begin{dmath}[compact,label={eq:CepsK}]
    C_{\epsilon_K} = \frac{\mathrm{Im}\,M_{12}^{K,\mathrm{full}}}
    {\mathrm{Im}\,M_{12}^{K,\mathrm{SM}}}\,.
  \end{dmath}
\end{dgroup*}
We can then immediately see how the observables entering the UT
analysis are affected:
\begin{align}
  \label{eq:dmbqnp} 
 \Delta m_{B_{q}} &= 
  C_{B_{q}} (\Delta m_{B_{q}})^{\mathrm{SM}} \\
  \label{eq:phidnp}
  \lambda_{J/\Psi K_{S}} &= e^{- 2 i (\beta + \phi_{B_{d}})}\,,\\
  \label{eq:phisnp}
  \lambda_{J/\Psi \phi} &= e^{2 i (\beta_{s} - \phi_{B_{s}})}\,,\\
  \label{eq:alphanp}
  \alpha_{\mathrm{exp}} &= \alpha - \phi_{B_{d}}\,,
\end{align}
where $\alpha_{\mathrm{exp}}$ denotes the value of $\alpha$ extracted
from $B_{d} \to \pi\pi$, $\rho\pi$ and $\rho\rho$ decays.\footnote{In
  the presence of NP contributions to loop-mediated SM processes, in
  the isospin or amplitude analysis one should allow the penguin
  contribution to have a phase different from the SM one
  \cite{Bona:2005eu}.}

As pointed out in ref.~\cite{Laplace:2002ik}, the presence of
$\phi_{B_{q}}$ can have a large impact on the semileptonic asymmetries
defined in eq.~(\ref{eq:asl}). As we have seen in
Sec.~\ref{sec:Bdmix}, in the SM the dominant contributions to $M_{12}$
and $\Gamma_{12}$ have the same CKM phase which drops in the ratio
$\Gamma_{12}/M_{12}$, so that Im$(\Gamma_{12}/M_{12})$ only arises from
subdominant GIM-suppressed contributions to  $\Gamma_{12}$. However,
if the mixing amplitude is affected by NP so that it gets an
additional phase $\phi_{B_{q}}$, the phase cancellation between $M_{12}$
and $\Gamma_{12}$ is spoiled and one gets a contribution to
Im$(\Gamma_{12}/M_{12})$ from the dominant term, proportional to
Re$(\Gamma_{12}/M_{12})^{\mathrm{SM}}/C_{B_{q}} \cos
2\phi_{B_{q}}$. It is then evident that the region in the third
quadrant in Fig.~\ref{fig:treeUT}, allowed at the tree-level, requires
a large value of $\phi_{B_{d}}$ which is ruled out at more than 95\%
probability by the experimental value of
$A_{\mathrm{SL}}^{B_{d}}$.\footnote{Also in this case when allowing
  for NP to be present in loop-mediated SM processes one should allow for penguin
  contributions to $\Gamma_{12}$ to have a phase different from the SM
  one \cite{Bona:2005eu}.}

\begin{figure}[!tb]
  \centering
  \includegraphics[width=0.4\textwidth]{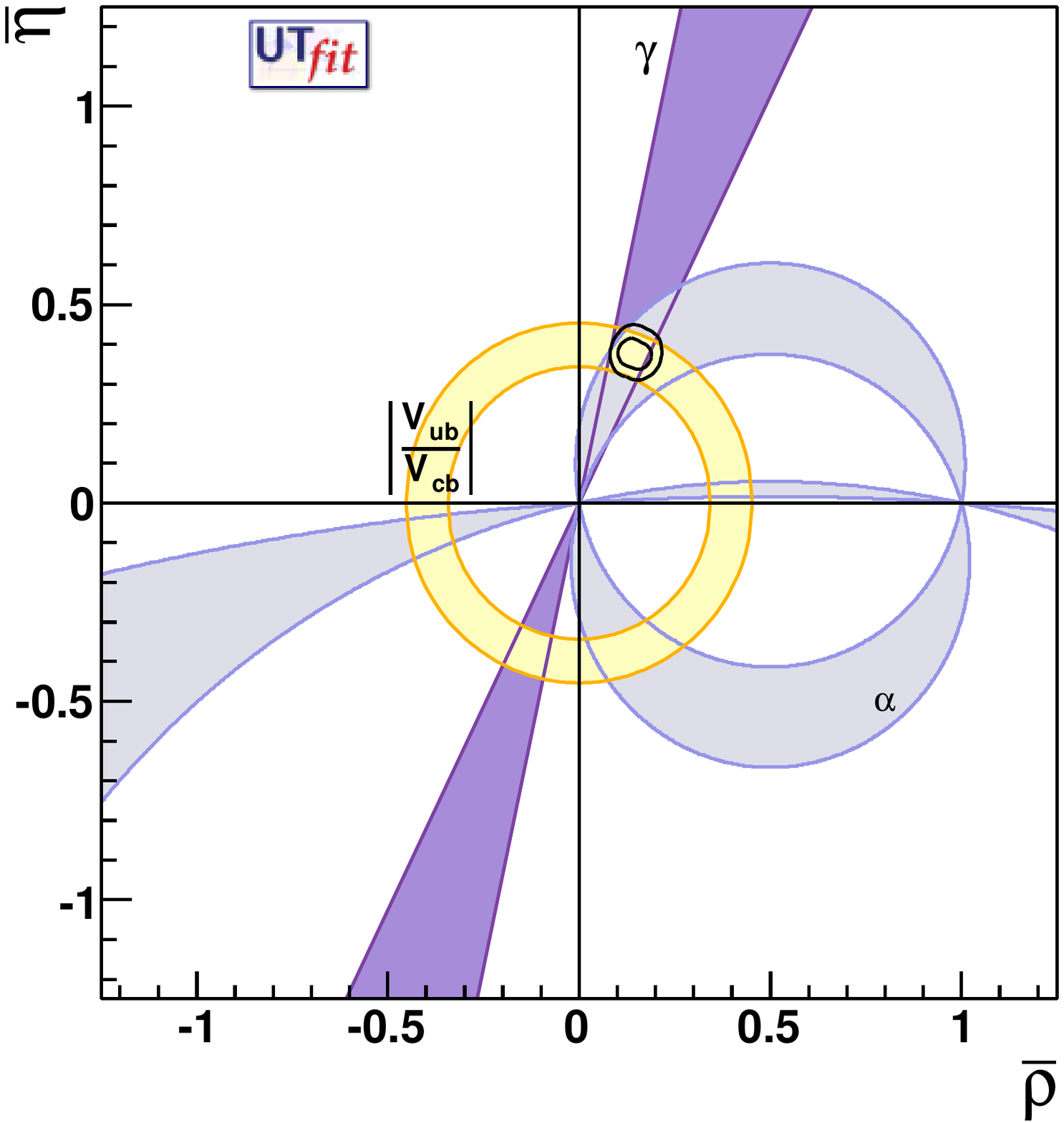}
  \includegraphics[width=0.4\textwidth]{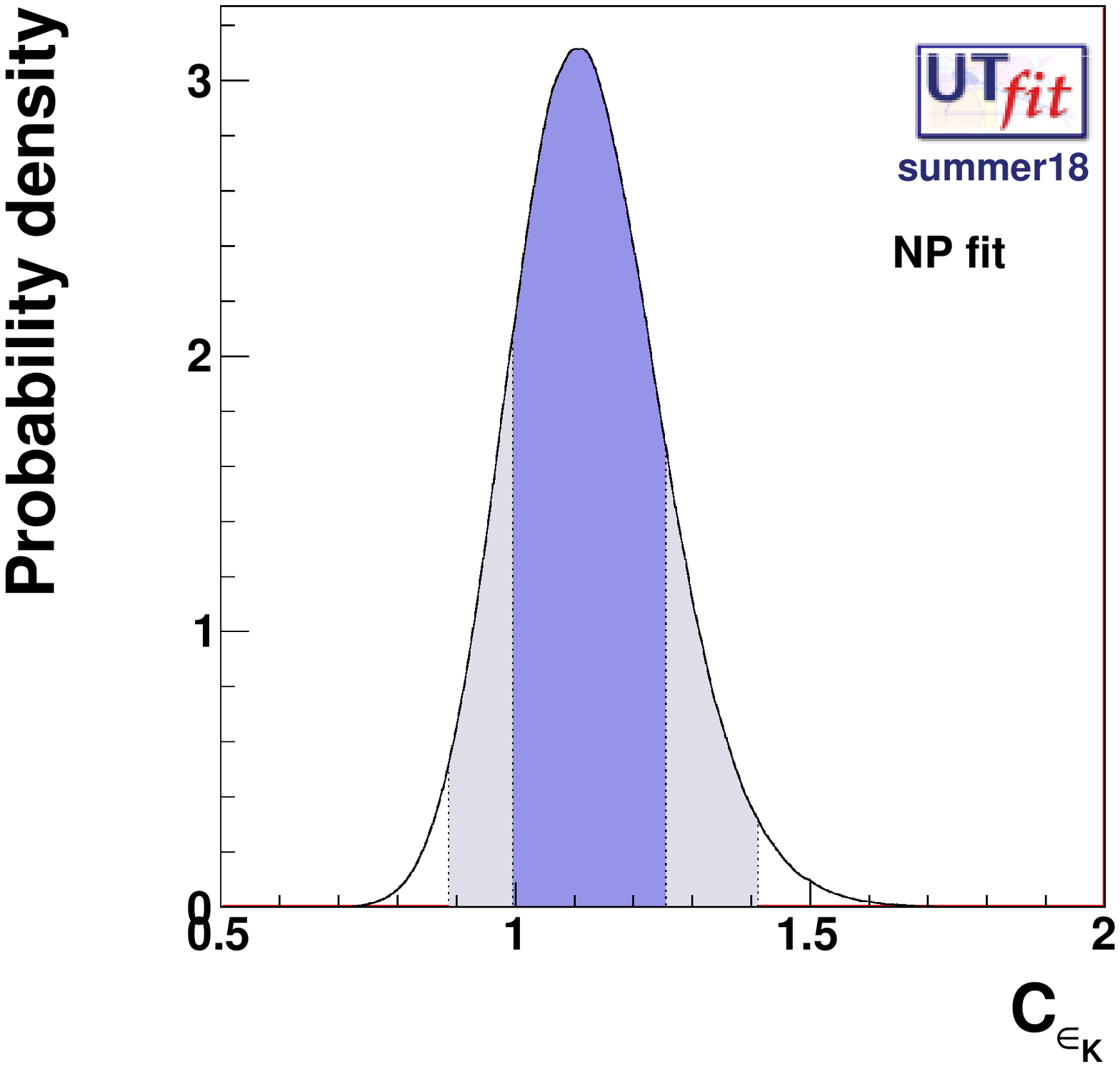}
  \includegraphics[width=0.4\textwidth]{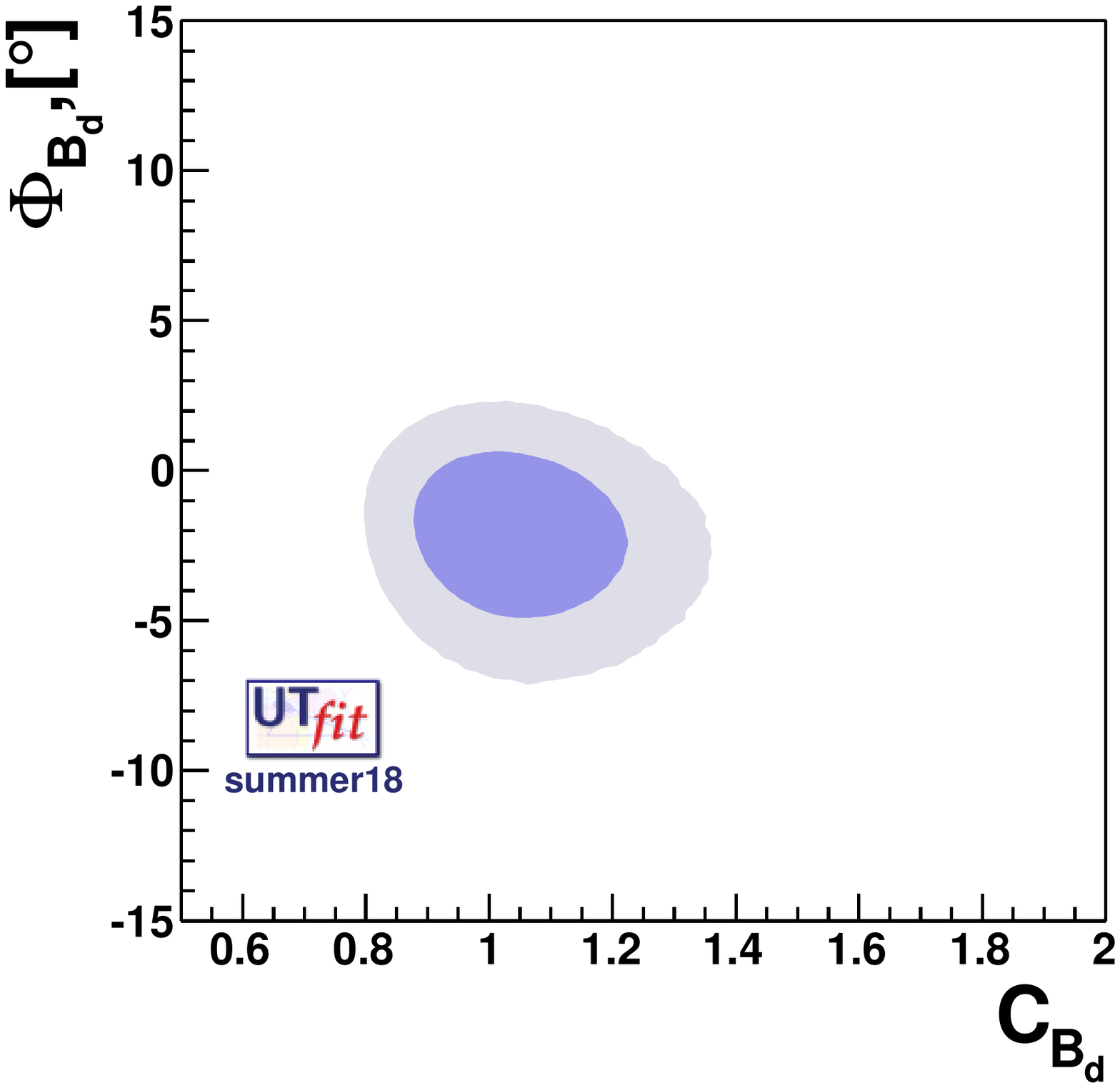}
  \includegraphics[width=0.4\textwidth]{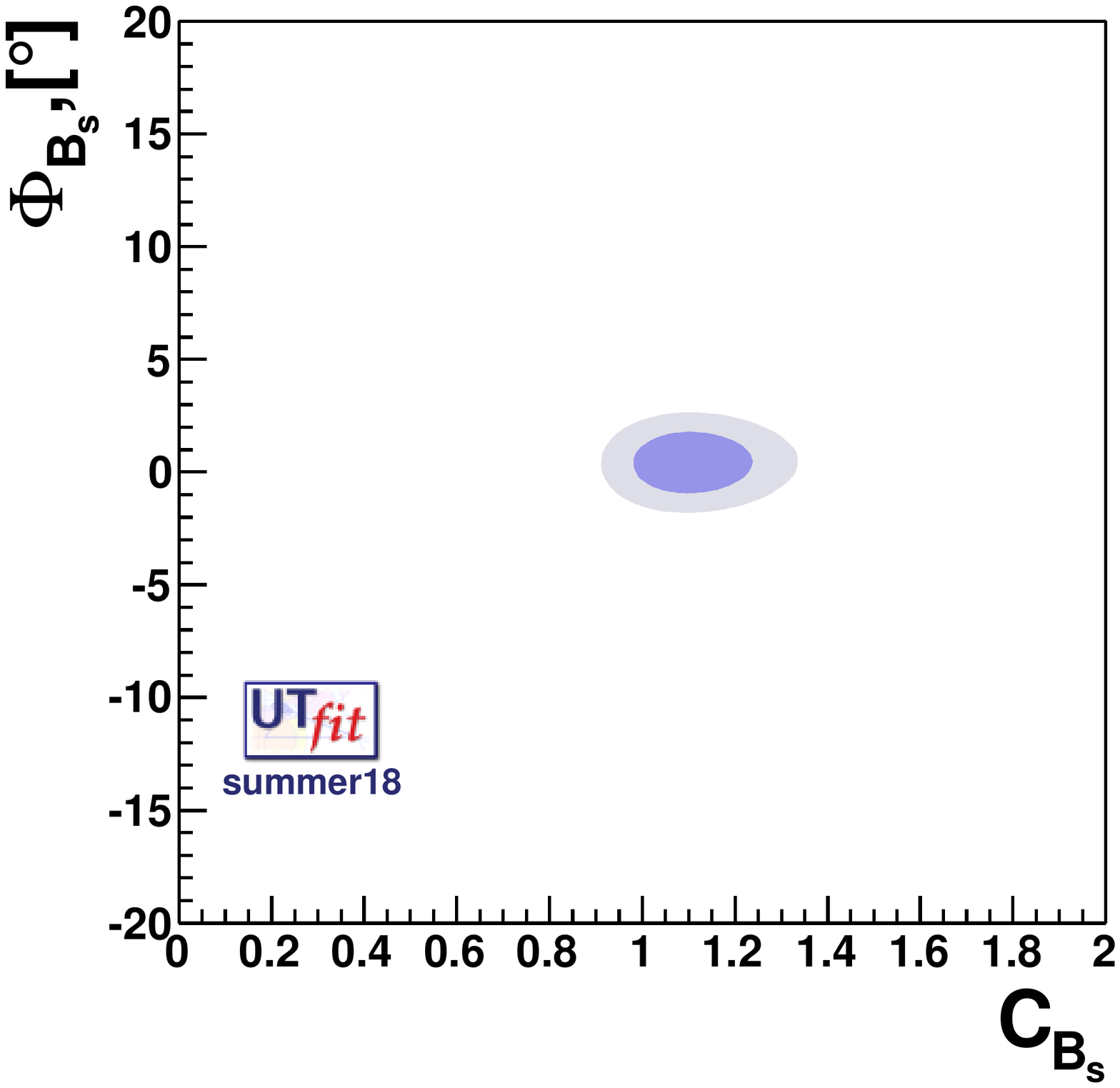}
  \caption{From left to right and from top to bottom: probability
    density functions for $(\bar{\rho},\bar{\eta})$,
    $C_{\epsilon_{K}}$, $(C_{B_{d}},\phi_{B_{d}})$,
    $(C_{B_{s}},\phi_{B_{s}})$. Darker (lighter) regions correspond to
    smallest $68\%$ $(95\%)$ probability regions.}
  \label{fig:NPUTA}
\end{figure}

\begin{figure}[!tb]
  \centering
  \includegraphics[width=0.4\textwidth]{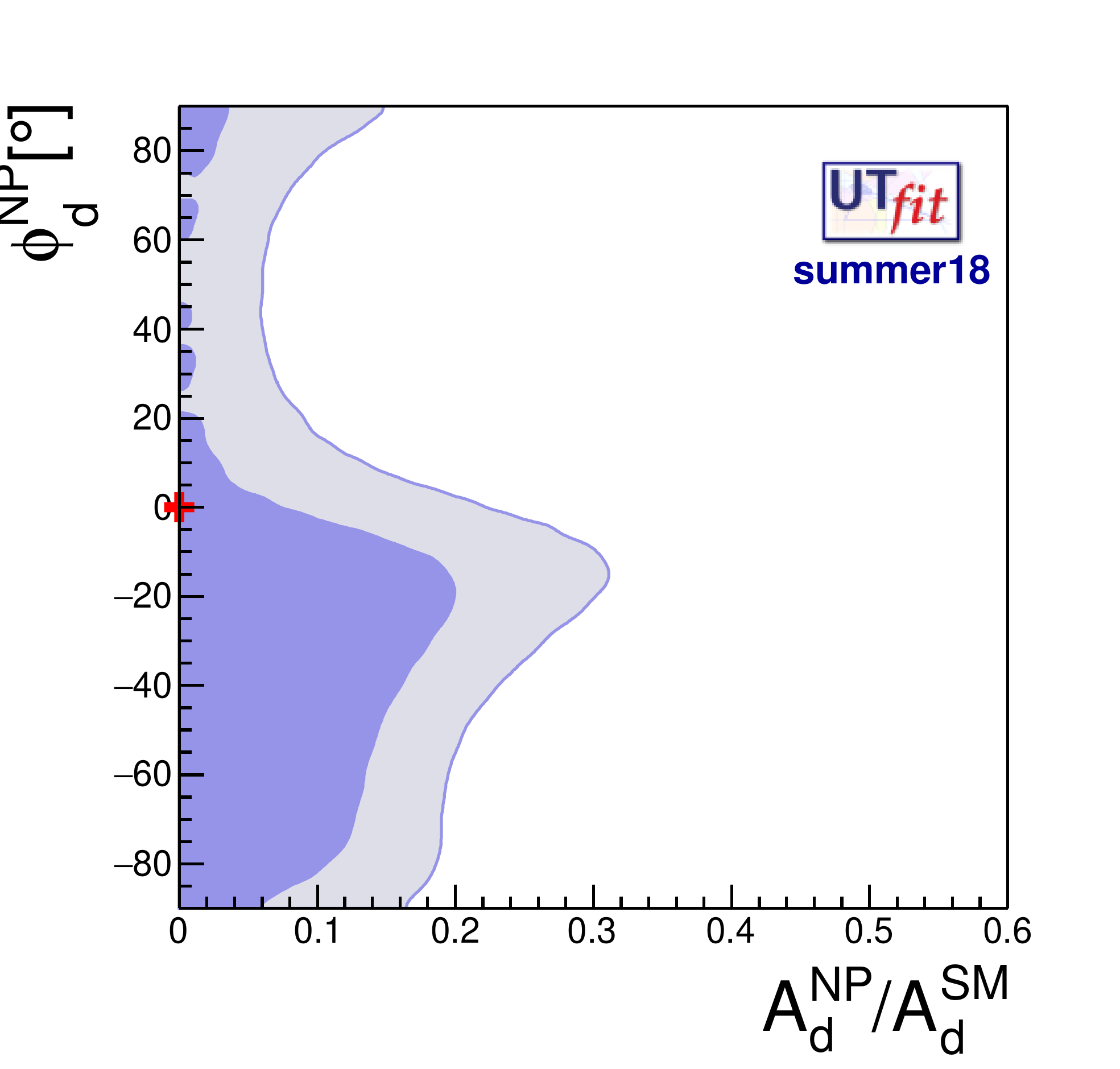}
  \includegraphics[width=0.4\textwidth]{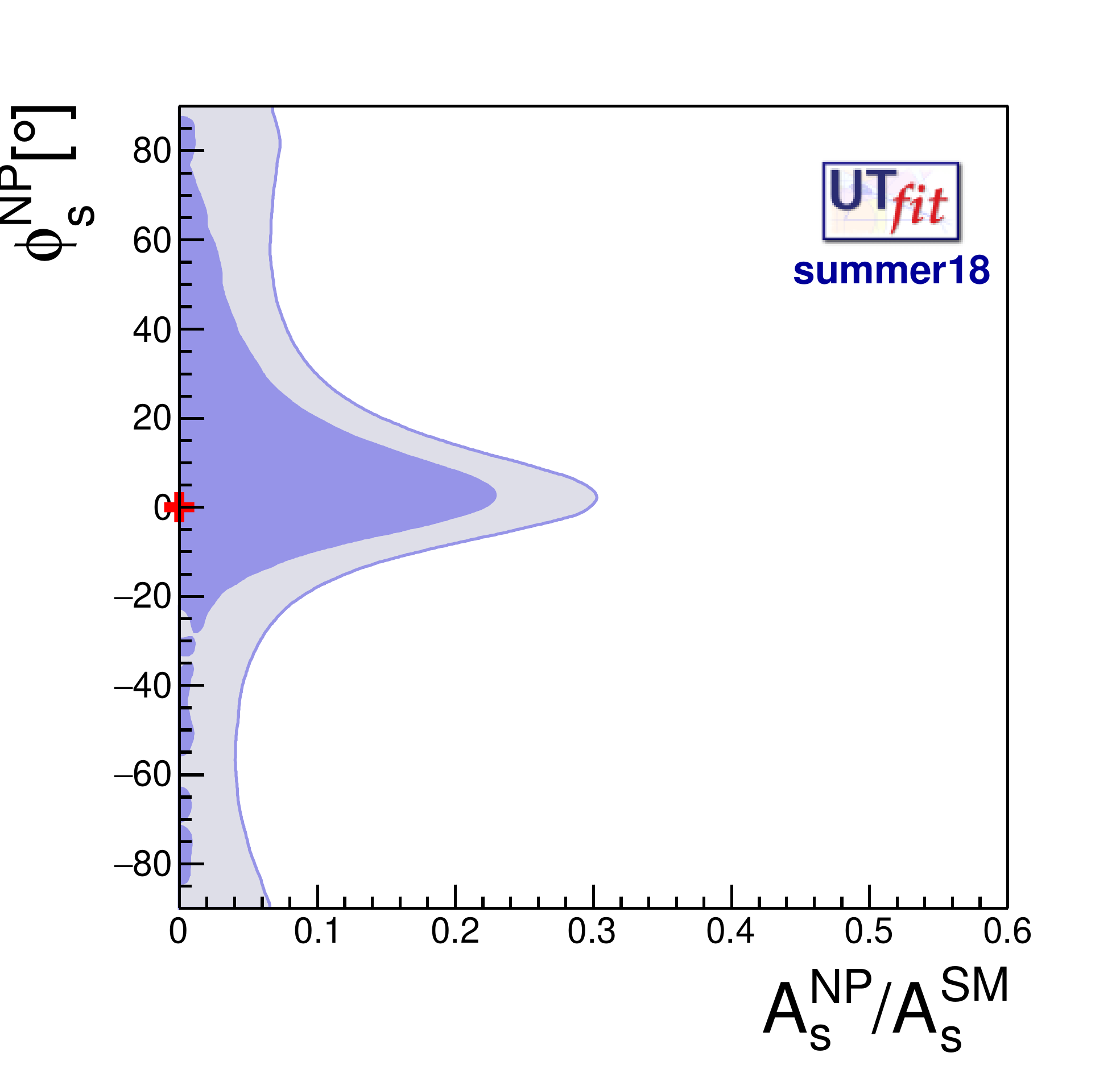}
  \caption{Probability density functions for
    $A^{\mathrm{NP}}_{q},\phi^{\mathrm{NP}}_{q}$. Darker (lighter)
    regions correspond to smallest $68\%$ $(95\%)$ probability
    regions.}
  \label{fig:achilleplots}
\end{figure}

We can therefore perform a simultaneous determination of the UT and of
the NP parameters introduced in eqs.~(\ref{eq:CBqphiBq}) and
(\ref{eq:CepsK}). The Summer 18 update from the UTfit collaboration is
reported in Fig.~\ref{fig:NPUTA}. It is instructive to extract from
the $C_{B_{q}}$ and $\phi_{B_{q}}$ parameters the absolute value and
phase of the NP contributions relative to the SM:
\begin{dmath}[compact,label={eq:achilleplots}]
  C_{B_{q}} e^{2 i \phi_{B_{q}}} = 1 + \frac{A^{\mathrm{NP}}_{q} e^{2
      i \phi^{\mathrm{NP}}_{q}}}{A^{\mathrm{SM}}_{q}}\,.
\end{dmath}
The current constraints on $A^{\mathrm{NP}}_{q}$ and
$\phi^{\mathrm{NP}}_{q}$ are reported in Fig.~\ref{fig:achilleplots}.
We see that our knowledge of the UT
in the presence of NP is roughly a factor of two worse than in the SM,
and that NP contributions to SM mixing amplitudes at the level of
$\approx 30-40\%$ are still allowed at $95\%$ probability, especially
if their phase does not differ too much from the SM one. This
shows that ample room is left for improvements, both from the
experimental and theoretical point of view, until we will be sensitive
to NP contributions in the flavour sector at the percent or
sub-percent level. However, given the combined loop and GIM
suppression of these observables in the SM, and given the hierarchical
structure of quark masses and mixings, already this relatively rough
sensitivity to NP contributions is able to provide us with the most
stringent constraints on the NP scale, as we will see below.

\subsection{Constraining the NP scale with
  \texorpdfstring{$\mathbf{\Delta F=2}$}{DeltaF=2} amplitudes}
\label{sec:NPscale}

We now combine the results on the $\Delta F=2$ effective Hamiltonian beyond
the SM in Secs.~\ref{sec:HeffDF2NP} and \ref{sec:DF2VIA} with the
constraints on NP contributions obtained in Secs.~\ref{sec:DDbar} and
\ref{sec:loopUT} to learn more on NP.

Assuming, as we did above, that NP has a negligible impact on
processes that arise at the tree-level in the SM, we write for
meson-antimeson mixing in all sectors
\begin{dmath}[compact,label={eq:npampc}]
  M_{12} = M_{12}^{\mathrm{SM}} + \frac{F_{i} L_{i}}{\Lambda^{2}}
  \langle M^{0} \lvert Q_{i} \rvert \overline{M}^{0} \rangle\,, \qquad
  \Gamma_{12} \simeq \Gamma_{12}^{\mathrm{SM}}\,,
\end{dmath}
where $F_{i}$ is a function of the (complex) NP flavour couplings, $L_i$
is a loop factor that is present in models with no tree-level Flavour
Changing Neutral Currents (FCNC), and $\Lambda$ is the scale of NP,
\emph{i.e.}  the typical mass of the new particles mediating the
$\Delta F=2$ transition. For a generic strongly-interacting theory
with arbitrary flavour structure, we expect $F_{i} \sim L_{i} \sim
\order{1}$ so that the allowed range for each of the NP contributions can be
immediately translated into a lower bound on $\Lambda$. Specific
assumptions on the flavour structure of NP, for example Minimal or
Next-to-Minimal Flavour Violation (MFV or NMFV), correspond to
particular choices of the $F_{i}$ functions, as detailed below. Notice
that in eq.~(\ref{eq:npampc}) the SM contribution
$M_{12}^{\mathrm{SM}}$ should be computed using for the CKM parameters
the results of the UT analysis in the presence of NP.

\begin{figure}[htbp]
\begin{center}      
  \includegraphics[width=0.9\textwidth]{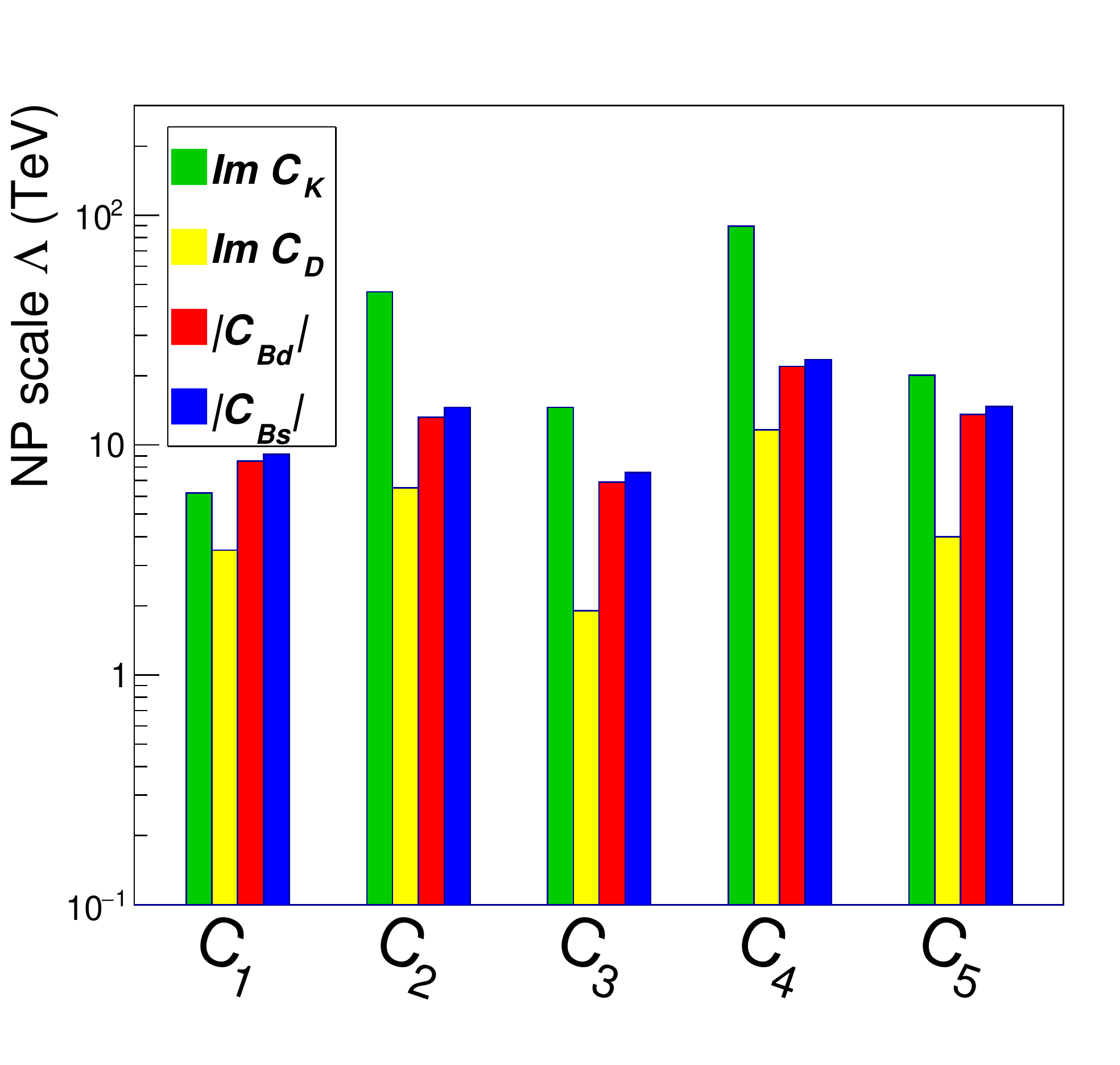}
\end{center}
\caption{Summary of the $95\%$ probability lower bound on the NP
    scale $\Lambda$ for NMFV. See the text for details}
\label{fig:NMFVbounds} 
\end{figure}

Switching on one operator at a time, assuming that
$F_i \sim L_i \sim \order{1}$, running its coefficient down from
the NP scale $\Lambda$ to the hadronic scale $\mu$ at which the
relevant matrix elements have been computed (see
refs.~\cite{Bouchard:2011xj,Bertone:2012cu,Boyle:2012qb,Bae:2013tca,Carrasco:2013zta,Carrasco:2014uya,Jang:2014aea,Jang:2015sla,Carrasco:2015pra,Garron:2016mva,Bazavov:2016nty,Boyle:2017skn}
for computations of the matrix elements for the full set of relevant
operators), computing its contribution according to
eq.~\ref{eq:npampc} and comparing it to the results presented in
Secs.~\ref{sec:DDbar} and \ref{sec:loopUT}, we obtain the $95\%$
probability lower bounds on $\Lambda$ presented in
Fig.~\ref{fig:df2bounds}. The bounds are dominated by CP violation in
$K-\bar{K}$ and $D-\bar{D}$ mixing, as expected from the extreme
suppression of these processes in the SM, and by the contributions of
the chirality-violating operators, which are enhanced both by the RG
evolution and by the matrix elements. These bounds are clearly beyond
the reach of any direct detection experiment, and strongly suggest us
that any NP close to the EW scale must have a hierarchical flavour
structure analogous to the SM one. One can then envisage the so-called
NMFV scenario, in which one has $F_{i} \simeq
F^{\mathrm{SM}}$, where $F^{\mathrm{SM}}$ is the CKM factor of the
relevant SM amplitude. The bounds on the NP scale in NMFV for
$L_{i}=1$ are reported in Fig.~\ref{fig:NMFVbounds}.
We see that the chiral and RG enhancement of $Q_{4}$ pushes the NP
scale to $O(100)$ TeV; to keep $\Lambda$ below $10$ TeV one must not
only enforce the same flavour structure of the SM, but also the same
chiral structure.

Requiring the same flavour and chiral structure of the SM corresponds
to the so-called MFV framework, initially formulated as the requirement of NP
contributions to FCNC observables being just a redefinition of the
loop function associated to the top-quark contribution, also known as
Constrained MFV (CMFV)
\cite{Gabrielli:1994ff,Buras:2000qz}. The CMFV hypothesis
allows for an improved determination of the UT in the presence of NP
and for several tests of consistency,
since it implies the independence on NP of ratios of observables in
which the top-mediated loop function drops, such as for example
$\Delta m_{B_{d}}/\Delta m_{B_{s}}$ \cite{Buras:2000dm,Buras:2003jf}.
More generally, one can observe that the requirement that NP has the
same flavour and chiral structure of the SM can be formulated in terms
of the flavour symmetry of the SM Lagrangian when the Yukawa couplings
are put to zero: the MFV hypothesis then amounts to the requirement
that Yukawa couplings are the only source of violation of the flavour
symmetry \cite{DAmbrosio:2002vsn}. This automatically leads to small
deviations from the SM for NP scales close to the EW one, provided that
Yukawa couplings are close to their ``SM'' value (\emph{i.e.}, to the
value they would take in the SM), while for example in two Higgs
doublets models with a large $v_{2}/v_{1}$ ratio of the two vacuum
expectation values one could face larger deviations enhanced by this
ratio.

While assuming that MFV holds exactly amounts to assuming that Yukawa
couplings are fundamental, thus giving up the hope of finding a dynamical
explanation of their hierarchical structure, it is certainly true that
a NP scale close to the EW one implies a flavour structure close to
MFV. This is clearly possible if the NP responsible for the origin of
the Yukawa couplings structure is much heavier than the EW scale.

\section{Conclusion and further reading}
\label{sec:concl}

The goal of these lectures is to allow the reader to get a first
idea of how we can probe NP with flavour observables. The level of
refinement of current forefront analyses in this field is clearly way
beyond the few basic elements here presented. Fortunately, several
excellent reviews are available on most of the topics sketched in the
previous sections. First of all, there are several other lectures on
the same topics which are more detailed, more general and more
inspired than ours, starting from the classic Les Houches lectures by
A.J. Buras \cite{Buras:1998raa}, SLAC and Trieste lectures by Y. Nir
\cite{Nir:1999mg} and from the excellent book by Branco, Lavoura and
Silva \cite{Branco:1999fs}, continuing to more recent lectures
\cite{Rosner:2000zk,Buras:2001pn,Nir:2001ge,Buras:2003jf,Buras:2003wd,Fleischer:2004xw,Silva:2004gz,Buras:2005xt,Fleischer:2006fx,Bigi:2007sq,Mannel:2008zza,Fleischer:2008uj,Nierste:2009wg,Nir:2010jr,Isidori:2013ez,Castelo-Branco:2014zua,Kou:2014twa,Grinstein:2015nya,Kamenik:2017znu,Blanke:2017ohr,Grossman:2017thq,Pich:2018njk,Zupan:2019uoi}.
For what concerns instead review articles, the NLO classic is
ref.~\cite{Buchalla:1995vs}, while among the many NP-oriented reviews
I find refs.~\cite{Grossman:1997pa,Isidori:2010kg} particularly
inspiring. Ref.~\cite{0907.3917} contains a remarkably complete
discussion of meson-antimeson mixing in the charm and bottom
sectors. Finally, while it was impossible to collect here all original
references for the topics we discussed, the reader is strongly
encouraged to read the original papers where all details can be found.

\section*{Acknowledgements}

It is a great pleasure to thank the organizers for providing such a
pleasant and stimulating environment, and the students for following
my lectures with undeserved interest. Any meaningful statement you may
find in this writeup originally comes from the teaching I received
from G. Martinelli, M. Ciuchini, E. Franco, L. Reina, A. Masiero,
R. Petronzio, A.J. Buras and A. Romanino (in chronological order, in
the form of lectures, notes and countless discussions). All confusing,
misleading or plainly wrong statements are instead my own
contribution. I am indebted to M. Ciuchini and to M. Valli for
carefully reading this manuscript, and to S. Davidson for her patience
and for her comments on the manuscript. This work has received funding from
the European Research Council (ERC) under the European Union's Horizon
2020 research and innovation program (grant agreement n$^o$ 772369).

\appendix

\section{Loops in Dimensional Regularization}
\label{sec:loops}

We collect here a few useful formul{\ae} for loop calculations in
dimensional regularization.

\subsection{Feynman parameters}
\label{sec:appFP}

Feynman parameters are useful to group denominators in loop
amplitudes. The basic formula is the following:
\begin{dmath}[compact,label={eq:Feyn1}]
  \frac{1}{AB} = \int_{0}^{1} \dd{x} \dd{y} \frac{\delta(x+y-1)}{
    \left[
      xA + y B
    \right]^{2}} = \int_{0}^{1} \dd{x} \frac{1}{
    \left[
      xA + (1-x) B
    \right]^{2}}\,.
\end{dmath}
It can be easily verified explicitly:
\begin{dmath}[compact,label={eq:Fen1proof}]
  \int_{0}^{1} \dd{x} \frac{1}{\left[
      xA + (1-x) B
    \right]^{2}} = - \frac{1}{A-B} 
  \left[
    \frac{1}{xA + (1-x)B}
  \right]_{0}^{1} = \frac{1}{A-B}
  \left(
    \frac{1}{A}-\frac{1}{B}
  \right)= \frac{1}{AB}\,.
\end{dmath}
We can raise the powers in the denominator by differentiating:
\begin{dmath}[compact,label={eq:Feyn1d}]
   \frac{1}{AB^{n}} =
   \frac{(-1)^{n-1}}{(n-1)!}\frac{\partial^{n-1}}{\partial B^{n-1}}
   \frac{1}{AB} = \frac{(-1)^{n-1}}{(n-1)!} \int_{0}^{1}
   \dd{x} \dd{y} \frac{\delta(x+y-1) n! y^{n-1} (-1)^{n-1}}{
    \left[
      xA + y B
    \right]^{n+1}} = \int_{0}^{1}
   \dd{x} \dd{y} \frac{\delta(x+y-1) n y^{n-1}}{
    \left[
      xA + y B
    \right]^{n+1}}\,.
\end{dmath}
We can add further terms in the denominator by iterating with
eqs.~(\ref{eq:Feyn1}) and (\ref{eq:Feyn1d}):
\begin{dmath}[compact,label={eq:Feyn2}]
  \frac{1}{ABC} = \frac{1}{AB} \frac{1}{C} = \frac{1}{C} \int_{0}^{1}
  \dd{x} \dd{y} \frac{\delta(x+y-1)}{ 
    \left[
      xA + y B
    \right]^{2}} = \int_{0}^{1} \dd{w} \dd{z} \dd{x}
  \dd{y}
  2w \frac{\delta(w+z-1)\delta(x+y-1)}{ 
    \left[
      zC + w(xA + y B)
    \right]^{3}} \xlongequal{\stackon{$y^\prime=wy$}{$x^\prime=wx$}}
  \int_{0}^{1} \dd{w} \dd{z}\delta(w+z-1) \int_{0}^{w} \dd{x^{\prime}}
  \dd{y^{\prime}}
  2 \frac{\delta(x^{\prime}+y^{\prime}-w)}{ 
    \left[
      zC + x^{\prime }A + y^{\prime} B
    \right]^{3}} =  \int_{0}^{1} \dd{z} \int_{0}^{1-z} \dd{x^{\prime}}
  \dd{y^{\prime}}
  2 \frac{\delta(x^{\prime}+y^{\prime}+z-1)}{ 
    \left[
      zC + x^{\prime }A + y^{\prime} B
    \right]^{3}} = \int_{0}^{1} \dd{x}\dd{y}\dd{z} \frac{2 \delta(x+y+z-1)}{\left[
      x A + y B + z C
    \right]^{3}}\,.
\end{dmath}
We thus obtain the general formula
\begin{dmath}[compact,label={eq:Feyngen}]
  \frac{1}{A_{1}^{m_{1}}A_{2}^{m_{2}}\ldots A_{n}^{M_{n}}} =
  \int_{0}^{1} \dd{x_{1}} \dd{x_{2}}\ldots \dd{x_{n}} \delta(\sum_{i}
  x_{i} -1) \frac{\prod_{i} x_{i}^{{m_{i}-1}}}{
    \left[
      \sum_{i}x_{i}A_{i}
    \right]^{\sum_i m_{i}}} \frac{\Gamma(\sum_{i} m_{i})}{\prod_{i} \Gamma(m_{i})}\,.
\end{dmath}

\subsection{Loop integrals}
\label{sec:appI}

\subsubsection{Momentum shift}

After grouping the denominators with Feynman parameters using
eq.~(\ref{eq:Feyngen}), the denominator will contain in general not
only the square of the loop momentum and constant terms, but also
terms linear in the loop momentum (from dot products with external
momenta). We get rid of linear terms in the denominator by performing
a shift of the loop momentum, which brings us to the general form
\begin{dmath}[compact,label={eq:shift}]
  \int \frac{\dd[d]{k}}{(2\pi)^{d}} \frac{k^{\mu_{1}}\ldots
    k^{\mu_{n}}}{(k^{2}-D + i\epsilon)^{m}}\,.
\end{dmath}
Then integrals with odd powers of $k$ in the numerator vanish by
symmetry, and we are left with even powers of $k$ only.

\subsubsection{Wick rotation}

The $i\epsilon$ term in eq.~(\ref{eq:shift}) is there to remind us
that we should be careful about the poles of the propagators entering
the diagram we are calculating. We can form a closed contour in the
complex $K^{0}$ plane by going from $-\infty$ to $+\infty$ on the real
axis, from $+\infty$ to $-\infty$ on the imaginary axis, and closing
the contour with two arcs at infinity from the real to the imaginary
axis.  Noting that the $i\epsilon$ prescription moves the poles to the
second and fourth quadrant, so that they are not inside the contour,
and neglecting the contribution at infinity, we see that the integral
on the real axis is equal to the integral on the imaginary axis. We
then go to the Euclidean with $k^{0}=ik^{0}_{E}$, and obtain
\begin{dmath}[compact,label={eq:Wick}]
  \int \frac{\dd[d]{k}}{(2\pi)^{d}} \frac{1}{(k^{2}-D)^{m}} = i
  (-1)^{m} \int \frac{\dd[d]{k_{E}}}{(2\pi)^{d}} \frac{1}{(k_{E}^{2}+D)^{m}}
\end{dmath}

\subsubsection{Angular integration}

Let us now split the integration pulling out the angular one:
\begin{dmath}[compact,label={eq:angularI}]
  \int \frac{\dd[d]{k_{E}}}{(2\pi)^{d}} \frac{1}{(k_{E}^{2}+D)^{m}} =
  \int \frac{\dd{\Omega_{d}}}{(2\pi)^{d}} \int k_{E}^{d-1} \dd{k_{E}}
  \frac{1}{(k_{E}^{2}+D)^{m}}\,.
\end{dmath}
We can obtain the angular term using a Gaussian integral:
\begin{dmath}[compact,label={eq:domegaint}]
  \left(
    \sqrt{\pi}
  \right)^{d} = 
  \left(
    \int_{-\infty}^{\infty} \dd{x} e^{-x^{2}} 
  \right)^{d} = 
  \int_{-\infty}^{\infty} \dd[d]{x} e^{-x^{2}} = \int \dd{\Omega_{d}}
  \int_{0}^{\infty} \dd{x} x^{d-1} e^{-x^{2}} =  \int \dd{\Omega_{d}}
  \frac{1}{2} \int_{0}^{\infty} \dd{x^{2}} (x^{2})^{\frac{d}{2}-1}
  e^{-x^{2}} = \int \dd{\Omega_{d}} \frac{1}{2} \Gamma
  \left(
    \frac{d}{2}
  \right)
\end{dmath}
so that
\begin{dmath}[compact,label={eq:domegares}]
  \int \dd{\Omega_{d}} = \frac{2 \pi^{\frac{d}{2}}}{\Gamma
    \left(
      \frac{d}{2}
    \right)}\,.
\end{dmath}

\subsubsection{Momentum integration}

We now turn to the integral over the absolute value of the Euclidean
momentum:
\begin{dmath}[compact,label={eq:Imom}]
  \int_{0}^{\infty} k_{E}^{d-1} \dd{k_{E}}
  \frac{1}{(k_{E}^{2}+D)^{m}} = \frac{1}{2} \int_{0}^{\infty}
  (k_{E}^{2})^{\frac{d}{2}-1}
  \dd{k^{2}_{E}}
  \frac{1}{(k_{E}^{2}+D)^{m}}\xlongequal{x=\frac{D}{k_{E}^{2}+D}}
  -\frac{1}{2} \int_{1}^{0} D x^{-2} \dd{x} \frac{(D/x)^{\frac{d}{2}-1}
  (1-x)^{\frac{d}{2}-1} }{(D/x)^{m}} = \frac{1}{2} D^{\frac{d}{2}-m}
\int_{0}^{1}  \dd{x} x^{m-\frac{d}{2}-1} (1-x)^{\frac{d}{2}-1} =
\frac{1}{2} 
\left(
  \frac{1}{D}
\right)^{m-\frac{d}{2}} B(m-\frac{d}{2},\frac{d}{2})\,,
\end{dmath}
where
\begin{dmath}[compact,label={eq:Bfunc}]
  \int_{0}^{1}  \dd{x} x^{\alpha-1} (1-x)^{\beta-1} = B(\alpha,\beta)
  = \frac{\Gamma(\alpha)\Gamma(\beta)}{\Gamma(\alpha+\beta)}\,. 
\end{dmath}

\subsubsection{Expansion for \texorpdfstring{$d = 4-2\epsilon$}{d4m2e}}

We now put together the results in eqs.~(\ref{eq:Wick}),
(\ref{eq:angularI}), (\ref{eq:domegares}) and (\ref{eq:Imom}):
\begin{dmath}[compact,label={eq:intfinal}]
  \int \frac{\dd[d]{k}}{(2\pi)^{d}} \frac{1}{(k^{2}-D)^{m}} =
  \frac{i}{(4\pi)^{\frac{d}{2}}} (-1)^{m}
  \left(
    \frac{1}{D}
  \right)^{m-\frac{d}{2}} \frac{\Gamma(m-\frac{d}{2})}{\Gamma(m)}
\end{dmath}
 and
 expand for $d$ close to four in the small parameter $\epsilon$:
\begin{dmath}[compact,label={eq:inteps}]
  \int \frac{\dd[d]{k}}{(2\pi)^{d}} \frac{1}{(k^{2}-D)^{m}} =
  \frac{i}{16\pi^{2}} 
  \left(-
    \frac{1}{D}
  \right)^{m-2} 
  \left(
    \frac{1}{4\pi D}
  \right)^{\epsilon} \frac{\Gamma(m-2+\epsilon)}{\Gamma(m)}\,.
\end{dmath}
Using the expansion of Euler $\Gamma$
\begin{align}
  \label{eq:EulerG}
  \Gamma(\epsilon) &= \frac{1}{\epsilon} - \gamma_{E} + \order{\epsilon}\,,\\
  \Gamma(-n + \epsilon) &= \frac{(-1)^{n}}{n!} 
  \left(
    \frac{1}{\epsilon} -\gamma_{E} + 1 + \ldots + \frac{1}{n} + \order{\epsilon}
  \right)\,,\nonumber
\end{align}
one obtains for example
\begin{dmath}[compact,label={eq:inteps2}]
  \int \frac{\dd[d]{k}}{(2\pi)^{d}} \frac{1}{(k^{2}-D)^{2}} =
  \frac{i}{16\pi^{2}} 
  \left(
    \frac{1}{\epsilon} -\gamma_{E} - \ln{4\pi} - \ln{D} + \order{\epsilon}
  \right) \equiv \frac{i}{16\pi^{2}} 
  \left(
    \frac{1}{\bar{\epsilon}} - \ln{D} + \order{\bar{\epsilon}}
  \right)\,,
\end{dmath}
where we introduced for convenience the parameter $\bar{\epsilon}$
defined in eq.~(\ref{eq:epsbar}).

\subsubsection{Some useful integrals}

The reader may find the following list of integrals useful:
\begin{dgroup*}
  \begin{dmath}[compact,label={eq:int0}]
    \int \frac{\dd[d]{k}}{(2\pi)^{d}} \frac{1}{(k^{2}-D)^{m}} =
  \frac{i}{(4\pi)^{\frac{d}{2}}} (-1)^{m}
  \left(
    \frac{1}{D}
  \right)^{m-\frac{d}{2}} \frac{\Gamma(m-\frac{d}{2})}{\Gamma(m)}\,,
  \end{dmath}
  \begin{dmath}[compact,label={eq:int1}]
    \int \frac{\dd[d]{k}}{(2\pi)^{d}} \frac{k^{2}}{(k^{2}-D)^{m}} =
  \frac{i}{(4\pi)^{\frac{d}{2}}} (-1)^{m-1} \frac{d}{2}
  \left(
    \frac{1}{D}
  \right)^{m-\frac{d}{2}-1} \frac{\Gamma(m-\frac{d}{2}-1)}{\Gamma(m)}\,,
  \end{dmath}
  \begin{dmath}[compact,label={eq:int11}]
    \int \frac{\dd[d]{k}}{(2\pi)^{d}} \frac{k^{\mu}k^{\nu}}{(k^{2}-D)^{m}} =
  \frac{i}{(4\pi)^{\frac{d}{2}}} (-1)^{m-1} \frac{g^{\mu\nu}}{2}
  \left(
    \frac{1}{D}
  \right)^{m-\frac{d}{2}-1} \frac{\Gamma(m-\frac{d}{2}-1)}{\Gamma(m)}\,,
  \end{dmath}
  \begin{dmath}[compact,label={eq:int2}]
    \int \frac{\dd[d]{k}}{(2\pi)^{d}} \frac{(k^{2})^{2}}{(k^{2}-D)^{m}} =
  \frac{i}{(4\pi)^{\frac{d}{2}}} (-1)^{m} \frac{d(d+2)}{4}
  \left(
    \frac{1}{D}
  \right)^{m-\frac{d}{2}-2} \frac{\Gamma(m-\frac{d}{2}-2)}{\Gamma(m)}\,,
  \end{dmath}
  \begin{dmath}[compact,label={eq:int22}]
    \int \frac{\dd[d]{k}}{(2\pi)^{d}}
    \frac{k^{\mu}k^{\nu}k^{\rho}k^{\sigma}}{(k^{2}-D)^{m}} = 
  \frac{i}{(4\pi)^{\frac{d}{2}}} (-1)^{m}
  \frac{g^{\mu\nu}g^{\rho\sigma} +g^{\mu\rho}g^{\nu\sigma} +
    g^{\mu\sigma} g^{\nu\rho}}{4}
  \left(
    \frac{1}{D}
  \right)^{m-\frac{d}{2}-2} \frac{\Gamma(m-\frac{d}{2}-2)}{\Gamma(m)}\,.
  \end{dmath}
\end{dgroup*}

\bibliographystyle{JHEP}
\bibliography{../../biblio/hepbiblio}

\end{document}